\documentclass{aa}  

\usepackage{graphicx}
\usepackage{amsmath}
\usepackage{amssymb}
\usepackage{amsfonts}
\usepackage{txfonts}

\usepackage{mathtools}

\begin{document}

    \title{Models for the 3-D axisymmetric gravitational potential \\ of the Milky Way Galaxy}
   
    \subtitle{A detailed modelling of the Galactic disk}

   \author{D. A. Barros
   		  \inst{1}\fnmsep\thanks{\email{douglas.barros@iag.usp.br}},                
   		  J. R. D. L\' epine\inst{1}
          \and
          W. S. Dias\inst{2}
          }
          
   \institute{Instituto de Astronomia, Geof\' isica e Ci\^encias Atmosf\' ericas, Universidade de S\~ao Paulo, Cidade Universit\' aria, S\~ao Paulo 05508-090, SP, Brasil
         \and
             UNIFEI, Instituto de Ci\^encias Exatas, Universidade Federal de Itajub\' a, Av. BPS 1303 Pinheirinho, 37500-903 Itajub\' a, MG, Brasil   
         }

   \date{Received ..... 2016; accepted ..... 2016}

 
  \abstract
  {} 
   {Galaxy mass models based on simple and analytical functions for the density and potential pairs have been widely proposed in the literature. Disk models constrained by kinematic data alone give information on the global disk structure only very near the Galactic plane. We attempt to circumvent this issue by constructing disk mass models whose three-dimensional structures are constrained by a recent Galactic star counts model in the near-infrared and also by observations of the hydrogen distribution in the disk. Our main aim is to provide models for the gravitational potential of the Galaxy that are fully analytical but also with a more realistic description of the density distribution in the disk component.}
   {From the disk model directly based on the observations (here divided into the thin and thick stellar disks and the H\,{\scriptsize I} and H$_2$ disks subcomponents),
we produce fitted mass models by combining three Miyamoto-Nagai disk profiles of any ``model order'' (1, 2, or 3) for each disk subcomponent. The Miyamoto-Nagai disks are combined with models for the bulge and ``dark halo'' components and the total set of parameters is adjusted by observational kinematic constraints. A model which includes a ring density structure in the disk, beyond the solar Galactic radius, is also investigated.}
   {The Galactic mass models return very good matches to the imposed observational constraints. In particular, the model with the ring density structure provides a greater contribution of the disk to the rotational support inside the solar circle. The gravitational potential models and their associated force-fields are described in analytically closed forms.}
   {The simple and analytical models for the mass distribution in the Milky Way and their associated three-dimensional gravitational potential are able to reproduce the observed kinematic constraints, and in addition, they are also compatible with our best knowledge of the stellar and gas distributions in the disk component. The gravitational potential models are suited for investigations of orbits in the Galactic disk.}

   
   \titlerunning{Models for the 3-D gravitational potential of the Galaxy}
   
   \authorrunning{D. Barros, J. L\' epine \& W. Dias}
   
   \maketitle
%

\section{Introduction}
\label{intro}

Reliable models for the gravitational potential of the Galaxy are mandatory when studies of the structure and evolution of the Galactic mass components rely upon the characteristics of the orbits of their stellar content. 
In this sense, Galaxy mass models are regarded as the simplest way of assessing and understanding the global structure of the main Galactic components, providing 
a great insight into their mass distribution  
once a good agreement between the model predictions and the observations is obtained.
A pioneer Galactic mass model was that of \citet{Schmidt1956}, contemporary of the early years of the development of radio astronomy and the first studies of the large-scale structure of the Milky Way. With the subsequent improvement of the observational data, updated mass models have been undertaken by several authors (e.g., \citealt[among others]{Bahcall_Soneira1980,Caldwell_Ostriker1981,Rohlfs_Kreitschmann1988}), and with the advent of the Hipparcos mission and large-scale surveys in the optical and near-infrared, new observational constraints have been adopted in the more recent Galaxy mass models (e.g., \citealt{Dehnen_Binney1998,Lepine_Leroy2000,Robin2003,Polido2013}).

In order to evaluate the capability of a given mass model in reproducing some observables, the force-field associated with the resulting gravitational potential has to be compared with available dynamical constraints such as the radial force in the plane given by the rotation curve, as well as the force perpendicular to the plane of the disk along a given range of Galactic radii. 
Regarding the latter one, the associated mass-surface density, to our better knowledge, is the one integrated up to the height of 1.1 kpc of the Galactic mid-plane (\citealt{Kuijken_Gilmore1991,Bovy_Rix2013}), and as pointed out by \citet{Binney_Merrifield1998}, this constraint 
is not able to provide much information about the mass distribution some kiloparsecs above the plane. Due to these shortcomings, a degeneracy in the set of best models is observed, which means that different mass models are able to reproduce the kinematic information of the observed data equally well. As stated by \citet{McMillan2011}, one possible way of circumventing such obstacles is by combining the kinematic data with star counts to improve the Galactic potential models and its force-field above the plane. 

Regarding the use of a Galactic potential model for the purpose of orbit calculations, one which has been widely adopted is that of \citet{Allen_Santillan1991}. Such model has the attractive characteristics of being mathematically simple and completely analytical, with closed forms for the potential and density, assuring both fast and accurate orbit calculations. \citet{Irrgang2013} have recalibrated the \citet{Allen_Santillan1991} model parameters using new and improved observational constraints.  
 
The main goal of the present work is to provide a fully-analytical, three-dimensional description of the gravitational potential of the Galaxy, but with the novelty of expending considerable efforts in a detailed modelling of the disk component.
The basic new aspects of the present Galactic mass model, all of which related to the disk modelling process, can be summarized in the following way:

\begin{itemize}
\item the structural parameters of the disk - scale-length, scale-height, radial scale of the central `hole' - are based on the Galactic star counts model in the near-infrared developed by \citet*[hereafter PJL]{Polido2013} for the case of the stellar disk component; for the gaseous disk counterpart, we adopt recent values returned by surveys of the distribution of hydrogen, atomic H\,{\scriptsize I} and molecular H$_2$, in the disk;

\item the density and potential of the disk components are modelled by the commonly used Miyamoto-Nagai disk profiles (equations 4 and 5 of \citealt{Miyamoto_Nagai1975}), but here we also attempt to make use of the higher ``model orders'' 2 and 3 of Miyamoto-Nagai disks (equations 6, 7, 8 and 9 of the above-referred paper) in order to better fit some of the disk subcomponents. The approach followed for the construction of the Miyamoto-Nagai disks is based on the one presented by \citet{Smith2015};

\item a model with a ring density structure added to the disk density profile is studied, with the ring feature placed somewhat beyond the solar Galactic radius. The inclusion of such ring structure is motivated by the attempt of modelling the local dip in the observed Galactic rotation curve also placed a little beyond the solar orbit radius. An explanation for the existence of such ring density structure is given by \citet*[hereafter BLJ]{Barros2013}.
\end{itemize}

The organization of this paper is as follows: in Sect.~\ref{MW_disk_mass_models}, we present the details of the mass models of the Galactic disk and the steps through the construction of Miyamoto-Nagai disks versions of the `observed' ones. In Sect.~\ref{MW_models}, we give the expressions for the bulge and dark halo components, as well as the functional form for the gravitational potential associated with the ring density structure. The group of observational constraints adopted for the fitting of the models are presented in Sect.~\ref{obs_constr}, while the fitting scheme and the estimation of uncertainties are presented in Sect.~\ref{fit_proc}. In Sect.~\ref{result_discuss}, we analyse the results of each mass model by a direct comparison with other models in the literature.  
Concluding remarks are drawn in the closing Sect.~\ref{conclusions}.


\section{Mass models for the disk of the Galaxy}
\label{MW_disk_mass_models}

We model the Milky Way's disk separating it into the stellar (thin and thick disks) and gaseous (H\,{\scriptsize I} and H$_2$ disks) components. In the following subsections, we present the observational basis taken as prior information to constrain the values of the parameters of the models, as well as the steps for the construction of the mass and potential disk models. In this paper, we use the cylindrical coordinates ($R$, $\phi$, $z$) for the density and potential expressions. The solar Galactic radius is denoted as $R_0$.

\subsection{The `observation-based' disk model}
\label{observ_basis}

\subsubsection{The stellar component}
\label{stellar_comp}

Our models for the density distribution in the thin and thick stellar disks of the Galaxy are based on the structural 
disk parameters presented by PJL. These authors have performed a star counts model of the Galaxy using near-infrared data of the 2MASS survey (\citealt{Skrutskie2006}), with lines of sight covering the entire sky and including the Galactic plane. The exploration of the parameter space and the estimation of its optimal values were done by the authors with the usage of statistical methods such as the Markov Chain Monte Carlo (MCMC) (\citealt{Gilks1996}) and the Nested Sampling (NS) algorithm (\citealt{Skilling2004}). PJL have modelled the radial profile of the density of each subcomponent of the stellar disk by a modified exponential law, based on the Galactic model of \citet{Lepine_Leroy2000}. Such profile is equivalent to the Freeman's Type II disk brightness profile, which contains a depletion in the center, with respect to a pure exponential law (\citealt{Freeman1970,Kormendy1977}). The stellar surface densities $\Sigma_{\mathrm{d}_{\bigstar}}$ for the thin and thick disks can then be written as:

\begin{equation}
\label{eq:sigma_thin_thick}
\Sigma_{\mathrm{d}_{\bigstar,\,i}}(R)=\Sigma_{0\mathrm{d}_{\bigstar,\,i}}\,\exp\left[-\,\frac{(R-R_{0})}{R_{\mathrm{d}_{\,i}}}-R_{\mathrm{ch}_{\,i}}\left(\frac{1}{R}-\frac{1}{R_{0}}\right)\right]\,,
\end{equation}
\noindent
where $\Sigma_{0\mathrm{d}_{\bigstar}}$ corresponds to the local disk stellar surface density (at $R=R_{0}$); $R_{\mathrm{d}}$ is the radial scale-length; and $R_{\mathrm{ch}}$ is the radial length of the `central hole' in the density of each stellar disk {\it i} subcomponent ({\it i} = thin, thick). The hypothesis that the Galactic disk is hollow in its center has been justified by some models that use observational data at infrared bands to describe the inner structure of the Galaxy, e.g. \citet{Freudenreich1998,Lepine_Leroy2000,Lopez-Corredoira2004,Picaud_Robin2004}. In the particular case of the PJL model, only the thin disk needs a density depression in its inner part; differently, the thick disk can be described by a simple radial exponential decay, i.e. $R_{\mathrm{ch}_{\,thick}}=0$.

For the stellar density variation perpendicular to the Galactic plane, PJL modelled the vertical profile of the thin and thick disks by exponential laws with scale-height $h_z$. In that case, the authors introduced the variation of the scale-height with the Galactic radius, $h_{z}=h_{z}(R)$, which is known as the flare of the disk. Recently, \citet{Kalberla2014} compiled some published results in the literature and found compelling evidence for the increase of the scale-heights with Galactocentric distance for different stellar distributions.
In the present study, however, we do not attempt to model such function for $h_{z}(R)$, and we consider the scale-height as a constant along the Galactic radius and with a value equal to the local scale-height $h_{z0}$ (at $R_0$) estimated by PJL, for each thin and thick disks. The reason for this approximation is justified by the fact that the introduction of the flaring of the disk requires a more careful analysis with respect to the form of the gravitational potential that would result by such distribution of density. 
The volume density for both thin and thick disks is written in the form:

\begin{equation}
\label{eq:rho_thin_thick}
\rho_{\mathrm{d}_{\bigstar,\,i}}(R,z)=\frac{\Sigma_{\mathrm{d}_{\bigstar,\,i}}(R)}{2\, h_{z_{\,i}}}\exp\left(-\frac{|z|}{h_{z_{\,i}}}\right)\,.
\end{equation} 
\noindent
The adopted values for the structural parameters of the thin and thick disks, i.e., the scale-lengths $R_{\mathrm{d}}$, radii of the central hole $R_{\mathrm{ch}}$, and scale-heights $h_{z}$, are, as mentioned before, the best-fitting values reported in the PJL model, which are listed in Table~\ref{tab:struct_diskparams}.

The local stellar surface densities for both thin and thick disks ($\Sigma_{0\mathrm{d}_{\bigstar}}$ in Eq.~\ref{eq:sigma_thin_thick}) are based on the model of \citet{Flynn2006} \citep[see also][]{Holmberg_Flynn2000,Holmberg_Flynn2004}. These authors discriminate the contributions for the total $\Sigma_{0\mathrm{d}_{\bigstar}}$ generated by different stellar components, namely: main-sequence stars of different absolute magnitudes; red giants and supergiants; stellar remnants (white dwarfs, neutron stars, black holes); and brown dwarfs. The main-sequence stars and giants contribute with $\Sigma_{\circ}=28.3$ M$_{\odot}$ pc$^{-2}$, which can be compared with the recent determination by \citet{Bovy2012a} of $\Sigma_{\circ}=30$ M$_{\odot}$ pc$^{-2}$ using the SEGUE spectroscopic survey data. The stellar remnants and brown dwarfs in the \citet{Flynn2006} model contribute with $\Sigma_{\bullet}=7.2$ M$_{\odot}$ pc$^{-2}$. Taking the combination of the Bovy et al. value for $\Sigma_{\circ}$ and the Flynn et al. value for $\Sigma_{\bullet}$ as a constraint to the local stellar surface mass density, we end up with $\Sigma_{0\mathrm{d}_{\bigstar}}=\Sigma_{\circ}+
\Sigma_{\bullet}=37.2$ M$_{\odot}$ pc$^{-2}$, the same value adopted by \citet{Read2014}. Separating this last value between the thin and thick disks, we take $\Sigma_{0\mathrm{d}_{\bigstar,\,thick}}=7.0$ M$_{\odot}$ pc$^{-2}$ for the thick disk, as in the \citet{Flynn2006} model, and $\Sigma_{0\mathrm{d}_{\bigstar,\,thin}}=30.2$ M$_{\odot}$ pc$^{-2}$ for the thin disk, where we have assigned the brown dwarfs and stellar remnants to the thin disk for practical purposes. These local surface densities along with the scale-heights result in the local volume densities in the mid-plane of the Galaxy: $\rho_{0\mathrm{d}_{\bigstar,\,thick}}=0.0055$ M$_{\odot}$ pc$^{-3}$ for the thick disk; $\rho_{0\mathrm{d}_{\bigstar,\,thin}}=0.0736$ M$_{\odot}$ pc$^{-3}$ for the thin disk (or $\rho_{0\mathrm{d}_{\bigstar,\,thin}}=0.0561$ M$_{\odot}$ pc$^{-3}$ considering only main-sequence stars and giants). The thick-to-thin disk density-ratio, $\rho_{0\mathrm{d}_{\bigstar,\,thick}}/\rho_{0\mathrm{d}_{\bigstar,\,thin}}\sim 10\%$ (neglecting the stellar remnants/brown dwarfs contribution), is close, given the errors, to the value measured by \citet{Juric2008} of $12\%$. The values adopted as constraints for $\Sigma_{0\mathrm{d}_{\bigstar,\,thin}}$ and $\Sigma_{0\mathrm{d}_{\bigstar,\,thick}}$ are listed in Table~\ref{tab:struct_diskparams}. In Table~\ref{tab:struct_diskparams}, we also give the total masses $M_{\mathrm{d}}$ calculated for the thin and thick disks, as well as the radial scale-length $R_{\mathrm{d_{exp}}}$ relative to the region of each disk subcomponent that presents the density exponential decay, and which will be used in the modelling process described in Sect.~\ref{MN-disks}. Since the thick disk is modelled by a single exponential, $R_{\mathrm{d}_{\,thick}}=
R_{\mathrm{d}_{\mathrm{exp}\,thick}}$.

\begin{table*}
\caption{Structural parameters, local surface densities and masses of the disk components taken as observational prior information to the Milky Way modelling.}             
\label{tab:struct_diskparams}      
\centering          
\begin{tabular}{c c c c c c c}
\hline\hline       

{\bf Component} & $R_{\mathrm{d}}$ & $R_{\mathrm{ch}}$ & $h_{z}$\tablefootmark{a} & $\Sigma_{0\mathrm{d}}$ & $M_{\mathrm{d}}$ & $R_{\mathrm{d_{exp}}}$\tablefootmark{b} \\
 & (kpc) & (kpc) & (kpc) & (M$_{\odot}$ pc$^{-2}$) & ($10^{10}$ M$_{\odot}$) & (kpc) \\ 
\hline                   
thin disk & 2.12 & 2.07 & 0.205 & 30.2 & 2.489 & 2.18 \\
thick disk & 3.05 & 0.00 & 0.640 & 7.0 & 0.568 & 3.05 \\
H\,{\scriptsize I} disk & 9.50 & 1.90 & 0.180 & 17.0 & 1.184 & 5.00 \\
H$_{2}$ disk & 1.48 & 4.20 & 0.100 & 3.0 & 0.227 & 1.52 \\
\hline                  
\end{tabular}
\tablefoot{
\tablefoottext{a}{For the H\,{\scriptsize I} and H$_2$ disks, the scale-heights are the $z_{1/2}$ parameters expressed in Eq.~\ref{eq:rho_HI_H2}.}
\tablefoottext{b}{$R_{\mathrm{d_{exp}}}$ corresponds to the radial exponential scale-length fitted to the region of the disk subcomponent where the surface density profile is dominated by the exponential decay.}
}
\end{table*}

\subsubsection{The gaseous component}
\label{gas_comp}

We consider the atomic H\,{\scriptsize I} and molecular H$_2$ gaseous disks as the major contributors to the density of the interstellar medium (ISM), with the proper correction factor for He. The radial profile of the surface density for these components is chosen as:

\begin{equation}
\label{eq:sigma_HI_H2}
\Sigma_{\mathrm{d_{\,HI,H_{2}}}}(R)=\Sigma_{0\mathrm{d_{\,HI,H_{2}}}}\,\exp\left[-\,\frac{\left(R^{\,n}-R_{0}^{\,n}\right)}{R_{\mathrm{d_{\,HI,H_{2}}}}^{\,n}}-R_{\mathrm{ch_{\,HI,H_{2}}}}^{2}\left(\frac{1}{R^{2}}-\frac{1}{R_{0}^{2}}\right)\right]\,,
\end{equation}
\noindent
with the exponents $n=3/2$ for the H\,{\scriptsize I} and $n=1$ for the H$_2$ gas disk components, respectively. Similarly to Eq.~\ref{eq:sigma_thin_thick}, $\Sigma_{0\mathrm{d}}$ corresponds to the local (H\,{\scriptsize I}, H$_2$) surface density; $R_{\mathrm{d}}$ is the radial scale-length; and $R_{\mathrm{ch}}$ is the radius of the `central hole' in the density distribution of H\,{\scriptsize I} and H$_2$ disks. The form for the gas surface densities presented in Eq.~\ref{eq:sigma_HI_H2}, together with the values for the structural parameters $R_{\mathrm{d}}$ and $R_{\mathrm{ch}}$ of the H\,{\scriptsize I} and H$_2$ disks listed in Table~\ref{tab:struct_diskparams}, are chosen to reproduce as closely as possible the density distributions observed in H\,{\scriptsize I} by \citet[cf. their Figure 3]{Kalberla_Dedes2008}, and in H$_2$ by \citet[cf. their Figure 9]{Nakanishi_Sofue2006}.

The volume density for both H\,{\scriptsize I} and H$_2$ disks is given in the form:

\begin{equation}
\label{eq:rho_HI_H2}
\rho_{\mathrm{d_{\,HI,H_{2}}}}(R,z)=\frac{\Sigma_{\mathrm{d_{\,HI,H_{2}}}}(R)}{2.12\,z_{1/2_{\,\mathrm{HI,H_{2}}}}}\exp\left[-\left(\frac{z}{1.18\,z_{1/2_{\,\mathrm{HI,H_{2}}}}}\right)^{2}\right]\,,
\end{equation}
\noindent
with the scale-heights $z_{1/2}$ corresponding to the half-width at half-maximum of the density peaks of H\,{\scriptsize I} and H$_2$ in the Galactic mid-plane. The Gaussian profile for the vertical distribution of hydrogen in the form presented in Eq.~\ref{eq:rho_HI_H2} has been widely adopted in the literature to represent the H\,{\scriptsize I} and H$_2$ distributions (e.g. \citealt{Sanders_Solomon_Scoville1984,Amores_Lepine2005}; PJL). As pointed out by \citet{Amores_Lepine2005}, the Gaussian vertical distribution for the gaseous disk correctly fits the observations, also being an expected solution for hydrostatic equilibrium considerations.
 
The scale-height of the H\,{\scriptsize I} distribution in the Milky Way has since long ago been observed to increase systematically with Galactic radius (e.g. \citealt[among others]{Lozinskaya_Kardashev1963,Burton1976,Kalberla_Dedes2008}). The flaring of the H\,{\scriptsize I} disk is naturally expected when one considers the fact that, for the case of hydrostatic equilibrium, the gravitational force perpendicular to the disk plane must balance the gas pressure-gradient force, and since the vertical velocity dispersion $\sigma_z$ of the H\,{\scriptsize I} gas is approximately constant with radius $R$ (e.g. \citealt{Spitzer1968}), the scale-height of this component must increase with $R$. The Galactic distribution of molecular gas also shows a flaring consistent with that observed in H\,{\scriptsize I} (\citealt[and references therein]{Kalberla2007}). Although being a very important feature to be included in any realistic mass model of the Galaxy, we do not attempt to model the flaring of the gaseous disk component for the same reason exposed in the case of the stellar disk. 
Therefore, here we consider the scale-heights of the H\,{\scriptsize I} and H$_2$ disk components as independent of radius and with values equal to the local ones (at $R_0$) calculated from the model of \citet[cf. their Equation 7]{Amores_Lepine2005}. Table~\ref{tab:struct_diskparams} lists the adopted prior values for the H\,{\scriptsize I} and H$_2$ disks scale-heights.

The model of \citet{Holmberg_Flynn2000} for the ISM component, which is based on the original multiphase model of \citet*{Bahcall1992}, discriminates the contributions between the molecular gas, the warm (ionized) gas, and a split of the neutral H\,{\scriptsize I} into cold and hot components. The total local surface density in the gas form, according to this model, is $\Sigma_{0\mathrm{d_{g}}}=13$ M$_{\odot}$ pc$^{-2}$, with an uncertainty of $\sim 50\%$. More recently, \citet{Hessman2015} has warned about the underestimation of these traditional determinations of the neutral and molecular gas densities. According to this author, substantial amounts of ``dark'' gas both in the form of optically thick cold neutral hydrogen and CO-dark molecular gas is known to contribute to the ISM density, which must raise the estimates of the local mid-plane gas densities by as much as $\sim 60\%$. Moreover, if local density features such as the Local Bubble or the local spiral arms structure are taken into account, the corrections for a larger $\Sigma_{0\mathrm{d_{g}}}$ are even higher. Therefore, as observational constraints to our gaseous disk model, we take the following values for the local surface gas densities: $\Sigma_{0\mathrm{d_{\,HI}}}=15$ M$_{\odot}$ pc$^{-2}$, as in the \citet{Hessman2015} updated estimate, which already takes into account the correction factor of 1.36 for the mass in helium (He); $\Sigma_{0\mathrm{d_{\,H_{2}}}}=3$ M$_{\odot}$ pc$^{-2}$, as in the \citet{Holmberg_Flynn2000} model. The warm gas disk component, which contributes locally with 2 M$_{\odot}$ pc$^{-2}$ (\citealt{Holmberg_Flynn2000,Hessman2015}), is here incorporated to the H\,{\scriptsize I} disk for practical purposes, increasing the value of $\Sigma_{0\mathrm{d_{\,HI}}}$ to 17 M$_{\odot}$ pc$^{-2}$. The adopted values for the local surface densities, total masses $M_{\mathrm{d}}$ and radial exponential scale-lengths $R_{\mathrm{d_{exp}}}$ of the H\,{\scriptsize I} and H$_2$ disks are listed in Table~\ref{tab:struct_diskparams}.

\begin{figure}
\centering
\includegraphics[scale=0.50]{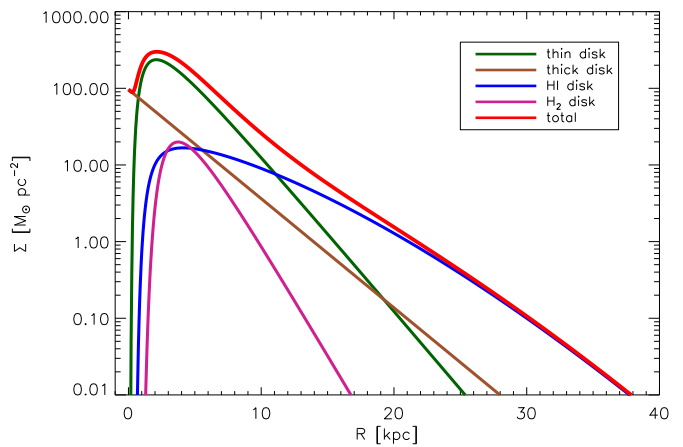}
\caption{Radial profile of the surface density of our `observation-based' model for the Milky Way's disk. The curves indicate the surface densities of the thin disk (green); thick disk (brown); H\,{\scriptsize I} disk (blue); H$_2$ disk (violet); and the total disk (red).}
\label{fig:disks_Sigs}
\end{figure}

As can be seen from Table~\ref{tab:struct_diskparams}, our `observation-based' disk model comprises a total mass $M_{\mathrm{d}}=4.47\times 10^{10}$ M$_{\odot}$, being $M_{\mathrm{d}_{\bigstar}}=3.06\times 10^{10}$ M$_{\odot}$ in stars and $M_{\mathrm{d_{g}}}=1.41\times 10^{10}$ M$_{\odot}$ in the gaseous form. The total local disk mass-surface density is $\Sigma_{0\mathrm{d}}=57.2$ M$_{\odot}$ pc$^{-2}$, which can be compared with the updated density estimates of 58 M$_{\odot}$ pc$^{-2}$ by \citet{Hessman2015}, and $54.2\pm 4.9$ M$_{\odot}$ pc$^{-2}$ by \citet{Read2014}. The local disk mid-plane volume density is $\rho_{0\mathrm{d}}=0.138$ M$_{\odot}$ pc$^{-3}$. This value is somewhat larger than the estimate of the local dynamical mass density of $0.102\pm 0.010$ M$_{\odot}$ pc$^{-3}$ by \citet{Holmberg_Flynn2000}, of which 0.095 M$_{\odot}$ pc$^{-3}$ is in visible matter. However, the density corrections discussed by \citet{Hessman2015} should increase these traditional estimates to local densities as large as $\sim 0.120$ or even $\sim 0.160$ M$_{\odot}$ pc$^{-3}$. Figure~\ref{fig:disks_Sigs} shows the radial distribution of the surface density resulted from our `observation-based' disk model, as well as the surface densities of each disk subcomponent. The surface densities of the thin and thick stellar disks drop to $\sim 1$ M$_{\odot}$ pc$^{-2}$ at the radius $R\sim 15$ kpc, while the H$_2$ disk reaches this value at $R\sim 10$ kpc. The H\,{\scriptsize I} disk subcomponent extends to larger radii, with a drop of its surface density to $\sim 0.1$ M$_{\odot}$ pc$^{-2}$ at $R\sim 30$ kpc, very similar to the mean surface density profile of H\,{\scriptsize I} derived by \citet{Kalberla_Dedes2008}.

The derivation of the gravitational potential directly from the density distributions in Eqs.~\ref{eq:rho_thin_thick} and~\ref{eq:rho_HI_H2}, through the Poisson equation $\nabla^{2}\Phi=4\pi G\rho$, must involve the use of some mathematical tools and/or numerical techniques, like the multipole expansion and numerical interpolation employed by \citet{Dehnen_Binney1998} in their determination of the potential and force-field associated with their Galactic mass models. Another way of handling such a task is to approximate the disk density distribution by an analytical functional form for which the associated potential is also known analytically, as is the case of the density-potential pairs of 
Miyamoto-Nagai disks. It is known that a single Miyamoto-Nagai disk only provides a rough approximation to a real galactic disk density profile. But we show later in the next sections that the combination of three 
Miyamoto-Nagai disks is able to give good approximations to the observed Galactic disk mass distribution, and so for the gravitational potential, for given ranges of Galactic radii and heights above the disk mid-plane. This is the approach that we adopt in the present paper: we adjust combinations of Miyamoto-Nagai disks to each component of our 'observation-based' disk model. The disk potential then described in a complete analytical form provides quicker computations of its related force-field, which is suitable for fast calculations of galactic orbits of large samples of stars or test-particles in a numerical simulation. This is one of the main advantages that we pursue for our future studies concerning stellar orbits in the Galactic disk. In the next subsection, we describe the method employed to the search for the best set of 
Miyamoto-Nagai disks (hereafter MN-disks) that reproduce the main features of the density distributions in Eqs.~\ref{eq:rho_thin_thick} and~\ref{eq:rho_HI_H2}.

\subsection{Reproducing the disk mass model with Miyamoto-Nagai disks}
\label{rep_MN-disks}

\subsubsection{The Toomre-Kuzmin disks}
\label{TK-disks}

As a first step, we make use of the family of disk models of infinitesimal thickness (the razor-thin disks) introduced by \citet{Toomre1963} and \citet{Kuzmin1956}, whose densities are written in the form $\rho(R,z)=\Sigma(R)\,\delta(z)$. The `model 1' of this family, of what here we refer to as Toomre-Kuzmin disks (TK-disks), is described by a surface density which is written as:

\begin{equation}
\label{eq:disk_TK1}
\Sigma_{\mathrm{TK}_{1}}(R)=\frac{a_{\mathrm{TK}_{1}}\,M_{\mathrm{TK}_{1}}}{2\,\pi}\,\frac{1}{\left[R^{2}+a_{\mathrm{TK}_{1}}^{2}\right]^{3/2}}\,,
\end{equation}
\noindent
where $M_{\mathrm{TK}_{1}}$ is the total disk mass and $a_{\mathrm{TK}_{1}}$ is related to the radial scale-length of the disk. The choice of use of TK-disks, in the first step, is due to the fact that all the information about the disk surface density can be recovered adjusting the only two parameters $M_{\mathrm{TK}}$ and $a_{\mathrm{TK}}$. As shown later, the three-dimensional structure of the disks is obtained after `inflating' the TK-disks with the introduction of the parameter $b$ related to the scale-height, in the same way as in the original method of \citet{Miyamoto_Nagai1975}.

From Eq.~\ref{eq:disk_TK1}, it can be seen that the density $\Sigma_{\mathrm{TK}_{1}}$ decreases with $R^{-3}$ at large radii. This is a slower decrease when compared to the observed exponential fall off of the brightness profiles in galaxy disks (\citealt{Binney_Tremaine2008}). Therefore, a single TK-disk (and the correspondent MN-disk) poorly matches the surface density profile of a radially exponential disk (\citealt{Smith2015}). This is the motivation to the use of a combination of three MN-disks (related to the `model 1' of TK-disks, Eq.~\ref{eq:disk_TK1}), firstly introduced by \citet{Flynn1996}, and since then used for modelling the disk of the Milky Way (\citealt[and references therein]{Smith2015}). In this procedure, each MN-disk has a different scale-length $a$ and mass $M$, being one of the masses with a negative value. The MN-disk with negative mass is also the one with the largest scale-length, a feature that helps to decrease the surface density at large radii (and improve the fit to an exponential disk), but that also leads to the undesirable occurrence of regions with negative densities near the mid-plane of the disk and at large radii (\citealt{Smith2015}).

It is also known that the density $\Sigma_{\mathrm{TK}}$ in Eq.~\ref{eq:disk_TK1} is just the first of a family of possible forms for the potential-density pair that obey the Poisson equation. \citet{Toomre1963} showed that, due to linearity between $\Phi$ and $\rho$ in the Poisson equation, new potential-density pairs can be obtained through the derivation of $\Phi/a$ $n$ times with respect to $a^2$ (\citealt{Binney_Tremaine2008}). For example, the densities for `models 2' and `3' of the TK-disks are described by

\begin{equation}
\label{eq:disk_TK2}
\Sigma_{\mathrm{TK}_{2}}(R)=\frac{3\,a_{\mathrm{TK}_{2}}^{3}\,M_{\mathrm{TK}_{2}}}{2\,\pi}\,\frac{1}{\left[R^{2}+a_{\mathrm{TK}_{2}}^{2}\right]^{5/2}}\,,
\end{equation}
\noindent
and
\begin{equation}
\label{eq:disk_TK3}
\Sigma_{\mathrm{TK}_{3}}(R)=\frac{5\,a_{\mathrm{TK}_{3}}^{5}\,M_{\mathrm{TK}_{3}}}{2\,\pi}\,\frac{1}{\left[R^{2}+a_{\mathrm{TK}_{3}}^{2}\right]^{7/2}}\,,
\end{equation}
\noindent
respectively (\citealt{Miyamoto_Nagai1975}). One can see that the densities $\Sigma_{\mathrm{TK}_{2}}$ and $\Sigma_{\mathrm{TK}_{3}}$ decrease with $R^{-5}$ and $R^{-7}$, respectively, at large radii. In this way, a combination of models 2 or 3 of TK-disks could, in principle, not only result in an equally good fit to an exponential disk but also circumvent the problem of the appearance of negative densities. In our present case, we also need to include a TK-disk with negative mass to the modelling of the central density depression that occurs in the thin stellar disk, the H\,{\scriptsize I} and the H$_2$ disk subcomponents. The difference here is that the scale-length of this component with negative mass does not need to be the largest one, thus avoiding the negative densities at large radii. Therefore, for each disk subcomponent (thin, thick, H\,{\scriptsize I}, H$_2$), we search for the combination of 3 TK-disks, of each model (1, 2, 3), that better reproduces the radial profile of the surface density of each subcomponent. Let us take, for example, the case of the thin disk. We search for the set of parameters $\left\lbrace M_{\mathrm{TK}_{1}^{1}},\;a_{\mathrm{TK}_{1}^{1}},\;M_{\mathrm{TK}_{1}^{2}},\;a_{\mathrm{TK}_{1}^{2}},\;M_{\mathrm{TK}_{1}^{3}},\;a_{\mathrm{TK}_{1}^{3}}\right\rbrace$\footnote{
$\left\lbrace M_{\mathrm{TK}_{j}^{i}},\;a_{\mathrm{TK}_{j}^{i}}\right\rbrace$, where $i$ denotes each one of the three TK-disk, and $j$ denotes the TK-disk model order.} of `model 1' of 3 TK-disks that generates the best fit to the radial distribution of surface density of the thin disk. These chosen 3 TK-disks result in the total surface density $\Sigma_{\mathrm{TK}_{1}^{\mathrm{tot}}}=\Sigma_{\mathrm{TK}_{1}^{1}}+
\Sigma_{\mathrm{TK}_{1}^{2}}+\Sigma_{\mathrm{TK}_{1}^{3}}$
. The same procedure is done with the 3 TK-disks in the forms of `model 2' and `model 3', which result in the total densities $\Sigma_{\mathrm{TK}_{2}^{\mathrm{tot}}}$ and $\Sigma_{\mathrm{TK}_{3}^{\mathrm{tot}}}$, respectively. We quantify $\xi^{2}$ as the sum, over a given radial range, of the squares of the residuals between the surface density of the thin disk (obtained from Eq.~\ref{eq:sigma_thin_thick} and parameters from Table~\ref{tab:struct_diskparams}) and the surface density resulted from the best-fitting combination of the three TK-disks, for each one of the models:

\begin{equation}
\label{eq:xi2_dens}
\xi_{k,j}^{2}=\frac{1}{N_{R}}\,\sum_{R}^{}\left[\Sigma_{\mathrm{d}_{\,k}}(R)-\Sigma_{\mathrm{TK}_{j}^{\mathrm{tot}}}(R)\right]^{2}\,, 
\end{equation}
\noindent
where $k$ refers to the disk subcomponent ($k$ = {\it thin} in the above example) and $j$ denotes the model order of TK-disks; $N_R$ is the number of radial bins over which the sum is calculated. The TK-disk model order that results in the lowest value for $\xi^{2}$ is the one which we consider as the model that better represents the surface density $\Sigma_{\mathrm{d}_{\,k}}$ of the disk subcomponent. The search for the best sets of parameters $\left\lbrace M_{\mathrm{TK}_{j}^{i}},\;a_{\mathrm{TK}_{j}^{i}}\right\rbrace$ corresponding to the $j$'s best TK-disk models were done by applying the global optimization technique based on the cross-entropy (CE) algorithm for parameters estimation. The CE algorithm (\citealt{Rubinstein1997,Rubinstein1999,Kroese2006}) provides a simple adaptive way of estimating the optimal set of reference parameters in the fitting process. We refer the reader to recent papers that have applied the CE technique to some astrophysical problems, e.g. \citet{Caproni2009,Monteiro_Dias_Caetano2010,Monteiro_Dias2011,
Martins2014,Dias2014,Caetano2015}, among others. Since we also employ the CE method in other occasions throughout this paper, in the next subsection we give a brief description of the main steps of the algorithm. Table~\ref{tab:disk_models_xi2} summarizes the model of 3 TK-disks found to better reproduce the surface density of each one of the disk subcomponents, as well as the corresponding $\xi^{2}$ (Eq.~\ref{eq:xi2_dens}) minimized within the CE algorithm.

\begin{table}
\caption{Models of 3 TK-disks that provide the best match to the surface density of each disk subcomponent.}             
\label{tab:disk_models_xi2}      
\centering          
\begin{tabular}{c c c}
\hline\hline       

{\bf Component} & {\bf Model}\,\tablefootmark{a} & $\xi^{2}$ \\
\hline                   
thin disk & 3 (Eq.~\ref{eq:disk_TK3}) & 4.35 \\
thick disk & 1 (Eq.~\ref{eq:disk_TK1}) & 1.12 \\
H\,{\scriptsize I} disk & 2 (Eq.~\ref{eq:disk_TK2}) & 0.25 \\
H$_{2}$ disk & 3 (Eq.~\ref{eq:disk_TK3}) & 0.74 \\
\hline                  
\end{tabular}
\tablefoot{
\tablefoottext{a}{The model order of 3 TK-disks that results in the lowest value for $\xi^{2}$ (Eq.~\ref{eq:xi2_dens}), which also corresponds to the associated model of 3 MN-disks.}
}
\end{table}

\subsubsection{The cross-entropy algorithm}
\label{CE_algorithm}

For a detailed presentation and description of the CE method, we refer the reader to the papers by \citet*{Monteiro_Dias_Caetano2010} and \citet{Dias2014}. Here, we briefly show an overview of the method and how it works.

Let us suppose that we have a set of data $D$ with individual points $d_1$, $d_2$, ..., $d_{N_{\mathrm{D}}}$ and we wish to study it in terms of an analytical model $\Theta$ with a vector of parameters $\theta_1$, $\theta_2$, ..., $\theta_{N_{\mathrm{p}}}$. 
The main goal of the CE continuous multi-extremal optimization method is to find a set of parameters $\theta_{i}^{*}\left(i=1,\,...,\,N_{\mathrm{p}}\right)$ for which the model provides the best description of the data, based on some statistical criterion. This is performed  by randomly generating $N$ independent sets of model parameters $\mathbf{X}=\left(\mathbf{x}_{1}, \mathbf{x}_{2}, \,...,\, \mathbf{x}_{N}\right)$, where $\mathbf{x}_{i}=\left(\theta_{1_{i}}, \theta_{2_{i}}, \,..., \,\theta_{N_{\mathrm{p}_{i}}}\right)$, under some chosen distribution, and the subsequent minimization of an objective function $S(\mathbf{X})$ used to transmit the quality of the fit during the run process. As the method  converges to the `theoretical' exact solution, then $S\rightarrow 0$, which means $\mathbf{x}\rightarrow \mathbf{x}^{*}=\left(\theta_{1}^{*}, \theta_{2}^{*}, \,..., \,\theta_{N_{\mathrm{p}}}^{*}\right)$.

The CE method performs an iterative statistical coverage of the parameter space,  
where the following is done in each iteration (\citealt{Monteiro_Dias_Caetano2010}):
\begin{description}
	\item[{\rm (i)}] Random generation of the initial parameter sample, respecting some underlying distributions and pre-defined criteria;
	\item[{\rm (ii)}] Selection of the best candidates based on some mathematical criterion, which will compose the elite sample array - samples with lower values for the objective function $S(\mathbf{X})$;
	\item[{\rm (iii)}] Random generation of updated parameter samples from the previous best candidates - the elite sample - to be evaluated in the next iteration;
	\item[{\rm (iv)}] Optimization process repeats steps (ii) and (iii) until a pre-specified stopping criterion is fulfilled.
\end{description}
In all the implementations of the CE algorithm done in this work, we followed the general step by step presented in section 2.2 of \citet{Monteiro_Dias_Caetano2010}. The tuning parameters used in the run processes were: $N=3\times 10^{3}$ sets of trial model parameters per iteration; a maximum number of 100 iterations; $N_{\mathrm{elite}}=100$, which is the number of sets of trial parameters that return the best solutions (the lowest values for the objective function) at a given iteration and that will be used to estimate the distribution parameters for the next iteration. For the smoothing parameters that reduce the convergence speed of the algorithm, preventing it from finding a non-global minimum solution, we have used: $\alpha=0.9$; $\alpha'=0.7$; and $q=7$. For more details about these parameters and how they are implemented in the algorithm, see \citet{Monteiro_Dias_Caetano2010}.


\subsubsection{The Miyamoto-Nagai disks - following the approach developed by Smith et al. (2015)}
\label{MN-disks}

Once the models of TK-disks have been found, the next step is to `inflate' them vertically to find the correspondent MN-disks. As firstly performed by \citet{Miyamoto_Nagai1975}, this is done by replacing the term $(a+|z|)$ in the potential function, associated to the density of a given model of TK-disk, with the term $\left[a+\sqrt{z^{2}+b^{2}}\right]$, where $b$ expresses the vertical scale-height of the disk. In this way, corresponding to the three models of TK-disks expressed by Eqs.~\ref{eq:disk_TK1}, ~\ref{eq:disk_TK2}, and~\ref{eq:disk_TK3}, we have the following three-dimensional density functions that compose the first three models of MN-disks:  

\begin{eqnarray}
\label{eq:rho_MN1}
\rho_{\mathrm{MN}_{1}}(R,z)=\left(\frac{b^{2}M}{4\pi}\right)\frac{\left[a R^{2}+\left(a+3\zeta\right)\left(a+\zeta\right)^{2}\right]}{\zeta^{3}\left[R^{2}+\left(a+\zeta\right)^{2}\right]^{5/2}}\,,
\end{eqnarray}

\begin{multline}
\label{eq:rho_MN2}
\rho_{\mathrm{MN}_{2}}(R,z)=\left(\frac{b^{2}M}{4\pi}\right)\frac{3\,\left(a+\zeta\right)}{\zeta^{3}\left[R^{2}+\left(a+\zeta\right)^{2}\right]^{7/2}}\left[R^{2}\left(\zeta^{2}-a\zeta+a^{2}\right)+\right.\\
+\left.\left(a+\zeta\right)^{2}\left(\zeta^{2}+4a\zeta+a^{2}\right)\right]\,,
\end{multline}

\begin{align}
\label{eq:rho_MN3}
\rho_{\mathrm{MN}_{3}}(R,z)&=\left(\frac{b^{2}M}{4\pi}\right)\frac{1}{\zeta^{3}\left[R^{2}+(a+\zeta)^{2}\right]^{9/2}}\,\left[3R^{4}\zeta^{3}+\right.\nonumber\\
&\quad+R^{2}(a+\zeta)^{2}\left(6\zeta^{3}+15a\zeta^{2}-10a^{2}\zeta+5a^{3}\right)+\nonumber\\
&\quad+\left.(a+\zeta)^{4}\left(3\zeta^{3}+15a\zeta^{2}+25a^{2}\zeta+5a^{3}\right)\right]\,,
\end{align}
\noindent
where $\zeta=\sqrt{z^{2}+b^{2}}$. In the above expressions, for the sake of good readability, we avoided the excessive use of subscripts in the parameters $M$, $a$, $b$ and $\zeta$ that could distinguish between each one of the models. Here, we make it clear that for each disk subcomponent modelled with a combination of 3 TK-disks of a given model, now we use the corresponding combination of 3 MN-disks of the equivalent model. 

The task now is to find the best values for the parameter $b$ that reproduce the vertical density distribution of each subcomponent of the Galactic disk. For this purpose, we follow the approach recently developed by \citet{Smith2015}. In that work, the authors create a procedure for modelling simple radially exponential disks from the combination of three MN-disks of `model 1', allowing the construction of disks of any mass, scale-length, and for an ample range of thicknesses, from infinitely thin disks to approximately spherical systems. The basic idea of the method consists in: starting from the model composed by the 3 TK-disks (for which $b=0$) that better match the radial profile of the disk surface density, one chooses the corresponding model composed by 3 MN-disks with a small value for the $b$ parameter (with the aim of remaining near the solution for the infinitesimal thickness disk), and whose parameters $M$ and $a$ of the MN-disks deviate from small amounts with respect to those of the TK-disks. With the continuous variation of $b$ in small steps, the parameters $M$ and $a$ are searched for at intervals close to the solution found in the previous step, ensuring a smooth variation of both $M$ and $a$ with the variation of the disk thickness $b$. In fact, \citet{Smith2015} analyze the variation of the ratios $M/M_{\mathrm{d}}$ and $a/R_{\mathrm{d}}$ as a function of the variation of the ratio $b/R_{\mathrm{d}}$, being $M_{\mathrm{d}}$ the mass and $R_{\mathrm{d}}$ the radial scale-length of the exponential disk to be modelled. Therefore, for each small variation of the ratio $b/R_{\mathrm{d}}$, one searches for the parameters $M$ and $a$ of the 3 MN-disks whose integral of the density $\rho(z)$ in the vertical direction better reproduces the radial profile of the disk surface density $\Sigma(R)$.

To relate the scale-height $b$ of the MN-disks with the scale-height $h_z$ of the vertically exponential disk, \citet{Smith2015} follow an approach analogous to that described above. In this case, the authors first sum up the fractional differences between the density of the 3 MN-disks combination and the exponential one, measured vertically from the mid-plane up to $z=5b$. They vary the ratio $b/R_{\mathrm{d}}$ in order to minimize the sum, then finding the best match to the exponential distribution.

In this section, we only present the values of the parameters $[M,\,a,\,b]$ calculated for each combination of 3 MN-disks that fit each disk subcomponent with their fixed values for $[M_{\mathrm{d}},\,R_{\mathrm{d_{exp}}},\,h_{z}]$ listed in Table~\ref{tab:struct_diskparams}. However, in Appendix~\ref{app1}, we also present the variations of the ratios $M/M_{\mathrm{d}}$ and $a/R_{\mathrm{d_{exp}}}$ as functions of $b/R_{\mathrm{d_{exp}}}$, as well as the relations between $b/R_{\mathrm{d_{exp}}}$ and $h_{z}/R_{\mathrm{d_{exp}}}$ for each disk subcomponent, calculated in the same way as in \citet{Smith2015}. Therefore, disks of any masses $M_{\mathrm{d}}$ and scales $R_{\mathrm{d_{exp}}}$ and $h_z$ can be built up using those relations, but remembering here that they are suited for models of disks with central density depressions.
Since the majority of our disk subcomponents are no longer modelled by a single radially exponential law (cf. Eqs.~\ref{eq:sigma_thin_thick} and ~\ref{eq:sigma_HI_H2}), coupled with the fact that we also use models of higher order (2 and 3) than the model 1 of MN-disks, our calculated relations between the aforementioned parameters are somewhat different from those presented in \citet{Smith2015}. But with equivalent utility, the relations in Appendix~\ref{app1} allow the interested user to construct models for disks with masses and sizes different from the ones modelled in this work.

In our search for the best values of the parameters $[M,\,a,\,b]$ of the MN-disks, we have also employed the CE technique described in the previous subsection, but now looking for the minimization of the sum of the squares of the residuals between the surface density of the disk subcomponent and the vertically integrated volume density of the 3 MN-disks combination

\begin{equation}
\label{eq:xi2_dens2}
\xi_{k,j}^{'2}=\frac{1}{N_{R}}\,\sum_{R}^{}\left[\Sigma_{\mathrm{d}_{\,k}}(R)-\int_z\rho_{\mathrm{MN}_{j}^{\mathrm{tot}}}(R,z)\,\mathrm{d}z\right]^{2}\,, 
\end{equation}
\noindent
for the relations of $M/M_{\mathrm{d}}$ and $a/R_{\mathrm{d_{exp}}}$ with $b/R_{\mathrm{d_{exp}}}$, and the minimization of the sum of the squares of the fractional differences between the volume densities of the 3 MN-disks combination and the disk subcomponent over a given vertical range and at $R=R_{0}$

\begin{equation}
\label{eq:xi2_dens3}
\xi_{k,j}^{''2}=\frac{1}{N_{z}}\,\sum_{z}^{}\left[\frac{\rho_{\mathrm{d}_{\,k}}(z)-\rho_{\mathrm{MN}_{j}^{\mathrm{tot}}}(z)}{\rho_{\mathrm{d}_{\,k}}(z)}\right]^{2}\,, 
\end{equation}
\noindent
for the relations between $b/R_{\mathrm{d_{exp}}}$ and $h_{z}/R_{\mathrm{d_{exp}}}$.

Let us take again the case of the thin stellar disk for exemplification. With the ratio $h_{z}/R_{\mathrm{d_{exp}}}$ calculated from the values listed in Table~\ref{tab:struct_diskparams} for the thin disk, and putting it into Eq.~\ref{eq:fit_poly4_thickness} with the coefficients for the thin disk given in Table~\ref{tab:coeff_poly_thickness}, both from Appendix~\ref{app1}, one finds the corresponding value for the ratio $b/R_{\mathrm{d_{exp}}}$. Substituting this last one into Eq.~\ref{eq:fit_poly4} with the coefficients given in Table~\ref{tab:coeff_poly_thin}, also from Appendix~\ref{app1}, one obtains the values of the parameters $M_i$ and $a_i$ for the combination of 3 MN-disks of model 3 (Eq.~\ref{eq:rho_MN3}) that better matches the density distribution $\rho_{\mathrm{d}_{\bigstar,\,thin}}$ of the thin stellar disk. We emphasize here that a single scale-height $b$ is used for all the three MN-disks of the combination, as originally done by \citet{Flynn1996} and followed by \citet{Smith2015}. All the procedure described above, which was illustrated with the thin disk, is repeated with the other subcomponents (thick disk, H\,{\scriptsize I} and H$_2$ disks) using their corresponding equations and tables in Appendix~\ref{app1}. Table~\ref{tab:disks-MN_params} lists the values of the parameters $[M,\,a,\,b]$ of each combination of 3 MN-disks that fit each disk subcomponent, as well as the correspondent models of MN-disks that had been previously found with the TK-disks modelling.

\begin{table*}
\caption{Parameters of the 3 MN-disks combination for modelling each subcomponent of the Galactic disk.}             
\label{tab:disks-MN_params}      
\centering          
\begin{tabular}{c c c c c c c c c}
\hline\hline       

{\bf Component} & $M_1$ & $a_1$ & $M_2$ & $a_2$ & $M_3$ & $a_3$ & $b$ & {\bf Model}\,\tablefootmark{a} \\
 & ($10^{10}$ M$_{\odot}$) & (kpc) & ($10^{10}$ M$_{\odot}$) & (kpc) & ($10^{10}$ M$_{\odot}$) & (kpc) & (kpc) & \\ 
\hline                   
thin disk & 2.106 & 3.859 & 2.162 & 9.052 & -1.704 & 3.107 & 0.243 & 3 (Eqs.~\ref{eq:rho_MN3} and~\ref{eq:Phi_MN3}) \\
thick disk & 0.056 & 0.993 & 3.766 & 6.555 & -3.250 & 7.651 & 0.776 & 1 (Eqs.~\ref{eq:rho_MN1} and~\ref{eq:Phi_MN1}) \\
H\,{\scriptsize I} disk & 2.046 & 9.021 & 2.169 & 9.143 & -3.049 & 7.758 & 0.168 & 2 (Eqs.~\ref{eq:rho_MN2} and~\ref{eq:Phi_MN2}) \\
H$_{2}$ disk & 0.928 & 6.062 & 0.163 & 3.141 & -0.837 & 4.485 & 0.128 & 3 (Eqs.~\ref{eq:rho_MN3} and~\ref{eq:Phi_MN3}) \\
\hline                  
\end{tabular}
\tablefoot{
\tablefoottext{a}{The model of 3 MN-disks combination correspondent to the TK-disks model that results in the lowest value for the quantity $\xi^{2}$ in Eq.~\ref{eq:xi2_dens}.}
}
\end{table*}

\begin{figure*}
\centering
\includegraphics[scale=0.50]{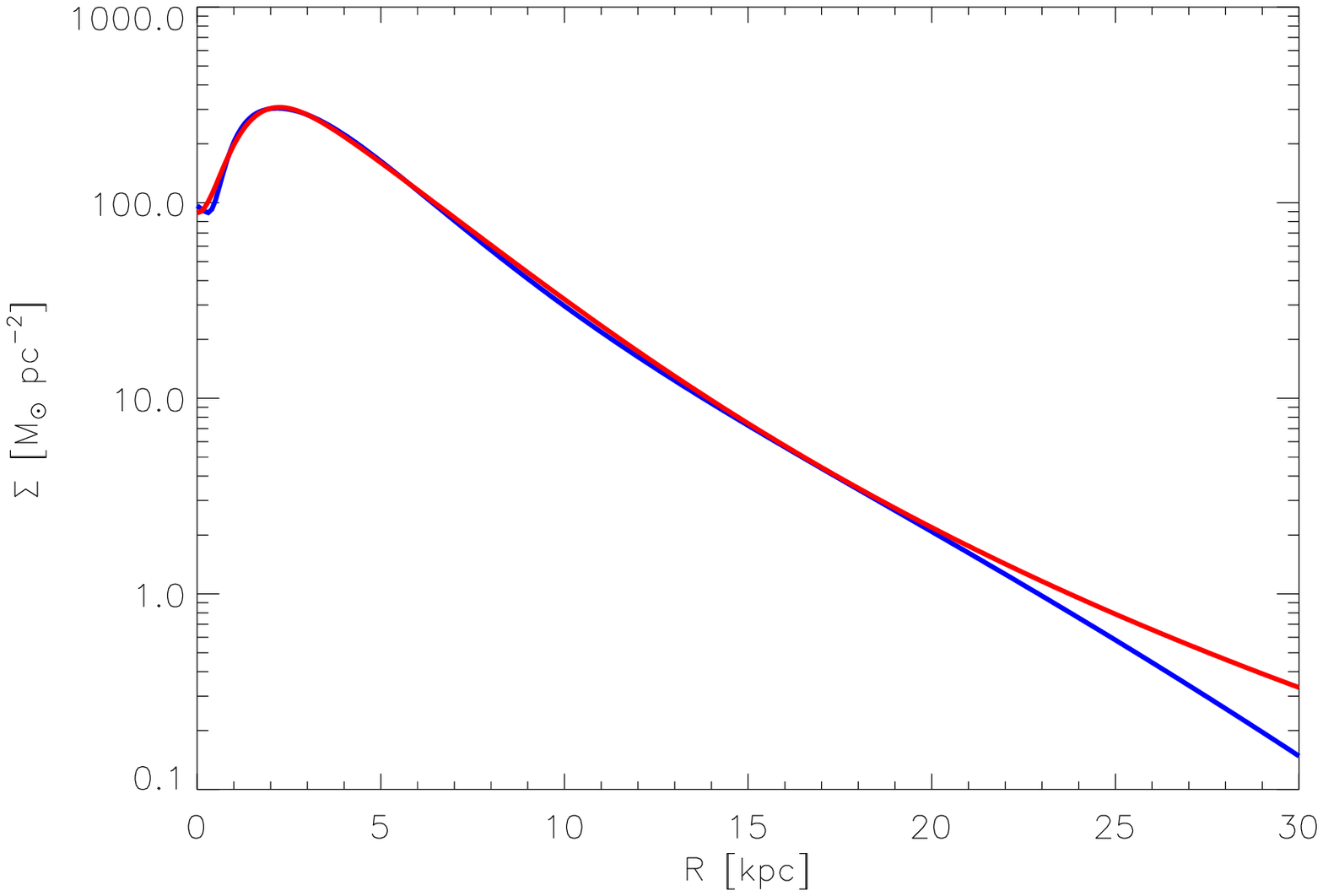}
\includegraphics[scale=0.50]{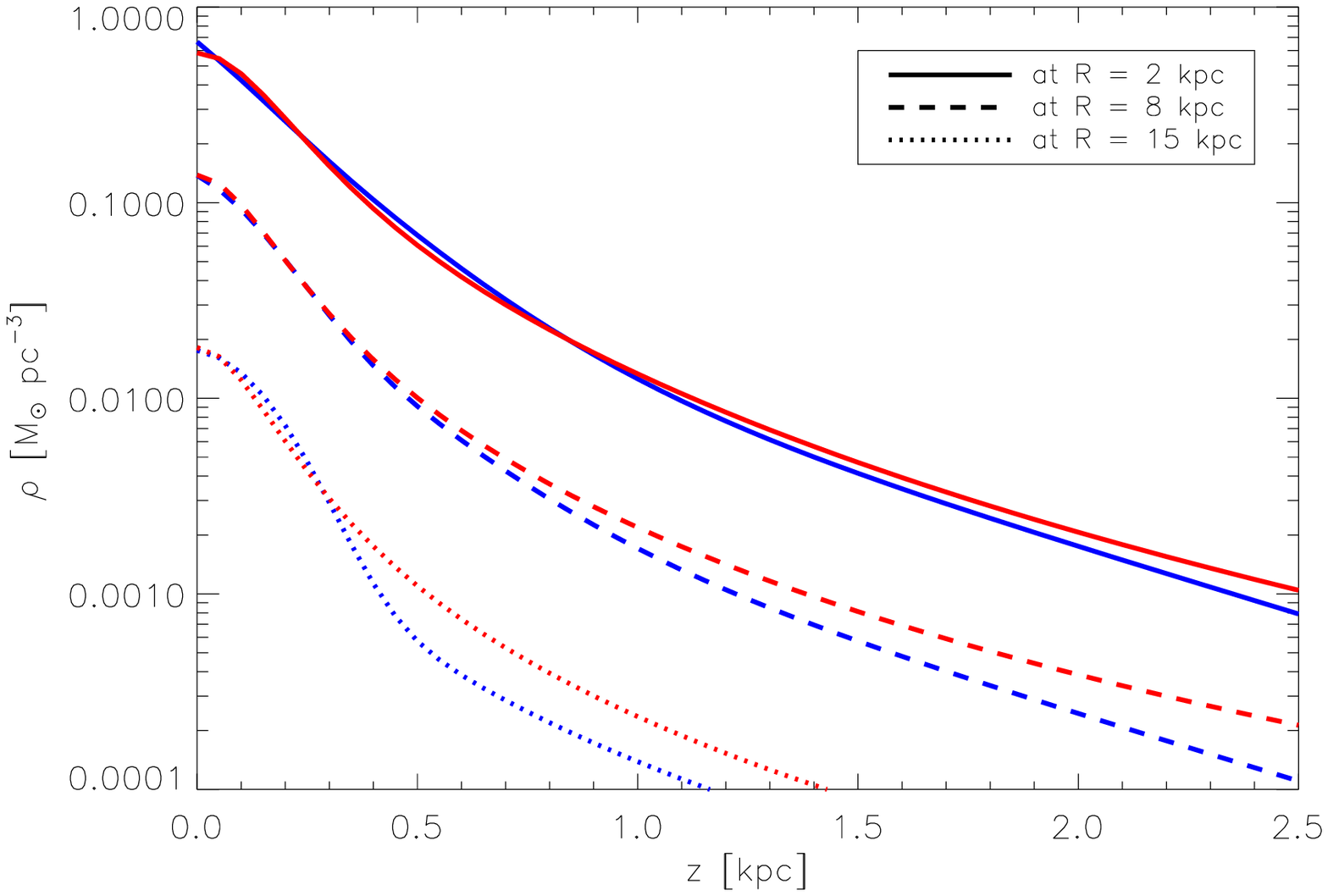}
\caption{{\it Left-hand panel}: radial distribution of the surface mass-density of the Galactic disk model - the `observation-based' disk model (blue curve), and the resultant from the sum of all the 3 MN-disks combinations fitted to each disk subcomponent (red curve). {\it Right-hand panel}: vertical profile of the volume density of the Galactic disk model as a function of the height $z$ from the mid-plane and at three different arbitrary radii: $R=2$ kpc (solid lines); $R=8$ kpc ($R_0$) (dashed lines); $R=15$ kpc (dotted lines). The blue curves are also related to the `observation-based' disk model, and the red curves to the total 3 MN-disks combinations.}
\label{fig:disk_Sig_rho}
\end{figure*}

Figure~\ref{fig:disk_Sig_rho} shows, on the left-hand panel, the radial profile of the surface mass-density for the `observation-based' disk model (blue curve), as well as the total surface density resultant from all the 3 MN-disks combinations fitted to each disk subcomponent (red curve). It can be seen a good agreement between these two curves, which denotes the great effectiveness of the method. The fractional differences between these surface density distributions are $<10\%$ at $1\,\mathrm{kpc}\lesssim R \lesssim 21.5\,\mathrm{kpc}$, reaching 50\% at $R\sim 26$ kpc and 100\% at $R\sim 29$ kpc, where the absolute values of the densities become lower than $\sim 0.5$ M$_{\odot}$ pc$^{-2}$. These features show the tendency of the MN-disks in returning larger densities at large radii.
The local disk surface density returned by the MN-disks fit model is $\Sigma_{0\mathrm{d}}=59.4$ M$_{\odot}$ pc$^{-2}$, which can be considered close to the prior value of 57.2 M$_{\odot}$ pc$^{-2}$ of the `observation-based' disk model, if we take the uncertainty of $\pm4.9$ M$_{\odot}$ pc$^{-2}$ over this quantity as estimated by \citet{Read2014}. The local surface density integrated within 1.1 kpc of the disk mid-plane returned by the MN-disks fit model is $\Sigma_{0\mathrm{d_{1.1\,kpc}}}=57.0$ M$_{\odot}$ pc$^{-2}$, in agreement with the values reported by \citet{Bienayme2006}. On the right-hand panel of Fig.~\ref{fig:disk_Sig_rho}, we show the distribution of the volume density as a function of the height $z$ from the disk mid-plane, taken at three arbitrary radii: $R=2$ kpc (solid line), $R=8$ kpc (dashed line), and $R=15$ kpc (dotted line). The blue and red curves also correspond to the density distributions of the `observation-based' disk model and the MN-disks fit model, respectively. The fractional differences between the volume density distributions are lesser than $25\%$ for $|z|\lesssim 2.3$ kpc at $R=2$ kpc, for $|z|\lesssim 1$ kpc at $R=8$ kpc, and for $|z|\lesssim 0.35$ kpc at $R=15$ kpc, reaching values greater than $25\%$ beyond these heights. These features denote the systematic trend of the MN-disks in returning higher densities at greater distances from the disk plane, while at relatively small $z$ the density distributions show roughly good matches at radii up to at least $R\sim 22$ kpc. As pointed out by \citet{Smith2015}, it is impossible for the 3 MN-disks model to perfectly reproduce the vertically exponential density profile, or even the $\mathrm{sech}^{2}(z)$ type profile, since they are mathematically distinct. The same can be stated on the vertical Gaussian profile adopted for the H\,{\scriptsize I} and H$_2$ disks in the present work; at heights lower than 1 kpc the 3 MN-disks start presenting great deviations from the Gaussian profile. This can be noticed analysing the distributions of $\rho(z)$ calculated at $R=15$ kpc (dotted lines in the right-hand panel of Fig.~\ref{fig:disk_Sig_rho}), where the density of the H\,{\scriptsize I} disk overtakes those of the other disk subcomponents (cf. Fig.~\ref{fig:disks_Sigs}).

\begin{figure*}
\centering
\includegraphics[scale=0.50]{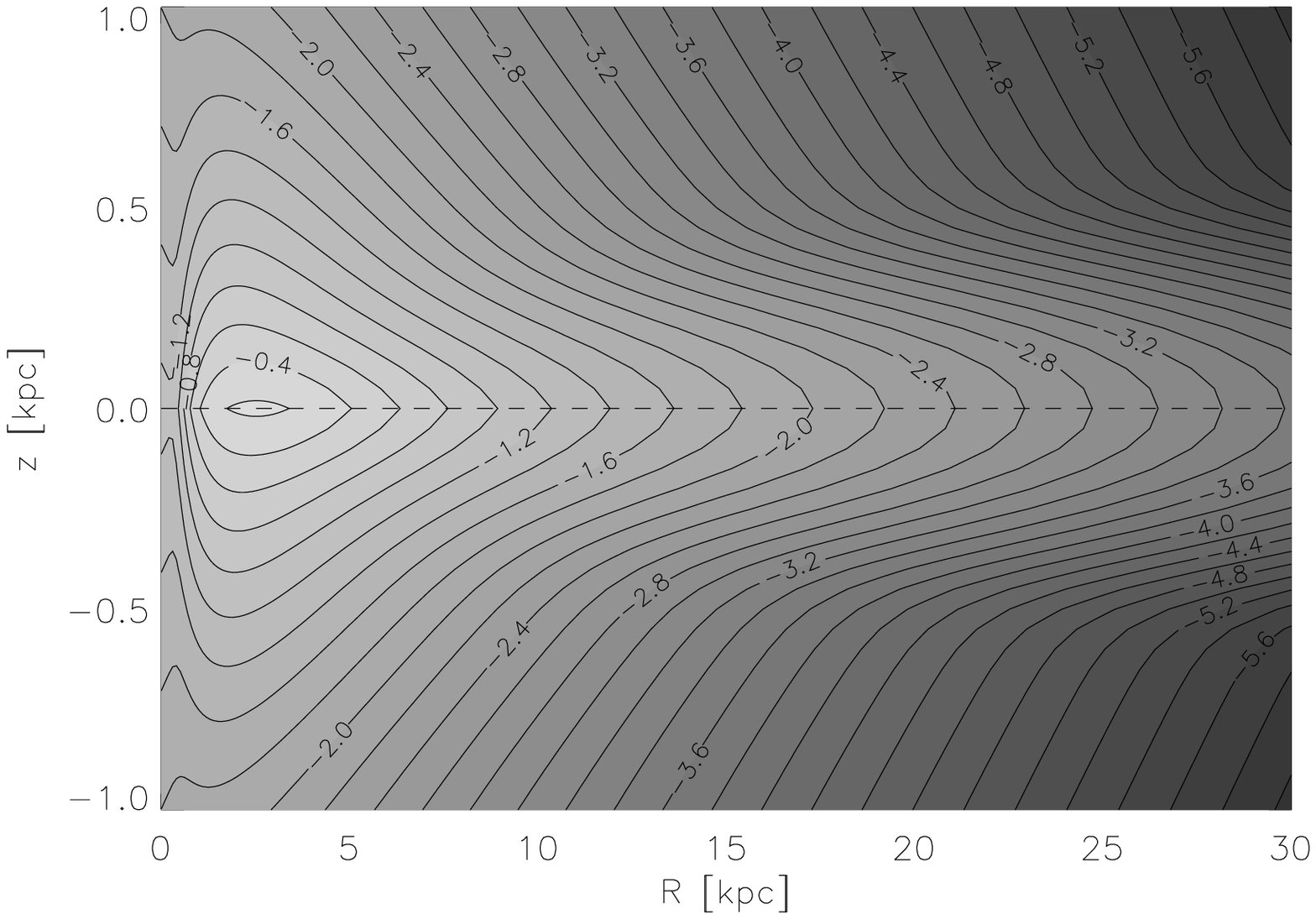}
\includegraphics[scale=0.50]{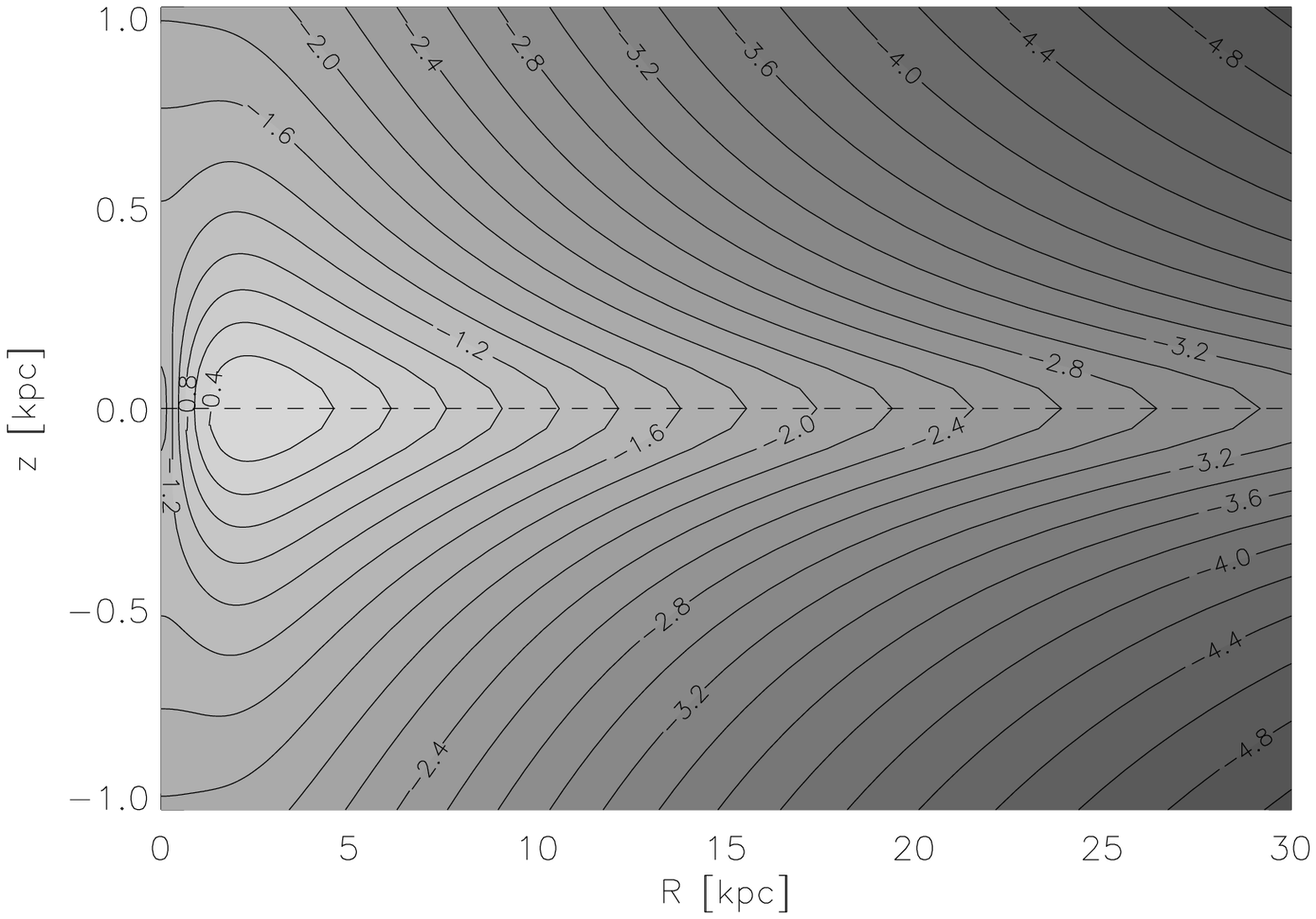}
\caption{Contours of iso-densities $\log \rho$ for the disk models built in the present work: {\it left-hand panel} - the `observation-based' disk model; {\it right-hand panel} - the sum of all the 3 MN-disks combinations fitted to each disk subcomponent. The values of the contours lie in the range $\log \rho=[-6;\,+1]$, with $\rho$ in M$_{\odot}$ pc$^{-3}$.}
\label{fig:disk_rho_contours}
\end{figure*}

Figure~\ref{fig:disk_rho_contours} shows the contours of iso-densities in the form $\log \rho$ ($\rho$ in M$_{\odot}$ pc$^{-3}$) in the meridional plane of the Galaxy: for the `observation-based' disk model (left-hand panel), and for the total 3 MN-disks combinations fit model (right-hand panel).

\subsection{The gravitational potential of the disk}
\label{pot_disk}
 
\begin{figure}
\centering
\includegraphics[scale=0.50]{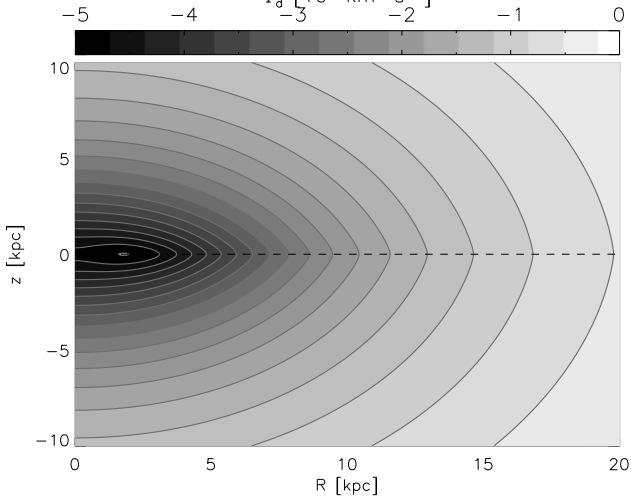}
\caption{Contours of equipotentials in the Galactic meridional plane for the gravitational potential resulting from the disk model composed by combinations of MN-disks, as described in Sect.~\ref{MN-disks}. The contours of $\Phi_{\mathrm{d}}$ are given in $10^4$ km$^2$ s$^{-2}$, as indicated in the colorbar.}
\label{fig:disk_Phi_contours}
\end{figure}

The gravitational potential expressions related through Poisson equation to the densities of the Miyamoto-Nagai disk models, given by Eqs.~\ref{eq:rho_MN1},~\ref{eq:rho_MN2} and~\ref{eq:rho_MN3}, are written as `model 1', `model 2' and `model 3', respectively, in the form:

\begin{equation}
\label{eq:Phi_MN1}
\Phi_{\mathrm{MN}_{1}}(R,z)=\frac{-GM}{\left[R^{2}+\left(a+\zeta \right)^{2}\right]^{1/2}}\,,
\end{equation}

\begin{equation}
\label{eq:Phi_MN2}
\Phi_{\mathrm{MN}_{2}}(R,z)=\frac{-GM}{\left[R^{2}+\left(a+\zeta \right)^{2}\right]^{1/2}}\left[1+\frac{a(a+\zeta)}{R^{2}+(a+\zeta)^{2}} \right]\,,
\end{equation}

\begin{multline}
\label{eq:Phi_MN3}
\Phi_{\mathrm{MN}_{3}}(R,z)=\frac{-GM}{\left[R^{2}+\left(a+\zeta \right)^{2}\right]^{1/2}}\left\{1+\frac{a(a+\zeta)}{R^{2}+(a+\zeta)^{2}}+\right.\\
-\left.\frac{1}{3}\frac{a^{2}\left[R^{2}-2(a+\zeta)^{2}\right]}{\left[R^{2}+(a+\zeta)^{2}\right]^{2}}\right\}\,,
\end{multline}
\noindent
where again $\zeta=\sqrt{z^{2}+b^{2}}$. Therefore, the gravitational potential of the thin stellar disk $\Phi_{\mathrm{d}_{\,thin}}$ is modelled with 3 components of the potential $\Phi_{\mathrm{MN}_{3}}$ (Eq.~\ref{eq:Phi_MN3}), and whose parameters $M_i$, $a_i$ and $b$ ($i=1,\,2,\,3$) are given in the first row of Table~\ref{tab:disks-MN_params}. Equivalently, the thick stellar disk potential $\Phi_{\mathrm{d}_{\,thick}}$ is modelled with 3 components of the potential $\Phi_{\mathrm{MN}_{1}}$ (Eq.~\ref{eq:Phi_MN1}), the H\,{\scriptsize I} disk potential $\Phi_{\mathrm{d_{\,HI}}}$ is modelled with 3 components of $\Phi_{\mathrm{MN}_{2}}$ (Eq.~\ref{eq:Phi_MN2}), and the H$_2$ disk potential $\Phi_{\mathrm{d_{\,H_{2}}}}$ with 3 components of the potential $\Phi_{\mathrm{MN}_{3}}$ (Eq.~\ref{eq:Phi_MN3}); the parameters $M_i$, $a_i$ and $b$ for these disks are given in the second, third and fourth rows, respectively, of Table~\ref{tab:disks-MN_params}. The total gravitational potential of the disk $\Phi_{\mathrm{d}}$ is then given by the sum of all the combinations of 3 MN-disks that fit each Galactic disk subcomponent:

\begin{equation}
\label{eq:Phi_tot}
\Phi_{\mathrm{d}}=\Phi_{\mathrm{d}_{\,thin}}+\Phi_{\mathrm{d}_{\,thick}}+\Phi_{\mathrm{d_{\,HI}}}+\Phi_{\mathrm{d_{\,H_{2}}}}\,.
\end{equation}
\noindent
Figure~\ref{fig:disk_Phi_contours} shows the contours of equipotential curves in the plane $R-z$ for the gravitational potential of the disk $\Phi_{\mathrm{d}}$, expressed in Eq.~\ref{eq:Phi_tot}.

\section{Galactic models}
\label{MW_models}


\subsection{Bulge and dark halo components}
\label{bulge_dark-halo}

For the mass density distribution in the spheroidal component of the Galaxy, here assumed as composed by the central bulge and the stellar halo as a smooth extension of the bulge itself, we adopt the profile proposed by \citet{Hernquist1990} which reproduces the $r^{1/4}$ \citet{deVaucouleurs1977} surface brightness law over a large range of galactic radius:

\begin{equation}
\label{eq:rho_bulge}
\rho_{\mathrm{b}}(R,z)=\frac{M_{\mathrm{b}}}{2\pi}\,\frac{a_{\mathrm{b}}}{\sqrt{R^{2}+z^{2}}}\,\frac{1}{\left(\sqrt{R^{2}+z^{2}}+a_{\mathrm{b}}\right)^{3}}\,,
\end{equation}
\noindent
where $M_{\mathrm{b}}$ is the total mass and $a_{\mathrm{b}}$ is the scale radius of the bulge. The gravitational potential associated with the density distribution in Eq.~\ref{eq:rho_bulge} was shown by \citet{Hernquist1990} to be in the form:

\begin{equation}
\label{eq:Phi_bulge}
\Phi_{\mathrm{b}}(R,z)=\frac{-G\,M_{\mathrm{b}}}{\sqrt{R^{2}+z^{2}}+a_{\mathrm{b}}}\,.
\end{equation}

In the PJL star counts model, the bulge is modelled as an oblate spheroid, which is described by a modified Hernquist profile with three free parameters: the oblateness parameter $\kappa$, the spheroid to disk local stellar density ratio for normalization of the density profile, and the spheroid scale radius $a_{H} = 0.4\pm 0.1$ kpc. Considering the local stellar density of our `observation-based' disk model, the mass of the spheroid of PJL's model inside $R=3$ kpc (radius which encompasses $\sim 90$\% of the total spheroid mass) results in $M_{\mathrm{sph}}(R<3\,\mathrm{kpc}) \approx 2.2\times 10^{10}$ M$_{\odot}$. We use this information to constrain the initial ranges of the parameters $M_{\mathrm{b}}$ and $a_{\mathrm{b}}$ of our bulge models for the search for their best values in the fitting procedure described in Sect.~\ref{fit_proc}. For the bulge mass we adopt the initial guess interval $M_{\mathrm{b}} = 2.4 - 2.8\times 10^{10}$ M$_{\odot}$; for the scale radius we use $a_{\mathrm{b}} = 0.3 - 0.5$ kpc. These intervals comprise masses inside $R=3$ kpc for our bulge models with possible values in the range $1.8 - 2.3\times 10^{10}$ M$_{\odot}$.

\

The fact that the observed rotation curve, which is adopted to constrain the Galactic models, is nearly flat in the interval $R_0 \lesssim R \lesssim 2R_0$ (see Sect.~\ref{obs_constr}), besides the fact that in our models (see Sect.~\ref{result_discuss}) the sole disk+bulge density distributions are not capable to give such a support to the rotation curve at these radii, lead us to take into consideration the contribution of an extra mass component to which we refer by the often-used term `dark halo'. Since there is still some tension in the explanations for the behaviour of the rotation curves at large radii of some external spiral galaxies, as for the Milky Way, here we do not discuss about the reality of such dark halo component, nor about its material content, although it contributes as an extra term of mass in our Galactic models. For this reason, we do not go deeper in the modelling of such component, and take simple forms for its density and potential functions.

We model the dark halo component with a logarithmic potential of the form (e.g. \citealt{Binney_Tremaine2008}):

\begin{equation}
\label{eq:Phi_halo}
\Phi_{\mathrm{h}}(R,z)=\frac{v_{\mathrm{h}}^{2}}{2}\,\ln \left(R^{2}+z^{2}\,q_{\phi}^{-2}+r_{\mathrm{h}}^{2}\right)\,,
\end{equation}
\noindent
where $r_{\mathrm{h}}$ is the core radius; $v_{\mathrm{h}}$ is the circular velocity at large $r$ (i.e., relative to the core radius); and $q_{\phi}$ is the axis ratio of the equipotentials. For simplicity, we consider a spherical dark halo, for which $q_{\phi}=1$. The density distribution corresponding to the potential $\Phi_{\mathrm{h}}$ is given by (for $q_{\phi}=1$):

\begin{equation}
\label{eq:rho_halo}
\rho_{\mathrm{h}}(R,z)=\frac{v_{\mathrm{h}}^{2}}{4\pi G}\frac{\left(R^{2}+z^{2}+3r_{h}^{2}\right)}{\left(R^{2}+z^{2}+r_{\mathrm{h}}^{2}\right)^{2}}\,.
\end{equation} 
\noindent
This profile yields a flat rotation curve at large radii. However, as we make clear in the next sections, we are interested in a description of the Galactic potential which can be suitable for the study of stellar orbits that do not reach great extensions of the outer disk, so the real behaviour of the rotation curve of the Galaxy at very large radii is not of our concern in the present study.

The values of the parameters $M_{\mathrm{b}}$ and $a_{\mathrm{b}}$ that describe the bulge component, and $r_{\mathrm{h}}$ and $v_{\mathrm{h}}$ for the dark halo, are found after fitting the contributions of such components to the Galactic rotation curve, as well as to other observational constraints as described in Sect.~\ref{obs_constr}.


\subsection{The disk component}
\label{disk_compnt}

The model constructed for the mass distribution in the disk of the Galaxy, and its associated gravitational potential, is described in details in the subsections of Sect.~\ref{MW_disk_mass_models}. In general, the overall disk is modelled by a superposition of distinct models of Miyamoto-Nagai disks (combinations of 3 MN-disks for each one of the 4 subcomponents: thin, thick, H\,{\scriptsize I} and H$_2$ disks $\rightarrow$ 12 MN-disks in total). Each MN-disk has 3 free parameters ($M$, $a$, $b$), but since each combination of 3 MN-disks is modelled with a single value for $b$ (see Sect.~\ref{MN-disks}), we have a total of 28 parameters for the construction of the disk (all of them are given in Table~\ref{tab:disks-MN_params}).

In the present work, differently from other studies, the values of the structural parameters of the disk (here represented by the length scales $a$ and $b$ of the MN-disks) are not constrained by the kinematic information from the rotation curve of the Galaxy. Here, we opt to constrain such parameters based only on the information brought by the star-counts model of PJL and the observed distribution of the gaseous component in the disk. 
It must be said that, however, some of the disk scale parameters in the Galactic model of PJL possess considerable uncertainties, as for instance the radial length of the central hole of the thin disk $R_{\mathrm{ch}_{\,thin}}=2.07_{-0.8}^{+2.0}$ kpc, which the kinematic data would help to reduce. 
Although this can represent a drawback of our approach, in Sect.~\ref{uncert_estim} we give an estimate of the uncertainties on the parameters $a$ and $b$ of the MN-disks based on the 
uncertainties on the scale parameters of the `observation-based' disk model. 
Our reason for leaving the scale parameters fixed at the values from Table~\ref{tab:disks-MN_params} is justified by the fact that, due to the way the MN-disks were constructed, the scale-lengths $a$ were found as functions of the scale-heights $b$, and they should preserve the relationships during the fitting of the kinematic data. However, a proper fitting of the kinematic constraints should vary these scale lengths in an independent way, in order to probe the correlations among the parameters. Moreover, it is well known in the literature that the kinematic data commonly used for the models fitting do not provide strong constraints on the vertical distribution of the Galaxy's mass (e.g. \citealt{Dehnen_Binney1998,McMillan2011}). Therefore, different sets of disks scale-heights would be found able to reproduce the kinematic data equally well.
Another problematic issue in varying the scales $a$ and $b$ of all the MN-disks in the fitting of the models, taking into account the uncertainties on the values of $R_{\mathrm{d}}$, $R_{\mathrm{ch}}$ and $h_{z}$ of all the disk subcomponents, is the high computing time involved in the whole process. We recognize this disadvantageous point of our approach in fitting MN-disks to the observed disk; other studies in the literature that do not make use of this approximation are able to vary the scale parameters of their disk potential models in a more straightforward way (e.g. \citealt{McMillan2011,Piffl2014b}). 

The total mass of the disk, otherwise, is left to be constrained not only by the rotation curve which gives information on the dynamical mass of the Galaxy, but also by prior informations on the local disk surface density in visible matter. Therefore, for the fitting of the disk model to the observational constraints described in Sect.~\ref{obs_constr}, we take the values of the scale-lengths $a_i$ and scale-heights $b$ of the MN-disks as fixed and equal to the ones given in Table~\ref{tab:disks-MN_params}. The total mass of the disk will be adjusted introducing the parameter $f_{mass}$, a scale factor by which the disk total mass is then multiplied. Once the best value for $f_{mass}$ is found, then the masses $M_i$ of all the MN-disks in Table~\ref{tab:disks-MN_params} have to be multiplied by this factor. In this way, in the fitting process of the disk model based on the dynamical constraints, only the disk total mass is treated as a free parameter. Although the individual masses $M_i$ of the MN-disks have already been found after the construction of the disk mass model in Sect.~\ref{MW_disk_mass_models}, in the next step we will use their single values in Table~\ref{tab:disks-MN_params} as first guesses for the optimal fit values to the dynamical constraints, but keeping unaltered their relative contributions to the disk total mass since all of them will be rescaled by the same factor $f_{mass}$.

\subsubsection{An alternative model - a disk with a ring density structure near the solar orbit radius}
\label{ring_dens_min}

There are reasons to believe that the Galactic disk presents a local minimum of density associated with the co-rotation resonance radius of the Galactic spiral structure. 
Such a minimum density would be present in both the stellar and gaseous disks components. A structure like this one has been observed in the hydrogen distribution of the disk, as discussed later in this section. A minimum in the density of Cepheids is seen at the same radius (BLJ, see their Figure 14). However, being relatively young, the Cepheids possibly trace the gas distribution, and therefore the recent past of star formation. There is a theoretical argumentation and simulations given in BLJ, predicting that the co-rotation resonance scatters out the stars from that radius, situated close to the solar radius, and produces a minimum in the stellar density. However, there is no direct strong evidence, from photometric studies of the disk, for such a ring of minimum density of stars. For this reason, our proposed ring density structure is still a speculative one, which we attempt to test in this work.

In \citet{Zhang1996}, based on N-body simulation results of a spiral galaxy evolution and also on theoretical predictions about secular processes of energy and angular momentum transfer between stars and the spiral density wave, the author showed that a local minimum in the stellar density of the disk is formed, centered at the co-rotation radius. BLJ performed numerical integrations of test-particles orbits with a representative model for the potential due to the axisymmetric density distribution in the Galactic disk and models for the gravitational perturbation due to the spiral arms. The authors verified that a minimum of stellar density with relative amplitude $\sim 30\%-40\%$ of the background density is formed at the co-rotation radius, in a time interval $\sim 3$ billion years of the evolution of the system. \citet*{Lepine2001}, based on a simulation of gas-cloud dynamics in the spiral gravitational field model of the Galaxy, also verified a gap in the ISM distribution at the co-rotation radius. The authors compared their results (cf. their Figure 4) with the observed H\,{\scriptsize I} radial distribution presented by \citet[see his Figure 6]{Burton1976} and found close similarities between them. Such results are compatible with evidences for a ring void of gas as observed by \citet*{Amores2009} as gaps in the H\,{\scriptsize I} density distributed in a ring-like structure with radius slightly outside the solar circle. Recent studies on the three-dimensional distribution of the neutral and molecular hydrogen in the Galactic disk by \citet{Nakanishi_Sofue2016} and \citet{Sofue_Nakanishi2016} also corroborate the existence of the minimum gas density slightly beyond the solar Galactic radius. Since the co-rotation resonance is observed to be at a slightly larger but very close radius to the solar orbit one (e.g. \citealt{Dias_Lepine2005}), then the solar orbit would be placed close to the minimum of the disk surface density. 

\citet{Lepine2001} associated the minimum in the stellar and gas density distributions with a local dip in the observed rotation curve of the Galaxy at $R\sim 9$ kpc. A common association between these two features was also modelled by \citet*{Sofue2009}. In the simulation results of BLJ, the minimum density at co-rotation is followed by a local maximum density at a slightly larger radius. BLJ modelled such density feature as a wavy ring superposed on the surface density profile of the disk, in a similar way as done by \citet{Sofue2009}. 

Since we still have no clue of how could be the three-dimensional distribution of this density feature, here we make the simplest assumption that it would be mainly detected in the density distribution very near the mid-plane of the disk, so we can take the zero-thickness disk approximation for the surface density of the ring.
Here we propose a tentative analytical function for the gravitational potential associated with the ring density (local minimum followed by a maximum density), which is written in the form:

\begin{eqnarray}
\begin{aligned}
\label{eq:pot_ring}
\Phi_{ring}(R,z)&=\varphi_{R}(R)\cdot\varphi_{z}(z),\, \quad \mathrm{where} \\
\\
\varphi_{R}(R)&=-A_{ring}\,\mathrm{sech}\left[\ln\left(\frac{R}{R_{ring}}\right)^{\beta_{ring}}\right]\,\tanh\left[\ln\left(\frac{R}{R_{ring}}\right)^{\beta_{ring}}\right] \\
\mathrm{and} \\
\varphi_{z}(z)&=\mathrm{sech}\left(\frac{z}{h_{z_{ring}}}\right)\,,
\end{aligned}
\end{eqnarray}
\noindent
where $A_{ring}$ is the amplitude of the potential related to the amplitude of the minimum and maximum densities of the ring, given in units of km$^2$ s$^{-2}$; $\beta_{ring}$ is a parameter related to the ring width; $R_{ring}$ is the ring node radius - where sits the inflection point between the minimum and maximum density features; and $h_{z_{ring}}$ is the scale-height of the ring potential. Solving the Poisson equation for the ring potential $\nabla^{2}\Phi_{ring}=4\pi G\Sigma_{ring}\,\delta z$ (within the zero-thickness assumption), we can find the corresponding expression for the ring surface density $\Sigma_{ring}$. 
 
In the model that includes the ring structure, the surface density $\Sigma_{ring}$ is added to the disk surface density $\Sigma_{\mathrm{d}}$, but since the ring is modelled by a minimum followed by a maximum density of similar amplitudes, which makes the net ring mass $M_{ring}\sim 0$, the disk total mass is then kept approximately unaltered. With the ring density structure, the Galactic models are adjusted with the inclusion of the ring free parameters: $A_{ring}$, $\beta_{ring}$ and $R_{ring}$; the scale-height $h_{z_{ring}}$ is chosen to be fixed at a predetermined value.

The inclusion of the ring density feature in the Galactic models is translated in a rescaling of the disk total mass towards larger values when compared to the disk models without the ring. 
This result is better discussed in Sect.~\ref{ring_feat}.  


\section{Observational constraints}
\label{obs_constr}

In this section we present the groups of observational data used to constrain the free parameters of the Galactic models introduced in Sect.~\ref{MW_models}. These groups basically comprise kinematic data from the rotation curve of the Galaxy, with tangent velocities at radii $R<R_{0}$ and rotation velocities at $R>R_{0}$, the local angular rotation velocity $\Omega_0$, and values for the estimated total local disk mass-surface density $\Sigma_{0\mathrm{d}}$ already discussed in Sect.~\ref{MW_disk_mass_models}, as well as the surface density integrated within 1.1 kpc of the disk mid-plane. In the following, we discuss each group of constraints.


\subsection{The local angular rotation velocity and the solar kinematics}
\label{ang_vel_sol_kinet}

In this work, we take the Galactocentric distance of the Sun $R_0$ from the statistical analysis performed by \citet{Malkin2013} on 53 $R_0$ measurements published in the literature over the last 20 years, which average value, for practical purposes, is recommended by the author as being

\begin{equation*}
\label{eq:r0}
R_{0}=8.0\pm 0.25\, \mathrm{kpc}\,.
\end{equation*}
\noindent

For the peculiar velocity of the Sun $\mathbf{v}_{\odot}$ relative to the local standard of rest (LSR), we take the re-evaluation proposed by \citet*[hereafter SBD]{Schonrich2010}:

\begin{eqnarray*}
\begin{aligned}
\label{eq:sun_vel}
\mathbf{v}_{\odot}&=(u_{\odot},\,v_{\odot},\,w_{\odot})\\
&=(-11.1,\,12.24,\,7.25)\pm (1,\,2,\,0.5)\,\mathrm{km}\,\mathrm{s}^{-1}\,,
\end{aligned}
\end{eqnarray*}
\noindent
where we use a left-handed system for ($U$, $V$, $W$), in which $U$ is positive towards the Galactic anti-center, $V$ is positive in the direction of Galactic rotation, and $W$ is positive towards the direction of the North Galactic Pole; $(u_{\odot},\,v_{\odot},\,w_{\odot})$ are the Sun's velocity components in such system. The above-quoted uncertainties are the systematic ones, since these dominate the total uncertainties as estimated by SBD.

To constrain the angular velocity of the Sun $\Omega_{\odot}$, we consider the direct measurement of Sgr A$^{*}$ proper motion along the Galactic plane carried out by \citet{Reid_Brunthaler2004}, whose value is $\mu_{\mathrm{Sgr\,A^{*}}}=6.379\pm 0.024$ mas yr$^{-1}$. Thus we have $\Omega_{\odot}=\mu_{\mathrm{Sgr\,A^{*}}}=30.24\pm 0.11$ km s$^{-1}$ kpc$^{-1}$. This last equality comes from the assumption that Sgr A$^{*}$ is at rest at the Galactic center, so its apparent proper motion can be thought to be solely due to the sum of the Galactic rotation at the LSR and the solar peculiar motion with respect to the LSR in the same direction, i.e. $\Omega_{\odot}\equiv \Omega_{0}+v_{\odot}/R_{0}$ (e.g. \citealt{Honma2012}). We then have

\begin{equation}
\label{eq:Omega0}
\Omega_{0}=\Omega_{\odot}-\frac{v_{\odot}}{R_{0}}\,,
\end{equation}
\noindent
for the local angular rotation velocity $\Omega_0$. With the above-quoted values for $\Omega_{\odot}$, $v_{\odot}$ and $R_0$, one obtains $\Omega_{0}=28.7$ km s$^{-1}$ kpc$^{-1}$. The uncertainty on $\Omega_0$ is calculated as $\sigma_{\Omega_{0}}=\sqrt{\left(\sigma_{\mu_{\mathrm{Sgr\,A^{*}}}}
\right)^{2}+\left(\frac{\sigma_{v_{\odot}}}{R_{0}}\right)^{2}+\left(\frac{v_{\odot}}{R_{0}^{2}}\sigma_{R_{0}}\right)^{2}}$, which returns the value $\sigma_{\Omega_{0}}=0.4$ km s$^{-1}$ kpc$^{-1}$; to the uncertainty $\sigma_{v_{\odot}}$ of 2 km s$^{-1}$ given by SBD, we have added 1 km s$^{-1}$ to allow for a possible peculiar motion of Sgr A$^{*}$ at the Galactic center (\citealt{McMillan_Binney2010}). The corresponding rotation velocity at the LSR, $V_{0}=\Omega_{0}R_{0}$, assumes the value $V_{0}=230 \pm 8$ km s$^{-1}$.


\subsection{The rotation curve}
\label{rot_curve}

\subsubsection*{Tangent velocities:} 
The tangent (or terminal) velocity $V_{\mathrm{term}}$ is usually assumed as the maximum velocity of the ISM gas along a given line-of-sight at Galactic coordinates $b=0^{\circ}$ and $-90^{\circ}\leq l \leq 90^{\circ}$, which can be related to the circular velocity $V_{c}(R)$ at the tangent point $R=R_{0}\sin l$ (or sub-central point), considering a circularly rotating gas. For the tangent velocity data, we use the CO-line data inside the solar circle from table 2 of \citet{Clemens1985}\footnote{The tangent velocity data on table 2 of \citet{Clemens1985} were corrected for 3 km s$^{-1}$ line width, and a correction of 7 km s$^{-1}$ for the LSR peculiar motion in the azimuthal direction was used in the calculation of the rotation velocities, as proposed by the author in the above-cited paper.}, which cover longitudes
in the first Galactic quadrant. We also use the H\,{\scriptsize I} tangent point data from table 2 of \citet*{Fich_Blitz_Stark1989}, which cover both the first and fourth Galactic quadrants. We converted the LSR tangent velocities of these compiled data to heliocentric velocities and then back to LSR tangent velocities using the components of the peculiar solar motion adopted in this work. Then the Galactocentric distances $R=R_{0}\sin l$ and the rotation velocities 
$V_{\mathrm{rot}} = V_{\mathrm{term}} + V_{0}\sin l$ were calculated using the Galactic constants adopted in this work ($R_0$, $V_0$) = (8 kpc, 230 km s$^{-1}$). We have propagated the uncertainties on both $R_0$ and $V_0$ to the uncertainties on $R$ and $V_{\mathrm{rot}}$ ($\sigma_R$ and $\sigma_{V_{\mathrm{rot}}}$), respectively.

It has often been argued that the central region of the Galaxy is strongly affected by non-axisymmetric structures like the bar, which can induce non-circular motions of the ISM. The true Galactic rotation curve would then be distorted by the measurement of non-uniform azimuthal velocities, making the tangent-point method inappropriate at these regions. Recently, \citet*{Chemin2015} attempted to quantify the asymmetries in the Galactic rotation curve derived by the tangent-point method.
Figure 3 of \citet{Chemin2015} shows that the largest discrepancies between the first and fourth quadrant rotation curves are in the interval $1 \lesssim R \lesssim 2$ kpc. Perhaps the most delicate feature in the Milky Way inner rotation curve, as derived from the tangent-point method, is the velocity peak at $R\sim 300$ pc. Based on a numerical simulation
of a disk galaxy similar to the Milky Way, \citet{Chemin2015} concluded that the tangent velocities lead to an inner velocity profile with a peak which is not present in the true rotation curve, when the bar major axis is viewed with angles $< 45^{\circ}$ with respect to the direction of the galactic center. However, we argue that the interpretation of this inner peak as being due to the contribution of the bulge can still be defensible: at $R\sim 300$ pc, a centrally concentrated bulge might dominate the rotation curve, while the bar, as a more extended structure, might influence the rotation curve at radii larger than that.

Also based on the simulated galaxy, \citet{Chemin2015} show that the resulting velocity profile strongly deviates from the true rotation curve in the region $R<4$ kpc; the tangent-point method in the inner regions systematically select high-velocity gas along the bar and spiral arms, or low-velocity gas in the less dense media. The authors state that the observed rotation curve of the Milky Way derived by the tangent-point method is expected to be close to the true one only for radii $4\lesssim R\leq 8$ kpc. The unreliable rotation curve at $2\lesssim R\lesssim 4$ kpc in the Chemin et al.'s simulation is attributed to effects associated with the co-rotation of the bar, which have, in their model, a pattern speed of the order of 59 km s$^{-1}$ kpc$^{-1}$. Other authors have obtained lower values for the bar pattern speed, e.g. $30-40$ km s$^{-1}$ kpc$^{-1}$ (\citealt{Rodriguez-Fernandez_Combes2008}), or 33 km s$^{-1}$ kpc$^{-1}$ (\citealt{Li2016}). These lower patterns would put the bar corotation at radii larger than 4 kpc. The tangent-point data from \citet{Fich_Blitz_Stark1989} show a velocity difference on the two sides of the Galactic center which is of the order of 12 km s$^{-1}$ on average, in the range $R=2-4$ kpc. We think that this amplitude of asymmetry, contrary to the one observed in the region $1< R< 2$ kpc, does not have influence on our models. 
Given the considerations presented above, we decided to restrict the tangent velocity data to $|\sin l| \geq 0.3$ (e.g. \citealt{Dehnen_Binney1998}), which in our case is equivalent to Galactic radii $R \geq 2.4$ kpc. We end up with a data set which totalizes 280 rotation velocity measurements in the inner solar circle region ($2.4 \leq R \leq 8.0$ kpc).

\subsubsection*{Maser sources data:}
From very long baseline interferometry techniques, several maser sources associated with high-mass star-forming regions (HMSFRs) have been studied and their positions, parallaxes and proper motions have been measured with high accuracy. Complementing these data with heliocentric radial velocities from Doppler shifts, we can have access to the full three-dimensional location of each source in the Galaxy, as well as their full space motion relative to the Sun. Since it is believed that the maser sources do not present large peculiar motions, we can use their velocity components in the direction of Galactic rotation as a proxy for the rotation curve of the Galaxy (\citealt{Irrgang2013}), taking for this the value of $v_{\odot}$ estimated by SBD. The data for HMSFRs with maser emission were obtained from table 1 of \citet{Reid2014}, where the authors list parallaxes, proper motions, and LSR radial velocities of 103 regions measured with Very Long Baseline Interferometry (VLBI) techniques from different surveys and projects (the Bar and Spiral Structure Legacy (BeSSeL) Survey\footnote{http://bessel.vlbi-astrometry.org} and the Japanese VLBI Exploration of Radio Astrometry (VERA)\footnote{http://veraserver.mtk.nao.ac.jp}). We converted the tabulated LSR radial velocities to heliocentric radial velocities by adding back the components of the standard solar motion (\citealt{Reid2009}). With the coordinates, parallaxes, proper motions, heliocentric radial velocities, and their respective errors, we calculated the heliocentric $U$, $V$, $W$ velocities and uncertainties $\sigma_{U}$, $\sigma_{V}$ and $\sigma_{W}$ for each source, following the formalism described by \citet{Johnson_Soderblom1987}. Correcting for the SBD's solar peculiar motion and for the LSR circular velocity $V_0$, we calculated the Galactocentric component $V_{\phi}$ of the space velocity of each maser source in the direction of Galactic rotation. The uncertainties on $V_{\phi}$, $\sigma_{V_{\phi}}$, were obtained by propagation from the uncertainties on the parallaxes, proper motions, heliocentric radial velocities, and the uncertainties of the solar motion. The rotation velocities and uncertainties are then defined by $V_{\mathrm{rot}}=V_{\phi}$ and $\sigma_{V_{\mathrm{rot}}}=\sigma_{V_{\phi}}$.
The Galactic radii were obtained directly from the positions and parallaxes of the sources, as well as their uncertainties. The distribution of Galactic radii comprises regions both inside and outside the solar circle. We have discarded the sources with radii $R<4$ kpc because most of them present values of $V_{\phi}$ with large deviations from the rotation curve traced by the tangent velocity data at these radii; a similar selection criterion was used by \citet{Reid2014}. This selection reduces the number of maser sources used to probe the rotation curve to 94 objects.

Although the rotation curve for $R > R_0$ traced by the maser sources is restricted to few data points, these measurements are the best ones that we can think of, considering the precision in the distances, proper motions, and line-of-sight velocities. The less accurate distances of H\,{\scriptsize II} regions, for instance, can produce false trends in the rotation curve traced by these objects, which is the main reason for not using them in the present study. Indeed, \citet{Binney_Dehnen1997} pointed out that the apparent rising rotation curve traced by the H\,{\scriptsize II} regions outside the solar radius can be explained by the objects
tending to be concentrated in a `ring' (which could be a segment of a spiral arm) with a mean radius larger than their estimated Galactic radii. We believe that these shortcomings disappear when we take the outer rotation curve traced by the maser sources data with well-determined distances. 
A more quantitative analysis on the correlations between the velocity uncertainties and distance uncertainties of the maser sources is presented in Sect.~\ref{ring_feat}.


\subsection{Local surface densities}
\label{surf_dens_local}

As presented in Sect.~\ref{observ_basis}, our `observation-based' disk model is constructed to give a total local disk mass-surface density of $\Sigma_{0\mathrm{d}}=57.2$ M$_{\odot}$ pc$^{-2}$, which is based on recent determinations in the literature about the local surface densities in both stellar and gaseous components. This is also the value that we adopt as constraint to the local dynamical disk surface density, which can be compared to the estimate by \citet{Holmberg_Flynn2004} of $56\pm 6$ M$_{\odot}$ pc$^{-2}$ for this quantity. We adopt the same uncertainty on $\Sigma_{0\mathrm{d}}$ of $\sigma_{\Sigma_{0\mathrm{d}}}=6$ M$_{\odot}$ pc$^{-2}$. 
\citet{Holmberg_Flynn2004} also estimated the local surface density integrated within 1.1 kpc of the disk mid-plane as being $\Sigma_{0_{1.1\mathrm{kpc}}}=74\pm 6$ M$_{\odot}$ pc$^{-2}$, which,  according to the authors, takes into account both disk and dark halo contributions. We take such value as observational constraint on $\Sigma_{0_{1.1\mathrm{kpc}}}$ and its uncertainty.


\section{Fitting procedure}
\label{fit_proc}

As exposed in Sect.~\ref{MW_models}, we attempt to construct two types of Galactic models: those which do not incorporate the ring density structure in the disk (Sect.~\ref{ring_dens_min}), and which we refer to as model M{\sc I}; and those which do incorporate the ring density structure and will be denoted by model M{\sc II}. The free parameters of model M{\sc I} are the bulge parameters $M_{\mathrm{b}}$ and $a_{\mathrm{b}}$, the dark halo parameters $r_{\mathrm{h}}$ and $v_{\mathrm{h}}$, and the disk mass scale factor $f_{mass}$. The free parameters of model M{\sc II} are the same of model M{\sc I}, plus the ring parameters $A_{ring}$, $\beta_{ring}$ and $R_{ring}$; the scale-height $h_{z_{ring}}$ is chosen to be fixed at the value 0.65 kpc. 

We search for the best-fit set of parameters for both models M{\sc I} and M{\sc II} using a $\chi^{2}$-minimization procedure, which is implemented through the cross-entropy (CE) algorithm (Sect.~\ref{CE_algorithm}). We first try initial guesses for the parameters sets $\left\{M_{\mathrm{b}},\,a_{\mathrm{b}},\,r_{\mathrm{h}},\,v_{\mathrm{h}},\,f_{mass}\right\}$ of model M{\sc I} and $\left\{M_{\mathrm{b}},\,a_{\mathrm{b}},\,r_{\mathrm{h}},\,v_{\mathrm{h}},\,f_{mass},\,A_{ring},\,\beta_{ring},\,R_{ring}\right\}$ of model M{\sc II} based on visual fits of the given model to the observed rotation curve. 
From these trial sets of parameters, we randomly generate $N=3\times 10^{3}$ independent sets of model parameters by considering initial uniform distributions centered on the given trial parameters and with half-widths equal to the initial uncertainties chosen for each parameter. We select the best 100 sets candidates that compose the elite sample of sets based on the $\chi^{2}$-minimization criterion (the $N_{\mathrm{elite}}=100$ sets that return the lowest values for $\chi^{2}$). From the mean $\mu$ and standard deviation $\sigma$ of each ensemble of 100 parameters of this elite sample, we generate $N$ other independent sets of trial parameters through Gaussian distributions $N(\mu,\,\sigma^{2})$ which will be evaluated in the next iteration. 
This process is repeated by several iterations until we obtain a stable set of parameters for each model M{\sc I} and M{\sc II}.

The total gravitational potential $\Phi$ is calculated as the sum of the gravitational potentials of each individual Galactic component: bulge, disk and dark halo, $\Phi(R,z)=\Phi_{\mathrm{b}}(R,z)+\Phi_{\mathrm{d}}(R,z)+\Phi_{\mathrm{h}}(R,z)$, in the case of model M{\sc I}, and the addition of the ring potential in the case of model M{\sc II}, $\Phi(R,z)=\Phi_{\mathrm{b}}(R,z)+\Phi_{\mathrm{d}}(R,z)+\Phi_{\mathrm{h}}(R,z)+\Phi_{ring}(R,z)$. Considering the balance between the centrifugal force and gravity, the circular velocity $V_{c}(R)$ of a given model, measured for instance in the Galactic mid-plane $z=0$, is linked to the total gravitational potential by the form

\begin{equation}
\label{eq:vel_circ}
V_{c}(R)=\sqrt{R\,\frac{\mathrm{d}\Phi(R,0)}{\mathrm{d}R}}\,.
\end{equation}
\noindent
Therefore, the gravitational potential $\Phi$ and the model rotation curve $V_{c}(R)$ are totally defined by the set of parameters of each model M{\sc I} and M{\sc II} described above. The radial derivative of the potential $\mathrm{d}\Phi(R,0)/\mathrm{d}R$ in Eq.~\ref{eq:vel_circ} is given by the sum of the radial derivatives of the potentials of each Galactic component, for which explicit expressions are presented in Appendix~\ref{app2}. Each data point $i$ of the rotation curve has a pair of values ($R_i$, $V_{\mathrm{rot}_{\,i}}$). In the fitting process of the rotation curve, we search for the minimization of the residuals between the rotation velocities $V_{\mathrm{rot}_{\,i}}$ of the observational data and the circular velocities $V_{c}(R_{i})$ resulted from the model at each radius $R_{i}$. We thus minimize the quantity

\begin{equation}
\label{eq:chisqr_velrot}
\chi_{\mathrm{rc}_{\,k}}^{2}=\sum\limits_{i}\left[\frac{V_{\mathrm{rot}_{\,i}}-V_{c}(R_{i})}{\sigma_{V_{tot_{\,i}}}}\right]^{2}\,,
\end{equation}
\noindent
where we denote $\chi_{\mathrm{rc}_{\,k}}^{2}$ as the chi-squared component relative to the rotation curve observational constraint.
We separate the rotation curve data into four different groups: group 1 corresponds to the CO tangent velocity data; group 2 corresponds to the H\,{\scriptsize I} tangent velocities in the first quadrant ($l>0^{\circ}$); group 3 is relative to the H\,{\scriptsize I} tangent velocities in the fourth quadrant ($l<0^{\circ}$); and group 4 comprises
the maser sources data. Then the $\chi_{\mathrm{rc}_{\,k}}^{2}$ in Eq.~\ref{eq:chisqr_velrot} is calculated for each group separately, where the subindex $k$ denotes the number of the group.

As the data points of the rotation curve present errors in both $R$ and $V_{\mathrm{rot}}$ ($\sigma_R$, $\sigma_{V_{\mathrm{rot}}}$), a proper fitting procedure has to take into account the effects of both groups of errors on the model to be adjusted. We deal with this issue by following the solution adopted by \citet{Irrgang2013}: the uncertainty $\sigma_R$ is converted to an error in $V_{\mathrm{rot}}$ by estimating its effect on the model rotation curve according to the relation $\sigma_{V_{c}} = (\mathrm{d}V_{c}/\mathrm{d}R)\,\sigma_{R}$, and then $\sigma_{V_{c}}$ is
added in quadrature to $\sigma_{V_{\mathrm{rot}}}$. The total velocity uncertainty $\sigma_{V_{tot}}$ in Eq.~\ref{eq:chisqr_velrot} is just
given by $\sigma_{V_{tot}} = \sqrt{\sigma_{V_{\mathrm{rot}}}^{2} + \sigma_{V_{c}}^{2}}$.

Regarding the other observational constraints discussed in Sect.~\ref{obs_constr}, namely, the local angular rotation velocity $\Omega_0$, the local disk mass-surface density $\Sigma_{0\mathrm{d}}$ and the surface density within $|z|\leq 1.1$ kpc $\Sigma_{0_{\mathrm{1.1kpc}}}$, they contribute to the total $\chi^{2}$ in the form

\begin{equation}
\label{eq:chisqr_others}
\chi_{\mathrm{other}}^{2}=\displaystyle\sum\limits_{j=1}^{3}
\left(\frac{\psi_{j,\,\mathrm{obs}}
-\psi_{j,\,\mathrm{model}}}{\sigma_{\psi_{j,\,\mathrm{obs}}}}\right)^{2}\,,
\end{equation}
\noindent
where $\psi_{j,\,\mathrm{obs}}$ and $\sigma_{\psi_{j,\,\mathrm{obs}}}$ refer to each one of the three $j$ above-mentioned observables and their associated uncertainties, respectively; $\psi_{j,\,\mathrm{model}}$ refers to the respective quantities resulted from the models. The total weighted $\chi^{2}$ is just the sum

\begin{equation}
\label{eq:chisqr_tot}
\chi_{\mathrm{tot}}^{2}=\displaystyle\sum\limits_{k=1}^{4}
\left(\frac{\chi_{\mathrm{rc}_{\,k}}^{2}}{N_{\mathrm{rc}_{\,k}}}\right)+\frac{\chi_{\mathrm{other}}^{2}}{N_{\mathrm{other}}}\,.
\end{equation}
\noindent
where $N_{\mathrm{rc}_{\,k}}$ is the number of data points in each group $k$ of rotation curve data, and $N_{\mathrm{other}}=3$. 
Each contribution of the groups of observational constraints to the total $\chi^{2}$ is divided by the number of observational data actually used in the group. This is the same procedure adopted by \citet{Dehnen_Binney1998} and \citet{Irrgang2013} to ensure that the fitting process will not be dominated by the rotation curve due to the larger number of individual data points in this group of observational constraints.


\subsection{Uncertainty estimates}
\label{uncert_estim}

We apply Monte-Carlo techniques to obtain the uncertainties on the fitting parameters of the Galactic models. From the original observational data set compiled for the rotation curve, namely the Galactic radii $R$ and rotation velocities $V_{\mathrm{rot}}$ of the sources, we create $N_{run}=100$ new data sets by re-sampling the original one with replacement of the data to perform a bootstrap-like procedure (e.g. \citealt{Monteiro_Dias_Caetano2010}).  
We then run the CE algorithm that implements the fitting procedure described above $N_{run}$ times to find the best-fitting parameter sets, each time with one of the ``new rotation curve'' re-sampled in the way explained above. We end up with 100 sets of parameters for each model M{\sc I} and M{\sc II}, from which the uncertainties on the parameters $\left(M_{\mathrm{b}},\,a_{\mathrm{b}},\,r_{\mathrm{h}},\,v_{\mathrm{h}},\,f_{mass},\,A_{ring},\,\beta_{ring},\,R_{ring}\right)$ are determined. It was found that the average values of the parameters obtained from these 100 parameter sets are very close to the best-fit values found after the fitting of the models using the original rotation curve data. However, we prefer to consider the latter ones as the best values since they were directly obtained from the original data set. 
   
The uncertainties on the scale-lengths $a$, scale-heights $b$, and masses $M$ of the Miyamoto-Nagai disks depend directly on the uncertainties on the structural parameters and local surface densities adopted for the `observation-based' disk model. Therefore, a proper estimation of such uncertainties should take into consideration the combination of all errors of the parameters used to constrain the MN-disks models. A rough calculation, using Monte-Carlo techniques, indicates that uncertainties on the $a_i$ and $b_i$ parameters ($\sigma_{a_{i}}$, $\sigma_{b_{i}}$) of $\sim10\%$ of their values listed in Table~\ref{tab:disks-MN_params} seem to be appropriate estimates. The uncertainties on the masses of the MN-disks are considered proportional to the uncertainties on the mass scale factor $\sigma_{f_{mass}}$, i.e. $\sigma_{M_{i}}=M_{i}\,\sigma_{f_{mass}}$, with $M_{i}$ also given in Table~\ref{tab:disks-MN_params}.


\section{Results and discussions}
\label{result_discuss}

The best-fitting values for the parameters of both Galactic models M{\sc I} and M{\sc II} are summarized in Table~\ref{tab:params_MI-MII}. Table~\ref{tab:deriv_quant_MI-MII} lists the values for the local dynamical properties resulting from the models, some of which can be compared to the values used as observational constraints and, given the uncertainties, a good overall agreement is observed.
Figure~\ref{fig:rotcurve_models-MI-MII} shows, in the top panels, the rotation curves resulting from each Galactic model (solid lines), as well as the curves relative to the contribution of each Galactic component. The data for the observed rotation curve are presented as: red points with orange error bars for the CO tangent-point data (\citealt{Clemens1985}) and for the H\,{\scriptsize I} tangent-point data (\citealt{Fich_Blitz_Stark1989}); blue points with light blue error bars for the maser sources data (\citealt{Reid2014}). The bottom panels present the residuals between the observed and the modelled rotation velocities at each radius of the data points. 
Figure~\ref{fig:Vterm-sinl} shows the tangent-point data in the plane of observables $V_{\mathrm{term}}\,-\,\sin l$, as well as the curves resultant from the models M{\sc I} and M{\sc II} calculated as $V_{\mathrm{term}}=V_{\mathrm{rot}}(R)-V_{0}\sin l$, with $\sin l=R/R_{0}$. A good agreement between the two curve models and the data can be observed.

In order to compare the goodness of fit between models M{\sc I} and M{\sc II}, we use the chi-squared per degree of freedom statistics, $\chi^{2}/\mathrm{d.o.f.}$.  
Model M{\sc I} returns $\chi^{2}/\mathrm{d.o.f.}=1.85$, while model M{\sc II} provides a fit with $\chi^{2}/\mathrm{d.o.f.}=1.46$. The relatively better fit provided by model M{\sc II} can be explained, in part, by its better match to the dip in the observed rotation curve at radii in the interval $\sim 8-10$ kpc. 
As mentioned before, the correspondent dip in the model rotation curve is a consequence of the ring density structure added to the disk model density profile. 

In Table~\ref{tab:corr_matrix_MI-MII} we give the correlation matrix of the fitted parameters for both models M{\sc I} (the lower-left triangle) and M{\sc II} (the upper-right triangle). In a given correlation matrix, the elements are distributed in the interval $[-1;\,1]$, where the value of 1 indicates a perfect correlation and -1 indicates a perfect anti-correlation, whereas 0 corresponds to no correlation. As can be seen from Table~\ref{tab:corr_matrix_MI-MII}, the strongest correlations are between parameters associated with a single component, as is the case of the bulge parameters $M_{\mathrm{b}}$ and $a_{\mathrm{b}}$, and also the halo parameters $r_{\mathrm{h}}$ and $v_{\mathrm{h}}$. An anti-correlation between the asymptotic circular velocity of the dark halo $v_{\mathrm{h}}$ and the disk mass scale factor $f_{mass}$ is observed from model M{\sc I}; since the dark halo mass must be proportional to $v_{\mathrm{h}}$, this anti-correlation means a competition between the disk and the dark halo for the contribution to the dynamical mass in the inner Galaxy. For model M{\sc II}, anti-correlations are also seen between $M_{\mathrm{b}}$ and $f_{mass}$, and between $f_{mass}$ and the ring parameter $R_{ring}$.


\subsection{Comparison with studies in the literature}
\label{compare_studies_literat}

The masses attributed to the bulge in our models are relatively large ($M_{\mathrm{b}}\sim 2.6\times 10^{10}$ M$_{\odot}$) compared to other estimates in the literature. This is one cause for the small contribution of the disk component of model M{\sc I} to the rotational support in the inner Galaxy, and also but to a less extent in model M{\sc II} (see next section). Most of the studies in the literature find bulge masses of the order of $1\times 10^{10}$ M$_{\odot}$. However, these studies also adopt simple exponential profiles for the disk component, with masses in the central regions larger than those obtained with our disk models with central density depletion. As pointed out by \citet{McMillan2011}, his best-fitting Galactic model presents a stellar mass within the inner 3 kpc of $\sim 2.4\times 10^{10}$ M$_{\odot}$ (the same result obtained in the \citealt{Flynn2006} model), which is in close agreement with the bulge mass found by \citet{Picaud_Robin2004} assuming a disk model with a central hole. This is also the same value for the mass of the bulge in the model of \citet{Lepine_Leroy2000}, which is also based in a disk with a central density depletion. 
As described in Sect.~\ref{bulge_dark-halo}, we based the initial guess values for the bulge parameters on the spheroidal component of PJL's model, which returns a mass inside 3 kpc of $\approx 2.2\times 10^{10}$ M$_{\odot}$, with a scale radius parameter of 0.4 kpc. The values for the bulge masses $M_{\mathrm{b}}$ and scale radii $a_{\mathrm{b}}$ found in our models (see Table~\ref{tab:params_MI-MII}) reflect the consequence of this choice. The relatively short scale radii cause the rise of the peak in the rotation curves within the inner 1 kpc, as can be seen from Fig.~\ref{fig:rotcurve_models-MI-MII}. Therefore, according to the PJL star counts model, and also reflected in our models, the bulge is an important mass component in the central region of the Galaxy, with its contribution to the model rotation curve being dominant at these radii.  
Our models return stellar masses enclosed within the inner 3 kpc of $2.7\times 10^{10}$ M$_{\odot}$ (model M{\sc I}, being $1.98\times 10^{10}$ M$_{\odot}$ due to the bulge and $0.72\times 10^{10}$ M$_{\odot}$ due to the stellar thin and thick disks), and $3\times 10^{10}$ M$_{\odot}$ (model M{\sc II}, being $1.98\times 10^{10}$ M$_{\odot}$ due to the bulge and $1.02\times 10^{10}$ M$_{\odot}$ due to the stellar thin and thick disks).
We can therefore argue about the tendency of the disk models with central holes in redistributing the mass in the central regions from the disk to the bulge, without significantly altering the total stellar mass in such regions. 

The resulting disk stellar mass (thin disk + thick disk) is $3.2\times 10^{10}$ M$_{\odot}$ from model M{\sc I} and $4.74\times 10^{10}$ M$_{\odot}$ from model M{\sc II}. As a comparison, \citet[and references therein]{Flynn2006} argue that models with disk scale-lengths in the range $2-2.6$ kpc, based on studies of the Galactic emission in the near-infrared or studies of the local stellar distribution, correspond to disk stellar masses in the range $3.6-5.4\times 10^{10}$ M$_{\odot}$. The total stellar mass (bulge + thin disk + thick disk) of $5.81\times 10^{10}$ M$_{\odot}$ from model M{\sc I} is close to the interval of $4.85-5.5\times 10^{10}$ M$_{\odot}$ estimated by \citet{Flynn2006}, and the total stellar mass of $7.37\times 10^{10}$ M$_{\odot}$ from model M{\sc II} is relatively close to the best-fitting value of $6.61\times 10^{10}$ M$_{\odot}$ in the model by \citet{McMillan2011}. The disk mass in the gaseous form, already corrected for the mass contribution of helium, is $1.42\times 10^{10}$ M$_{\odot}$ from model M{\sc I} and $2.1\times 10^{10}$ M$_{\odot}$ from model M{\sc II}. The total baryonic mass (stellar + gas) from the models are: $7.23\times 10^{10}$ M$_{\odot}$ (model M{\sc I}) and $9.47\times 10^{10}$ M$_{\odot}$ (model M{\sc II}).
These values are somewhat larger than the ``back of the envelope'' estimate by \citet{Flynn2006} of $6.1\pm 0.5\times 10^{10}$ M$_{\odot}$, but are compatible with the masses (bulge + disk) of $7-8.2\times 10^{10}$ M$_{\odot}$ found in the models by \citet{Irrgang2013}, for instance.

Table~\ref{tab:deriv_quant_MI-MII} also lists the values of the Oort's constants $A$ and $B$ resulting from the models. Although these constants were not used as observational constraints in the modelling process, their returned values from model M{\sc I} agree with those estimated in the literature (e.g. \citealt{Feast_Whitelock1997}). The steeper gradient of the rotation curve at the solar Galactic radius $R_0$ causes the deviation of the Oort's constants of model M{\sc II} from their common estimated values. 
The local escape velocities of 452 km s$^{-1}$ from model M{\sc I} and 550 km s$^{-1}$ from model M{\sc II} are, respectively, close and within the $90\%$-confidence interval of $492-587$ km s$^{-1}$ as determined by \citet{Piffl2014}.


\subsection{The disk support to the rotation curve}
\label{disk_maximal}

The disk mass scale factor $f_{mass}$ of model M{\sc I} implies no increase of the originally modelled disk mass. Although returning local surface densities that are compatible, within the adopted uncertainties, with the observed ones used as constraints, the model M{\sc I} returns a dynamical mass for the Galactic disk which moderately contributes to the total rotation curve in the inner Galaxy. 
\citet{Sackett1997}, based on a set of Galactic observational constraints, verified that the `maximal disk hypothesis', commonly applied to external spiral galaxies, also gives a maximal disk when applied to the Milky Way. According to this definition, to be maximal, the exponential disk must provide $85\%\pm 10\%$ of the total rotation velocity of the galaxy at the radius $R=2.2 R_{\mathrm{d}}$, where the rotation curve of the disk presents a peak ($R_{\mathrm{d}}$ is the exponential disk scale-length). Since our disk models are no longer simple exponentials, we just compare the peak rotation speed of the disk with the total rotation speed at the same radius. From Fig.~\ref{fig:rotcurve_models-MI-MII}, the peak in the rotation curve of the disk happens at $R\sim 7$ kpc, for both models M{\sc I} and M{\sc II}. We estimate that the disk of model M{\sc I} is responsible for only $64\%$ of the total circular velocity $V_{c}(R=7\,\mathrm{kpc})$, which puts it in the condition of a ``sub-maximal'' disk, however. 

On the other hand, the mass scale factor of model M{\sc II} increases the disk mass by $49\%$ of its original value. This increase in the disk mass is a consequence of the addition of the ring density structure to the density profile of the disk (see Sect.~\ref{ring_feat}). Once the solar orbit radius is sitting inside the valley of the ring density feature (near the minimum density, see Fig.~\ref{fig:Sigma-MII}), to maintain the local disk surface density value within the range determined by the observations, the underlying surface densities at the other radii are proportionally increased. For instance, if it would not be by the presence of the ring density giving a local disk surface density of $\Sigma_{0\mathrm{d}}=61$ M$_{\odot}$ pc$^{-2}$ (cf. Table~\ref{tab:deriv_quant_MI-MII}), the disk mass of model M{\sc II} would return $\Sigma_{0\mathrm{d}}=90$ M$_{\odot}$ pc$^{-2}$ instead. We estimate that the disk of model M{\sc II} contributes with $78\%$ of the total circular velocity at $R=7$ kpc, putting it within the limit for a ``maximal disk'' condition.
Therefore, compared with model M{\sc I}, the disk of model M{\sc II} contributes more expressively to the inner rotation curve of the Galaxy, giving rise to a less important dark halo component at such regions. It is still important to note that \citet{Sackett1997}'s conclusion about the Galaxy supporting a maximal disk was based, among several observational constraints, on a local disk surface density with a value similar to that used in the present work, but adopting a local circular velocity of $V_{0}=210\pm 25$ km s$^{-1}$, while our adopted $V_0$ of 230 km s$^{-1}$ sits closer to the high side of such interval.  
Considering now the total rotation supplied by the mass from both bulge and disk, $\sim 82\%$ and $93\%$ of the total circular velocity at $R=7$ kpc are provided by the baryonic matter from models M{\sc I} and M{\sc II}, respectively.   


\subsection{The ring structure}
\label{ring_feat}

\begin{figure*}
\centering
\includegraphics[scale=0.51]{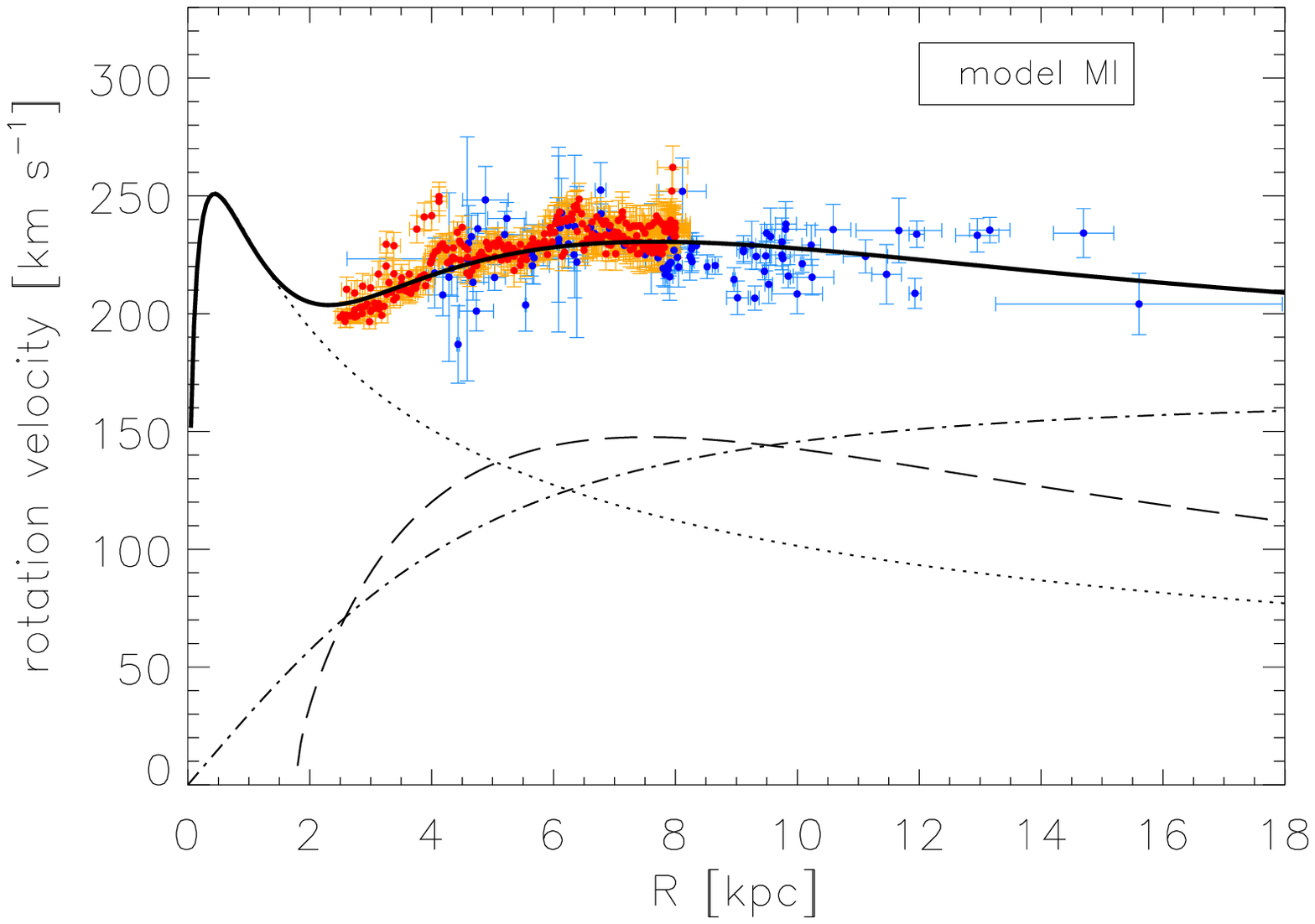}
\includegraphics[scale=0.51]{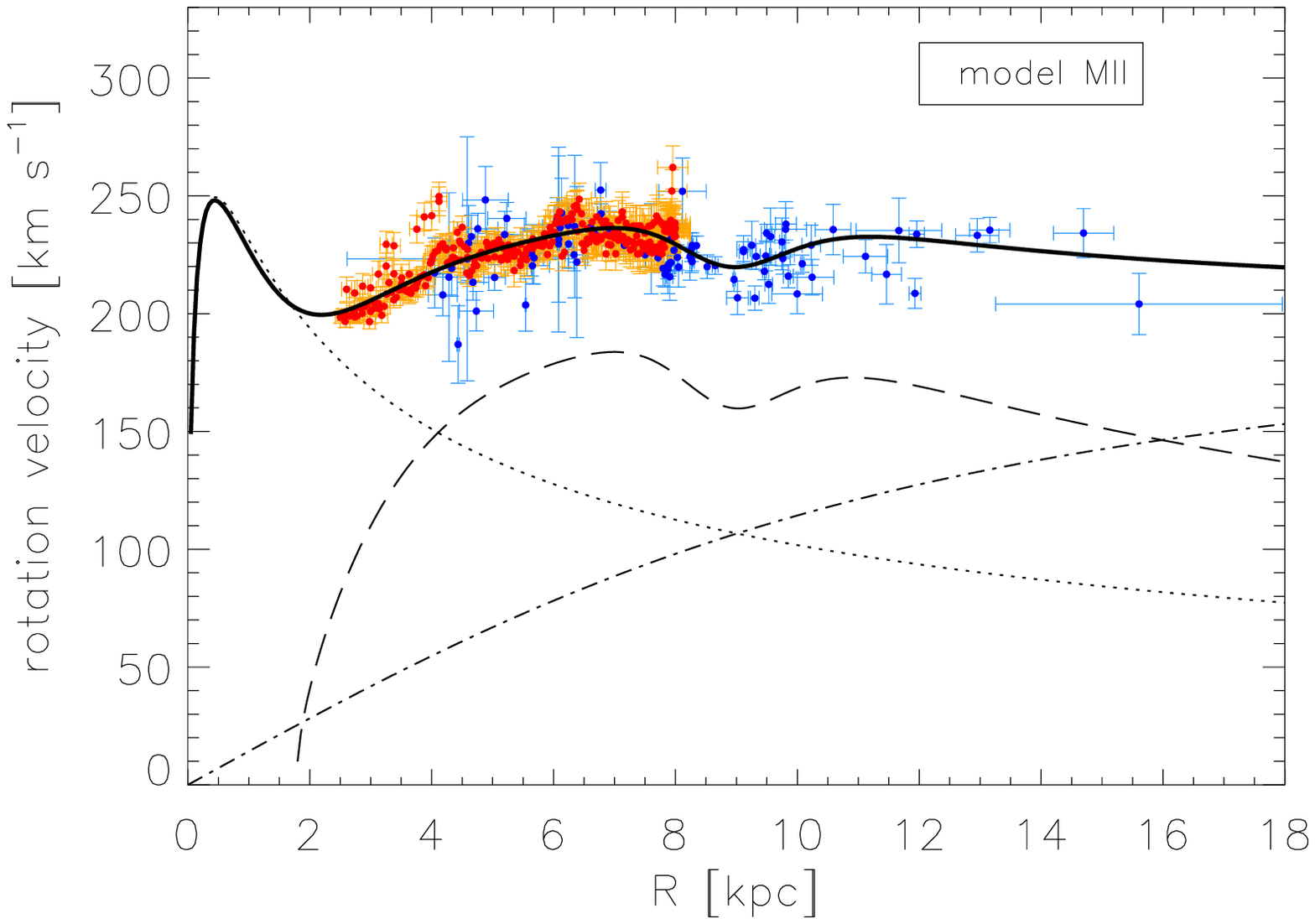}
\includegraphics[scale=0.51]{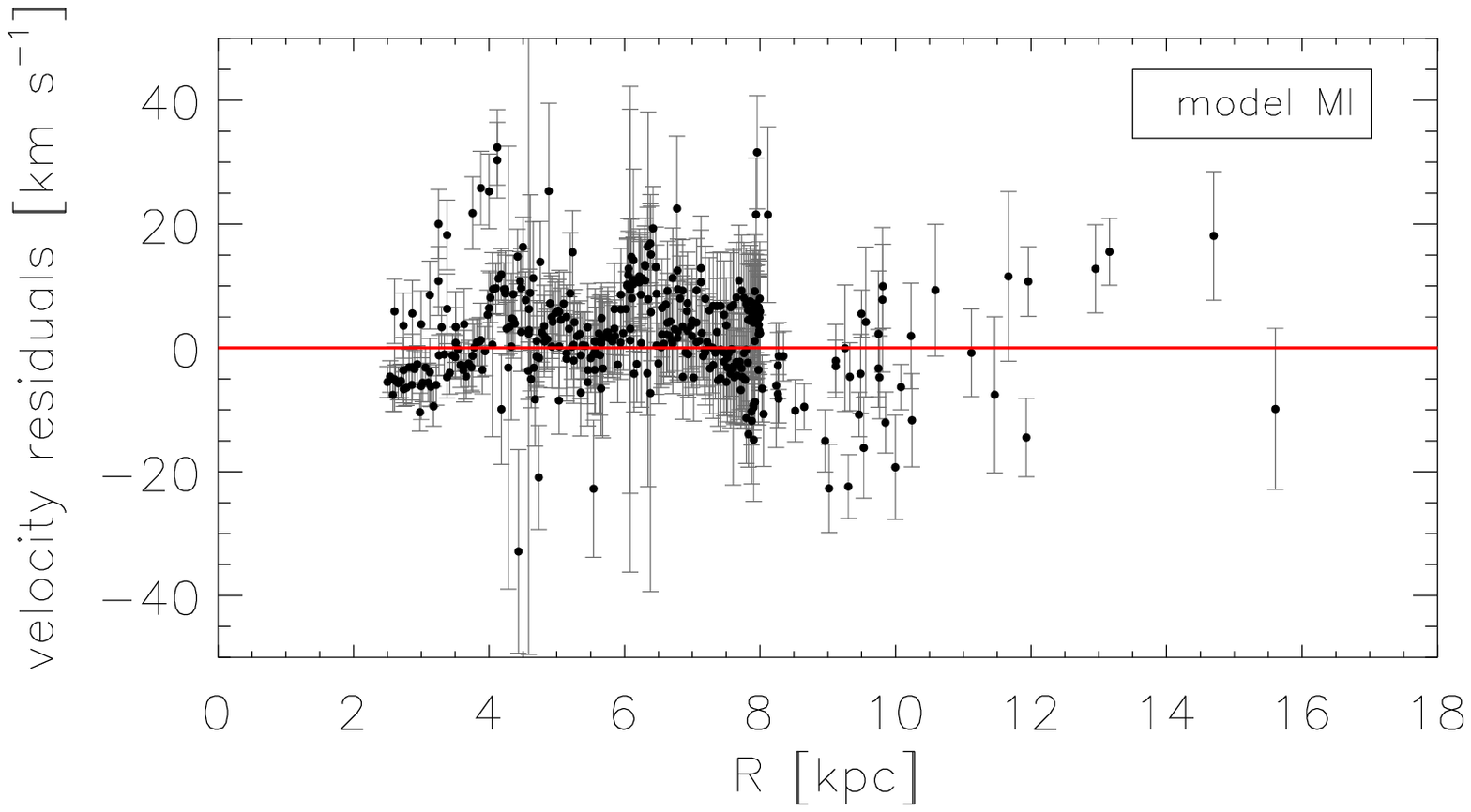}
\includegraphics[scale=0.51]{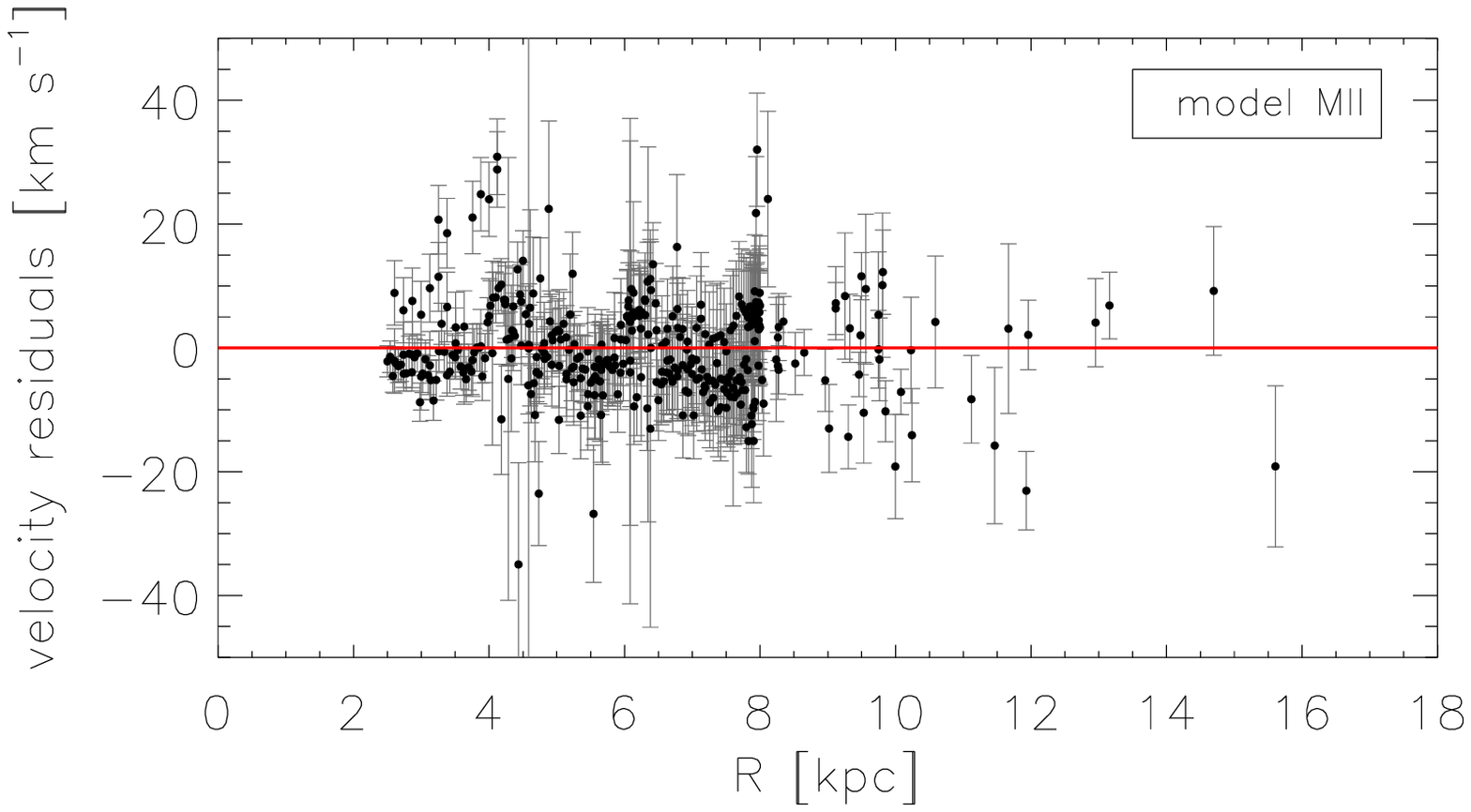}
\caption{{\it Left-hand column:} Top panel - rotation curve resulting from the Galactic model M{\sc I}. The circular velocities due to the three mass components are depicted as: dotted lines for the bulge, dashed lines for the disk, and dash-dotted lines for the dark halo. The total circular velocity is represented by the solid curve. 
The data for the observed rotation curve are presented as: red points with orange error bars for the CO tangent-point data (\citealt{Clemens1985}) and the H\,{\scriptsize I} tangent-point data (\citealt{Fich_Blitz_Stark1989}); and blue points with light blue error bars for the maser sources data (\citealt{Reid2014}). 
Bottom panel - velocity residuals between the observed rotation velocities and the total circular model velocities calculated at each radius of the data points. {\it Right-hand column:} the correspondent of the left-hand column but for the Galactic model M{\sc II}.
}
\label{fig:rotcurve_models-MI-MII}
\end{figure*}

\begin{figure}
\centering
\includegraphics[scale=0.51]{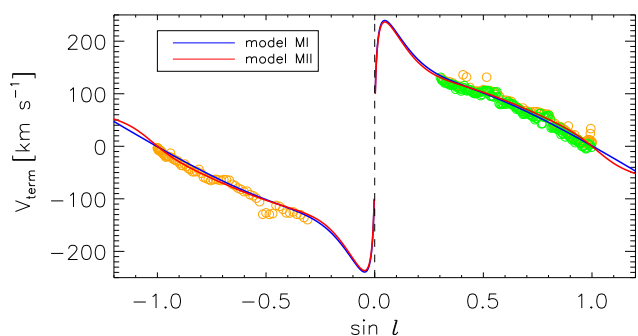}
\caption{Tangent-point data from \citet{Clemens1985} (green circles) and \citet{Fich_Blitz_Stark1989} (orange circles). The tangent velocity curves resultant from model M{\sc I} (blue curve) and model M{\sc II} (red curve) are also plotted.}
\label{fig:Vterm-sinl}
\end{figure}

\begin{figure}
\centering
\includegraphics[scale=0.51]{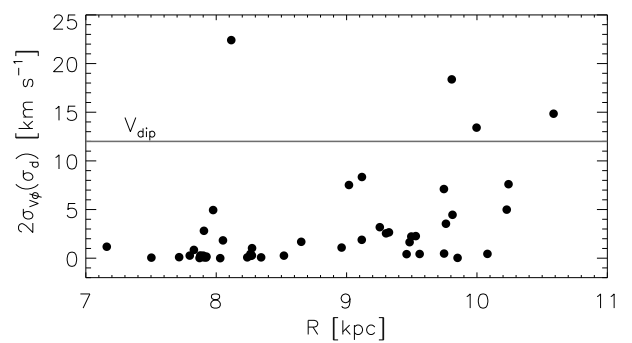}
\caption{Comparison of the rotation velocity difference $2\sigma_{V_{\phi}}(\sigma_d)$ (black circles $-$ see text for details) with the velocity dip $V_{dip}$ (horizontal line), for the maser sources in the radial interval comprised by the dip of the rotation curve.}
\label{fig:sigd_Vdip}
\end{figure}

In Sect.~\ref{ring_dens_min}, we give some arguments about our reasons for considering an alternative disk model which incorporates a ring density structure near the solar Galactic radius. In terms of the observational data treated in this work, we argue that the apparent velocity dip in the rotation curve of the maser sources, between radii of $\sim 7$ kpc and $\sim 11$ kpc (compared to a flat rotation curve $V_{c}(R)=V_0$, for example), can be associated with a ring-like density structure of the type that is being considered here. But before modelling such a ring structure, it is recommended that we check the reliability of the velocity dip above-mentioned. Indeed, correlations between distance uncertainties and velocity uncertainties could create false trends in the plot of the rotation curve. Since the velocity dip of the rotation curve is traced by the maser sources, we restrict to these data in the following analysis. For each maser source, we decompose the uncertainty on the azimuthal velocity $V_{\phi}$, $\sigma_{V_{\phi}}$, into two parts: one component is formed by the combination of the uncertainty on the heliocentric radial velocity, the uncertainties on the two components of the proper motion, and the uncertainties on the solar motion; the second component is solely due to the uncertainty on the distance of the source $\sigma_d$, derived from the uncertainty on its parallax. We express this second component as $\sigma_{V_{\phi}}(\sigma_d)$. To verify how much the correlations between the uncertainties on the velocities $V_{\phi}$ and the uncertainties on the distances $d$ affect the velocity dip of the rotation curve, we compute, for each maser source, the difference in velocities $(V_{\phi}+\sigma_{V_{\phi}}(\sigma_d))-(V_{\phi}-\sigma_{V_{\phi}}(\sigma_d))=2\sigma_{V_{\phi}}(\sigma_d)$, and compare it to the magnitude of the velocity dip $V_{dip}$. From the rotation curve in Fig.~\ref{fig:rotcurve_models-MI-MII}, we estimate $V_{dip}\approx 12$ km s$^{-1}$. The comparison between $2\sigma_{V_{\phi}}(\sigma_d)$ and $V_{dip}$ for the maser sources in the interval $7\,{\mathrm{kpc}}\leq R \leq 11$ kpc is shown in Fig.~\ref{fig:sigd_Vdip}. About $80\%$ of the sources present $2\sigma_{V_{\phi}}(\sigma_d)\leq 5$ km s$^{-1}$, much smaller than $V_{dip}$. This result seems to indicate that the correlations between distance uncertainties and velocity uncertainties are not able to create a false trend in the rotation curve which would be making us seeing it as a dip in the velocity profile. In other words, the velocity dip in the radial interval $7\,-\,11$ kpc seems to be a real feature of the rotation curve.

\begin{figure}
\centering
\includegraphics[scale=0.51]{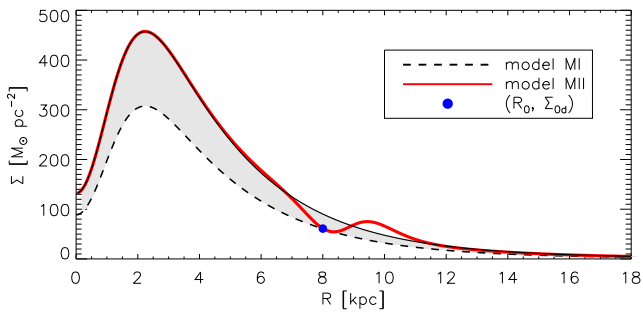}
\caption{Surface density radial profile for the disks of model M{\sc I} (dashed curve) and model M{\sc II} (red solid curve). The black solid curve represents a disk with mass equivalent to that from model M{\sc II} but without the ring density structure. The shaded area in light grey color emphasizes the resulting difference in mass between the disks from both models. The blue point denotes the pair ($R_{{0}}$; $\Sigma_{0\mathrm{d}}$).}
\label{fig:Sigma-MII}
\end{figure}

The ring density structure associated with the velocity dip in the rotation curve of model M{\sc II}, whose parameters ($A_{ring}$, $\beta_{ring}$, $R_{ring}$) are given in Table~\ref{tab:params_MI-MII}, is depicted in Fig.~\ref{fig:Sigma-MII} along with the radial profile for the surface density of the disk from the same model (red curve). The surface density of a disk with equivalent mass to that from model M{\sc II}, but without the ring structure, is shown by the curve in black solid line. The ring is formed by a minimum density at the radius of 8.3 kpc, a maximum at 9.8 kpc, and the node radius $R_{ring} = 8.9$ kpc, which is also the radius of the minimum point in the dip of the modelled rotation curve. The amplitude of the minimum is $\approx -0.34$ times the density at the same radius of the equivalent disk without the ring. The blue point in Fig.~\ref{fig:Sigma-MII} indicates the approximately common value of $\Sigma_{0\mathrm{d}}$ returned from both models M{\sc I} (indicated by the curve in dashed line) and model M{\sc II} (red curve), at the solar radius $R_0$. As anticipated in Sects.~\ref{ring_dens_min} and~\ref{disk_maximal}, the ring structure causes the rescale of the disk total mass from model M{\sc II} to a value as large as $\sim 48\%$ of the disk mass from model M{\sc I}. This increase in mass can be checked out by observing the area highlighted in light grey color between the $\Sigma(R)$ curves of model M{\sc I} and the equivalent disk of model M{\sc II} in Fig.~\ref{fig:Sigma-MII}.
We warn that the above result is dependent on the scale-length measured for the disk of Model M{\sc II}. The ring structure does not alter the global scale-length of the disk, i.e., the global disk scale-length of Model M{\sc II} is the same of model M{\sc I}. Then the above-quoted increase in the disk mass due to the ring density is satisfied. However, the ring tends to decrease the scale-length of the $\Sigma(R)$ distribution mainly in the radial range of its influence, that is $7\lesssim R\lesssim 11$ kpc. Once the disk scale-length is estimated using this shorter interval of radius, the above considerations of mass increase might not be appropriate.

The density bump at $R\sim 9.8$ kpc induced by the ring (see Fig.~\ref{fig:Sigma-MII}) can also be a matter of debate. 
In fact, the amplitude of this local maximum of density may be too exaggerated, which is a consequence of the form chosen for the ring potential in Eq.~\ref{eq:pot_ring}. Regardless of the true amplitude of the density bump, the question about its plausibility in the observational point of view still needs to be answered.   
We argue here that since the ring structure was modelled over both stellar and gaseous components, part of such density bump must be associated with the hydrogen distribution of the disk. And indeed, recent works by \citet{Nakanishi_Sofue2016} and \citet{Sofue_Nakanishi2016} clearly show an increase in the hydrogen density of the disk at radii slightly beyond 10 kpc. A stellar counterpart of the density bump, if it existed, would be more difficult to notice, for some reasons like the relative position of the Sun in the ring structure, the beginning of the flare of the stellar disk just after the solar radius, etc. 
We believe that estimates of the stellar surface density in a great coverage of the Galactic disk exterior to the solar circle (not only in radius but also in azimuth and height from the plane) are still needed to confirm or refute the existence of the stellar ring structure. In a positive case, our ring model could serve as a starting point to the study of the properties of such a structure in the Galactic disk. These are some aspects that make model M{\sc II} an `alternative model', in the sense treated in this work.

\begin{table*}
\caption{Best-fitting values for the parameters of the Galactic models M{\sc I} and M{\sc II}. The given uncertainties are the standard deviations of the distributions of each parameter, estimated after the bootstrapping procedure described in Section~\ref{uncert_estim}.}             
\label{tab:params_MI-MII}      
\centering          
\begin{tabular}{c c c c c c c c c}
\hline\hline       

{\bf Model} & $M_{\mathrm{b}}$ & $a_{\mathrm{b}}$ & $r_{\mathrm{h}}$ & $v_{\mathrm{h}}$ & $f_{mass}$ & $A_{ring}$ & $\beta_{ring}$ & $R_{ring}$ \\
 & ($10^{10}$ M$_{\odot}$) & (kpc) & (kpc) & (km s$^{-1}$) &  & (km$^{2}$ s$^{-2}$) &  & (kpc) \\ 
\hline                   
M{\sc I} & $2.61$ & $0.44$ & $5.4$ & $166$ & $1.00$ & ... & ... & ... \\
 & $\pm 0.05$ & $\pm 0.03$ & $\pm 0.6$ & $\pm 6$ & $\pm 0.03$ & & & \\
\hline
M{\sc II} & $2.63$ & $0.45$ & $13.4$ & $191$ & $1.49$ & $1053$ & $5.67$ & $8.9$ \\
 & $\pm 0.09$ & $\pm 0.03$ & $\pm 0.9$ & $\pm 6$ & $\pm 0.07$ & $\pm 300$ & $\pm 0.05$ & $\pm 0.2$ \\
\hline               
\end{tabular}
\end{table*}

\begin{table*}
\caption{Local properties resulting from the models M{\sc I} and M{\sc II} for the mass distribution and gravitational potential of the Galaxy.}             
\label{tab:deriv_quant_MI-MII}      
\centering          
\begin{tabular}{l c c}
\hline\hline       

 & {\bf Model M{\sc I}} & {\bf Model M{\sc II}} \\
{\bf Derived quantity} & {\bf value} & {\bf value} \\ 
\hline                   
local circular velocity, $V_0$ (km s$^{-1}$) & 230.5 & 229.5 \\
local angular velocity, $\Omega_0$ (km s$^{-1}$ kpc$^{-1}$) & 28.8 & 28.7 \\
Oort's constant $A$ (km s$^{-1}$ kpc$^{-1}$) & 14.6 & 21.1 \\
Oort's constant $B$ (km s$^{-1}$ kpc$^{-1}$) & -14.2 & -7.6 \\
local disk surface density, $\Sigma_{0\mathrm{d}}$ (M$_{\odot}$ pc$^{-2}$) & 60.8 & 61.3 \\
local surface density within $|z|\leq 1.1$ kpc, $\Sigma_{0_{\mathrm{1.1kpc}}}$ (M$_{\odot}$ pc$^{-2}$) & 78.6 & 73.7 \\
local escape velocity, $V_{\mathrm{esc}}$ (km s$^{-1}$) & 452 & 550 \\
\hline                  
\end{tabular}
\end{table*}

\begin{table*}
\caption{Correlation matrix for the model parameters: the lower-left triangle for model M{\sc I} and the upper-right triangle for model M{\sc II}.}             
\label{tab:corr_matrix_MI-MII}      
\centering          
\begin{tabular}{c | c c c c c c c c}
\hline      

 & $M_{\mathrm{b}}$ & $a_{\mathrm{b}}$ & $r_{\mathrm{h}}$ & $v_{\mathrm{h}}$ & $f_{mass}$ & $A_{ring}$ & $\beta_{ring}$ & $R_{ring}$ \\
\hline                 
$M_{\mathrm{b}}$ &  & 0.735 & -0.017 & -0.333 & -0.325 & -0.068 & -0.300 & 0.231  \\
$a_{\mathrm{b}}$ & 0.804 &  & -0.029 & -0.146 & -0.144 & -0.125 & -0.118 & -0.021  \\
$r_{\mathrm{h}}$ & 0.461 & 0.256 &  & 0.441 & 0.501 & 0.229 & 0.436 & 0.127  \\
$v_{\mathrm{h}}$ & 0.025 & -0.057 & 0.873 &  & 0.067 & -0.303 & 0.303 & -0.398  \\
$f_{mass}$ & 0.220 & 0.130 & -0.482 & -0.729 &  & 0.304 & 0.613 & -0.449  \\
$A_{ring}$ & ... & ... & ... & ... & ... &  & -0.458 & 0.438  \\
$\beta_{ring}$ & ... & ... & ... & ... & ... & ... &  & -0.515  \\
$R_{ring}$ & ... & ... & ... & ... & ... & ... & ... &   \\
\hline                  
\end{tabular}
\end{table*}


\section{Conclusions}
\label{conclusions}

We have developed models for the axisymmetric mass distribution of the Galaxy with the aim to derive fully-analytical descriptions of its associated three-dimensional gravitational potential. We have followed an approach which intentionally expends more efforts in a 
detailed modelling of the disk component. Based on photometric constraints for the stellar distribution in the disk given by the Galactic infrared star counts model of PJL, as well as on the observed distribution of atomic and molecular hydrogen gas, we derived an ``empirical basis" for the structural parameters (scale-length, scale-height, and radial scale of the disk central hole) of the thin and thick stellar disks and the H\,{\scriptsize I} and H$_2$ disks subcomponents. With {\it a priori} values for the masses of each disk, based on the most recent determinations of the local stellar and gaseous disk surface densities, we have constructed versions of Miyamoto-Nagai disks for each disk subcomponent mass model. The method follows the approach developed by \citet{Smith2015}, but here with the allowance of using the Miyamoto-Nagai disk models of higher orders 2 and 3 beside the commonly used one of order 1 (\citealt{Miyamoto_Nagai1975}). Along with parametric models for the bulge and an extra unknown spherical mass component to which for ``convenience'' we refer by the often-used term {\it dark halo}, we searched for the dynamical mass of each Galactic component 
by fitting the models to the kinematic constraints given by the observed rotation curve and some local Galactic measured properties. 

We have shown that a disk model which includes a ring density pattern beyond but very close to the solar orbit radius is able to better reproduce an observed local dip in the Galactic rotation curve centered at $R\sim 9.0$ kpc; such dip is naturally explained by a ring density structure composed by a minimum followed by a maximum density of similar amplitudes.
Furthermore, the model with the ring structure allows a more massive disk to increasingly contribute to the rotational support of the Galaxy inside the solar circle, helping the Milky Way to satisfy the condition to be considered a ``maximal disk'' galaxy.  
The model with the ring structure also relies on a numerical study of the stellar disk where a ring density structure develops as a consequence of interactions between the stars and the galactic spiral arms in resonance at the co-rotation radius (BLJ). Since we still have no information about the three-dimensional structure of this ring density feature, we had to make crude approximations for the vertical profile of its associated gravitational potential. We believe that, with the forthcoming data of the GAIA mission (\citealt{Perryman2001}), the ring structure in the disk may become an object of examination, and in the case of confirmation of its existence, constraints on its three-dimensional distribution in the global disk structure will help us to create more realistic disk models.     

The method we have applied for the construction of the disk mass model and its associated gravitational potential 
is quite flexible in the sense that for any set of structural parameters, disks of different masses can be generated.  
We emphasize at this point that the models of Galactic gravitational potential presented in this work are aimed for being used in studies of orbits in the disk that do not extend too far away in Galactic radii as well as do not reach great heights above the disk mid-plane. These limitations are imposed by the spatial coverage of the observational data used to constrain the models, in the sense that for radii greater than $\sim 2R_{0}$ (the maximum radius of the data used for the rotation curve) and heights $|z|\gtrsim 3$ kpc, we do not guarantee that our models return confident representations of the Galactic potential, but maybe reasonable ones at least.  
Also in this respect, no constraint on the mass at large radii is adopted here, as has been made by some studies that derive properties of the dark halo by requiring distant halo stars to be bound to the Milky Way potential.
We have included a density component associated with a spherical logarithmic potential just to explain the observed rotation curve data at radii $R_{0}\lesssim R \lesssim 2R_{0}$. However, as pointed out by \citet{Dehnen_Binney1998}, instead of a distinct physical component, it is possible that we are measuring the dynamical effects of a disk and/or a bulge in which the mass-to-light ratio of their content strongly increases from the center to the Galactic outskirts.

The models for the three-dimensional gravitational potential of the Galaxy presented in this work, being fully analytical and easy to obtain the associated gravitational force-field at any point, are suited for fast and accurate calculations of orbits of stellar-like objects belonging to the main populations of the Galactic disk.  
In a forthcoming study, we aim to present an application of these new potential models to a description of the distributions of orbital parameters for samples of Galactic open clusters and some expected correlations with their chemical abundance patterns.


\begin{acknowledgements}

We acknowledge the referee whose comments and suggestions have significantly improved the present paper.
DAB received financial support for this work from the Brazilian research agency CAPES (Coordena\c c\~ao de Aperfei\c coamento de Pessoal de N\' ivel Superior), through the PNPD postdoctoral program.
\end{acknowledgements}






\appendix

\section{Parameters and relations for Miyamoto-Nagai disks models}
\label{app1}

In this appendix, we present the variations of the ratios $M/M_{\mathrm{d}}$ and $a/R_{\mathrm{d_{exp}}}$ as functions of $b/R_{\mathrm{d_{exp}}}$ ($M$, $a$ and $b$ are the free parameters of the Miyamoto-Nagai disks), as well as the relations between $b/R_{\mathrm{d_{exp}}}$ and $h_{z}/R_{\mathrm{d_{exp}}}$ for each disk subcomponent, calculated in the same way as in \citet{Smith2015}. Here, for simplicity, we denote the exponential scale-length $R_{\mathrm{d_{exp}}}$ simply as $R_{\mathrm{d}}$. As stated in the main text, these relations are intended to help in the construction of Miyamoto-Nagai disks models for disks of any mass $M_{\mathrm{d}}$, scale-length $R_{\mathrm{d}}$ and scale-height $h_{z}$, that can be different from the ones modelled in the present work. The relations presented for the thin stellar disk, the H\,{\scriptsize I} and the H$_2$ disks are for models of disks with density depressions in their central regions, while for the thick stellar disk the relations are correspondent to simple radially exponential disks. Regarding the vertical density distributions, the models for the thin and thick stellar disks consider exponential density decays with the increase of the height above the mid-plane, and for the H\,{\scriptsize I} and H$_2$ disks a Gaussian density profile is adopted for the models. 
We emphasize here that such relations are constructed for models of 3 MN-disks combinations which are: model 3 for the thin disk and the H$_2$ disk (Eqs.~\ref{eq:rho_MN3} and~\ref{eq:Phi_MN3}); model 1 for the thick disk (Eqs.~\ref{eq:rho_MN1} and~\ref{eq:Phi_MN1}); and model 2 for the H\,{\scriptsize I} disk (Eqs.~\ref{eq:rho_MN2} and~\ref{eq:Phi_MN2}). 

\begin{figure}
\centering
\includegraphics[scale=0.50]{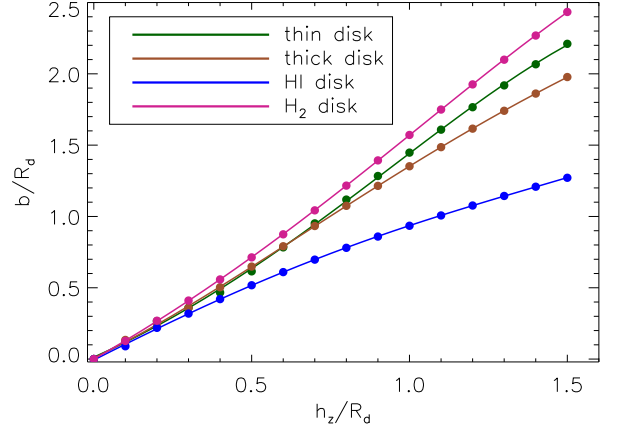}
\caption{Variation of the thickness ratio $b/R_{\mathrm{d}}$ as a function of the ratio $h_{z}/R_{\mathrm{d}}$, for the thin and thick stellar disks and the H\,{\scriptsize I} and H$_2$ disks subcomponents. Filled circles represent the optimal solutions found after the application of the cross-entropy algorithm, as described in Sect.~\ref{CE_algorithm} of the main text. Solid lines represent 4th-order polynomial fits to the distribution of points associated with each calculated relation.}
\label{fig:disksMN-hzxb}
\end{figure}

Figure~\ref{fig:disksMN-hzxb} shows the variation of the thickness ratio $b/R_{\mathrm{d}}$ as a function of the ratio $h_{z}/R_{\mathrm{d}}$, for the thin and thick stellar disks and the H\,{\scriptsize I} and H$_2$ disks subcomponents. Filled circles in the figure represent the calculated relations, in the way described in Sect.~\ref{MN-disks} of the main text. The solid lines represent 4th-order polynomial fits to the distribution of points associated with each calculated relation, and are given in the form:

\begin{equation}
\label{eq:fit_poly4_thickness}
\frac{b}{R_{\mathrm{d}}}=\sum_{j=0}^{4}{k_{j}\,\left(\frac{h_{z}}{R_{\mathrm{d}}}\right)^{j}}\,.
\end{equation}
\noindent
The values of the coefficients $k_0$ - $k_4$ of the above relation, and for each disk subcomponent, are given in Table~\ref{tab:coeff_poly_thickness}. 

\begin{figure*}
\centering
\includegraphics[scale=0.50]{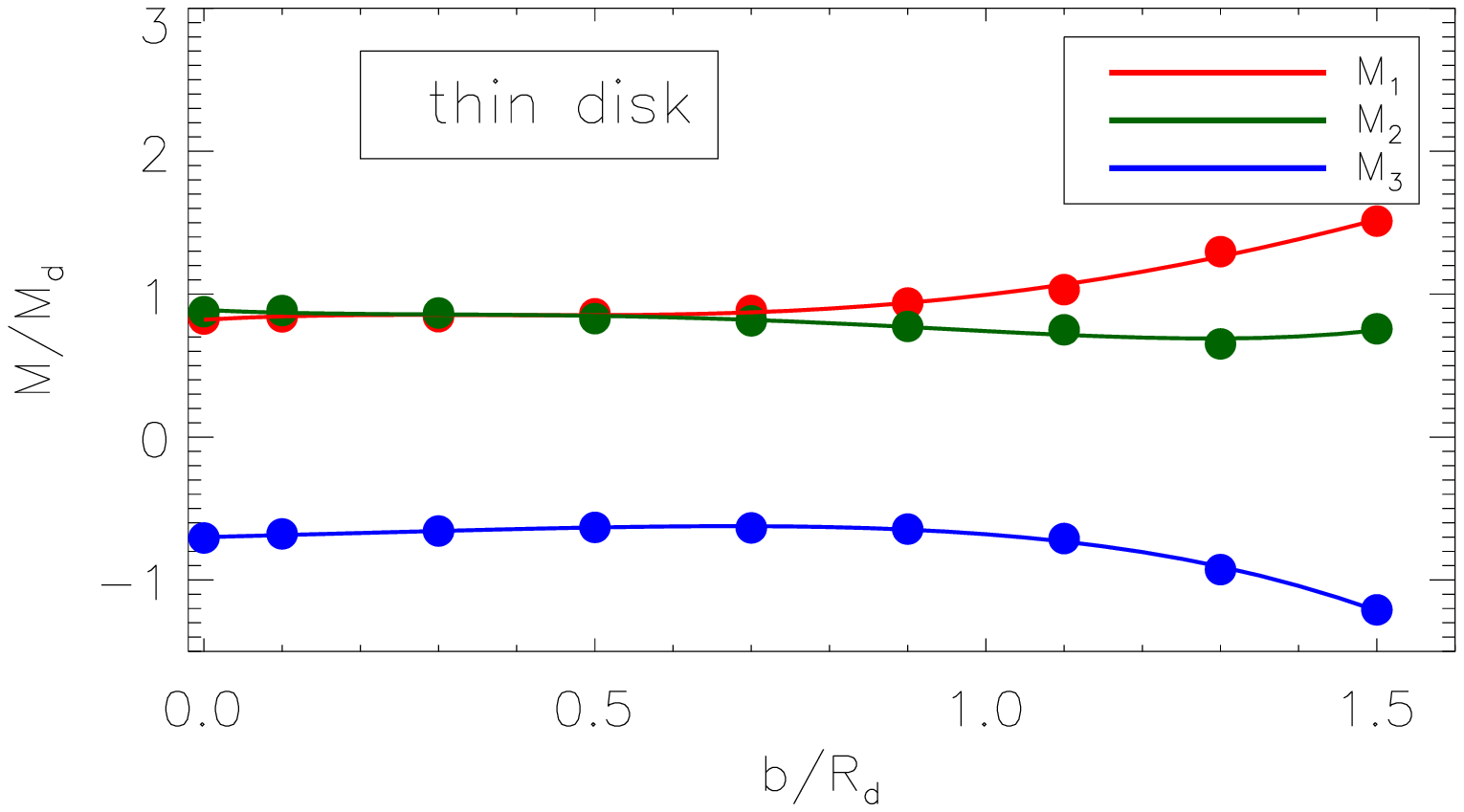}
\includegraphics[scale=0.50]{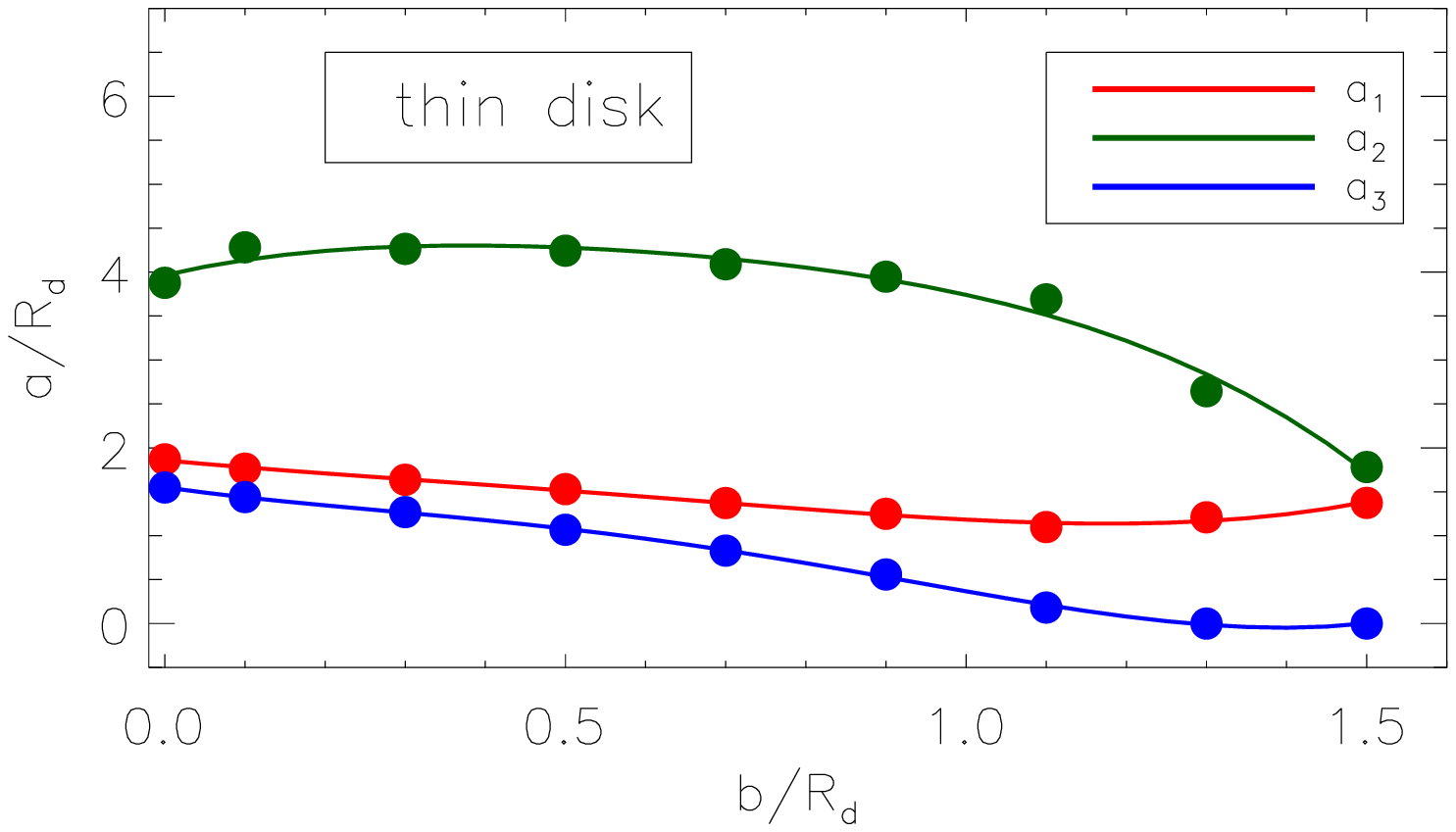}
\caption{{\it Left-hand panel}: Variation of the three mass parameters of the 3 MN-disks model as a function of the disk thickness ratio $b/R_{\mathrm{d}}$, for the modelling of the thin stellar disk. {\it Right-hand panel}: Variation of the three scale-length parameters as a function of the ratio $b/R_{\mathrm{d}}$. In the panels, the points represent the optimal solutions found after the application of the cross-entropy algorithm. Solid lines are fourth-order polynomial fits to the distribution of points.}
\label{fig:disksMN_thin}
\end{figure*}

Figure~\ref{fig:disksMN_thin} shows the variations of the ratios $M_{i}/M_{\mathrm{d}}$ (left-hand panel) and $a_{i}/R_{\mathrm{d}}$ (right-hand panel) as a function of the variation of the ratio $b/R_{\mathrm{d}}$ (filled circles), calculated for the modelling of the thin stellar disk. The subscript $i=1,\,2,\,3$ denotes each one of the 3 MN-disks used in the combination. The points corresponding to the ratio $b/R_{\mathrm{d}}=0$ denote the values of $M_{i}$ and $a_{i}$ which are the best-fitting solutions for the three Toomre-Kuzmin disks combination. The solid lines in each panel of Fig.~\ref{fig:disksMN_thin} represent 4th-order polynomial fits to the points that describe the variations of $M/M_{\mathrm{d}}$ and $a/R_{\mathrm{d}}$ as a function of $b/R_{\mathrm{d}}$. Each parameter $p=\frac{M}{M_{\mathrm{d}}}$ or $p=\frac{a}{R_{\mathrm{d}}}$ can then be written as a function of $b/R_{\mathrm{d}}$ in the form:

\begin{equation}
\label{eq:fit_poly4}
p=\sum_{j=0}^{4}{c_{j}\,\left(\frac{b}{R_{\mathrm{d}}}\right)^{j}}\,.
\end{equation}
\noindent
The values of the coefficients $c_0$ - $c_4$ for the 3 MN-disks fit models to the thin stellar disk, for the range of thicknesses $b/R_{\mathrm{d}}$ from 0 to 1.5, are given in Table~\ref{tab:coeff_poly_thin}.

\begin{figure*}
\centering
\includegraphics[scale=0.50]{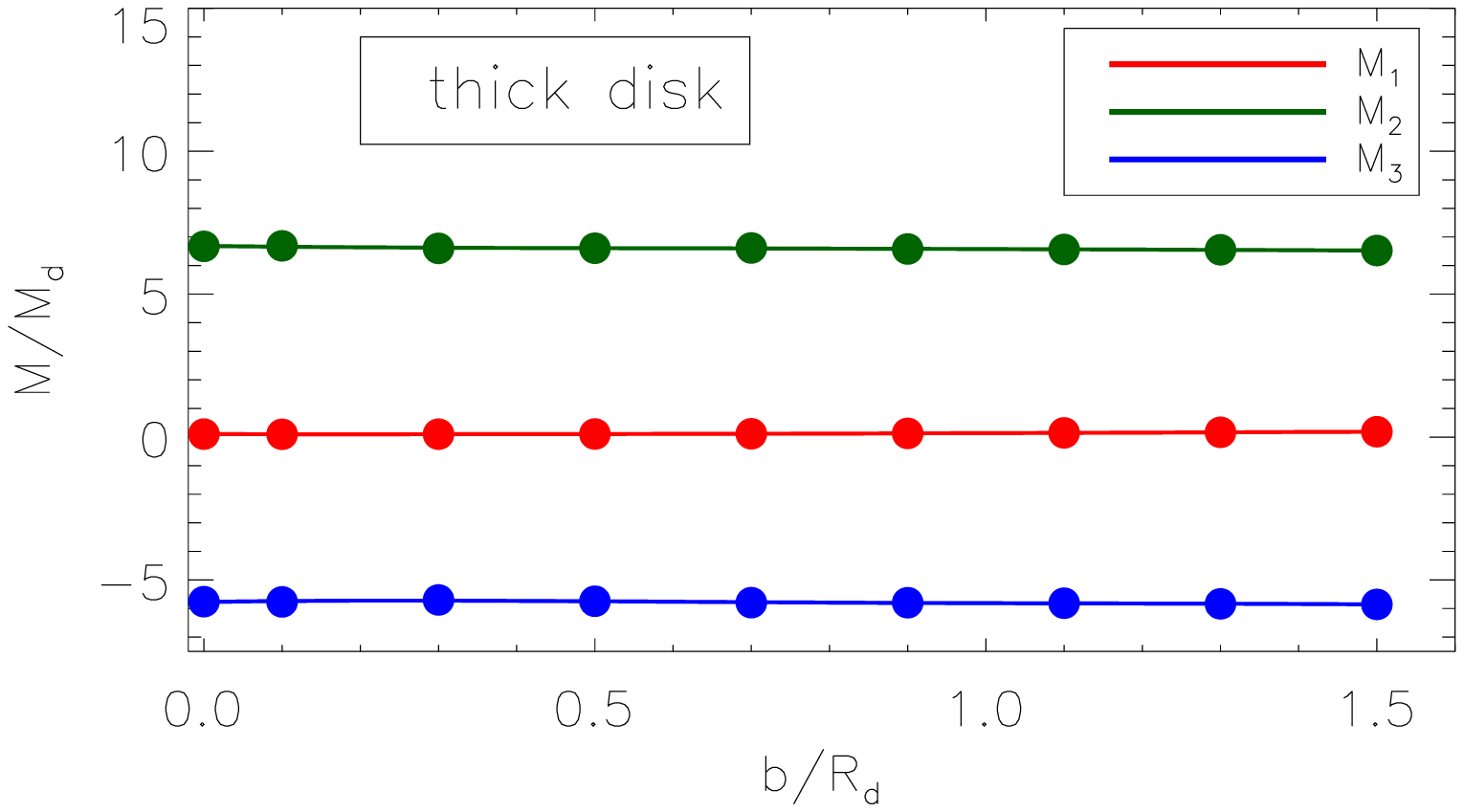}
\includegraphics[scale=0.50]{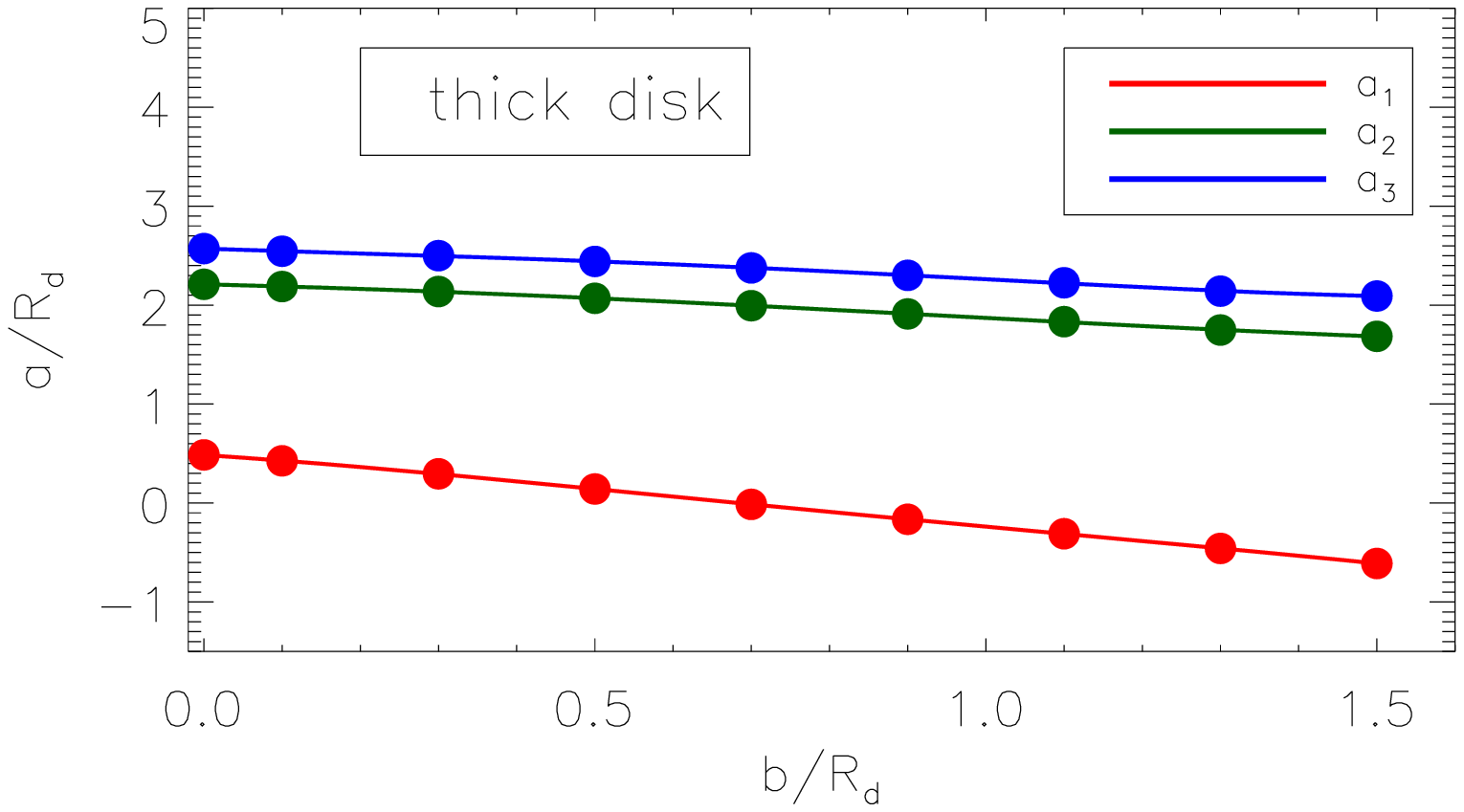}
\caption{The same as Fig.~\ref{fig:disksMN_thin}, but for the modelling of the thick stellar disk.}
\label{fig:disksMN_thick}
\end{figure*}

\begin{figure*}
\centering
\includegraphics[scale=0.50]{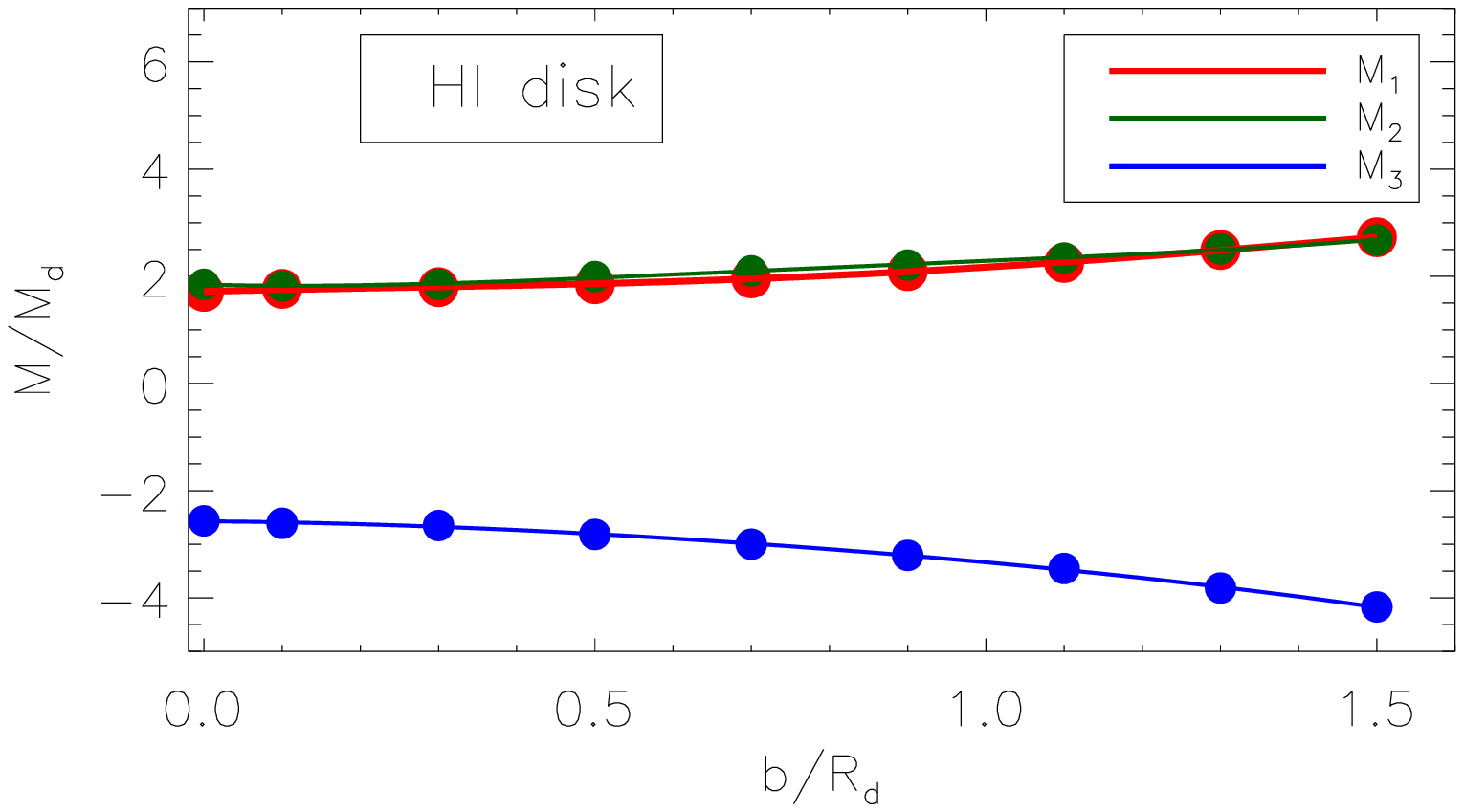}
\includegraphics[scale=0.50]{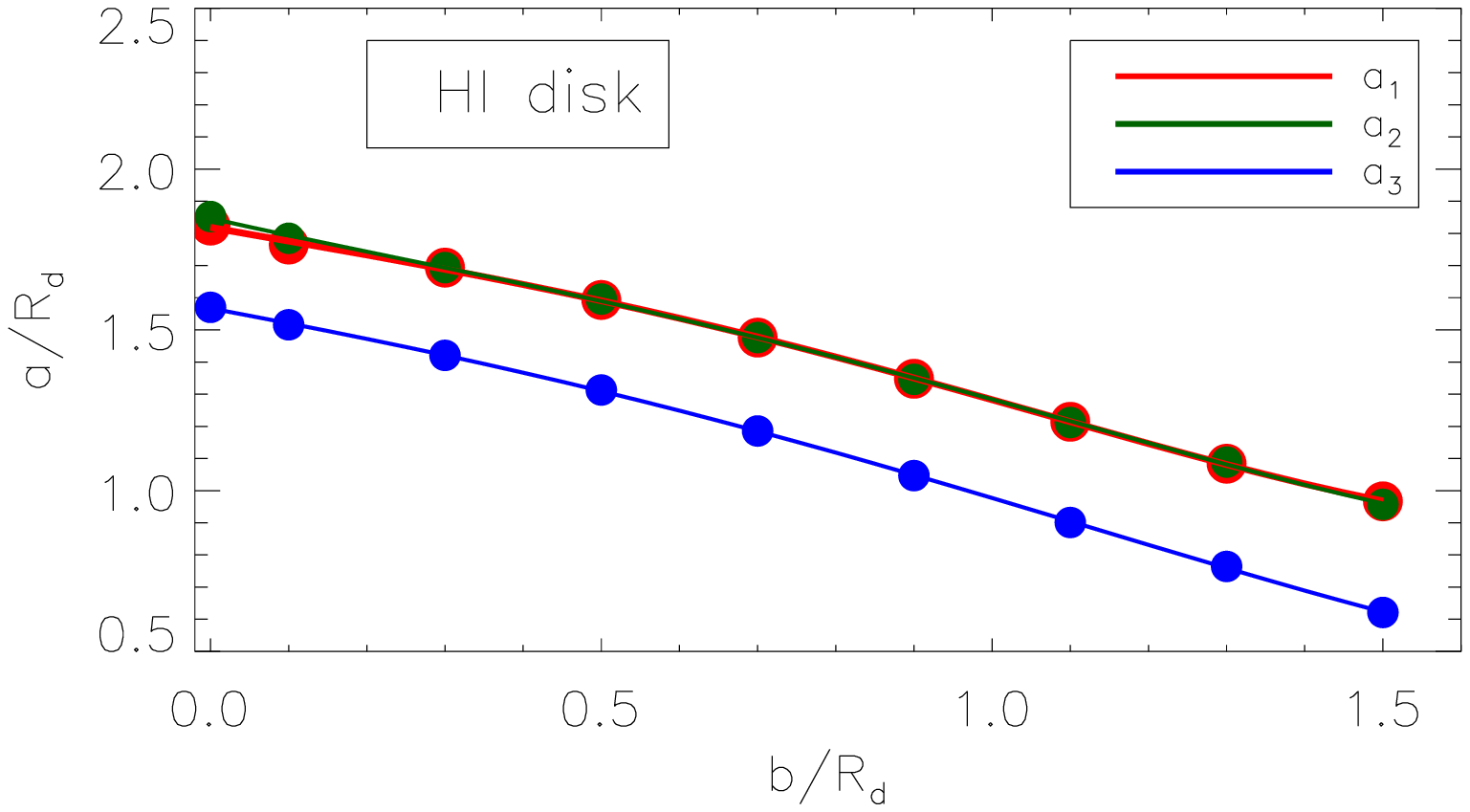}
\caption{The same as Fig.~\ref{fig:disksMN_thin}, but for the modelling of the H\,{\scriptsize I} disk.}
\label{fig:disksMN_HI}
\end{figure*}

\begin{figure*}
\centering
\includegraphics[scale=0.50]{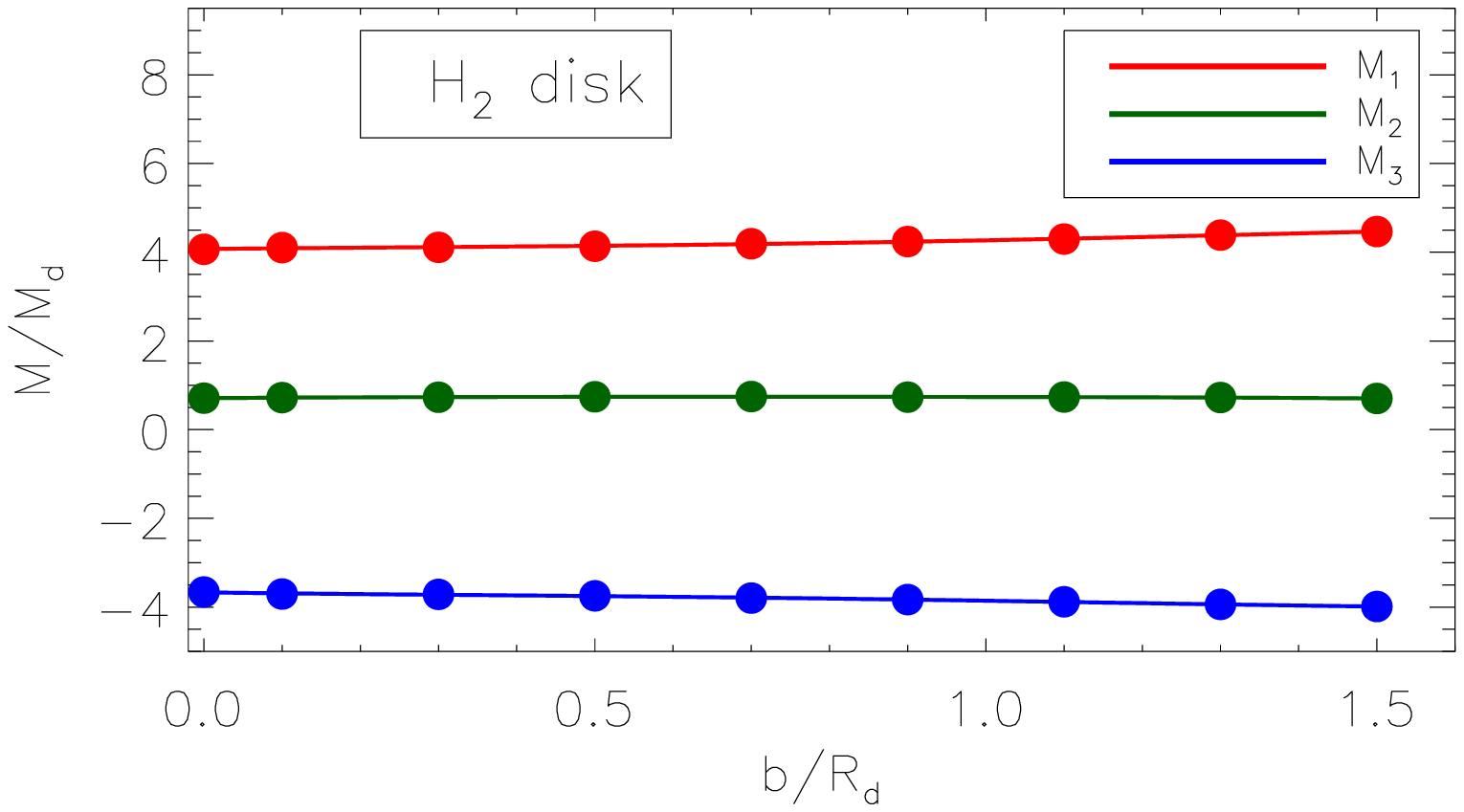}
\includegraphics[scale=0.50]{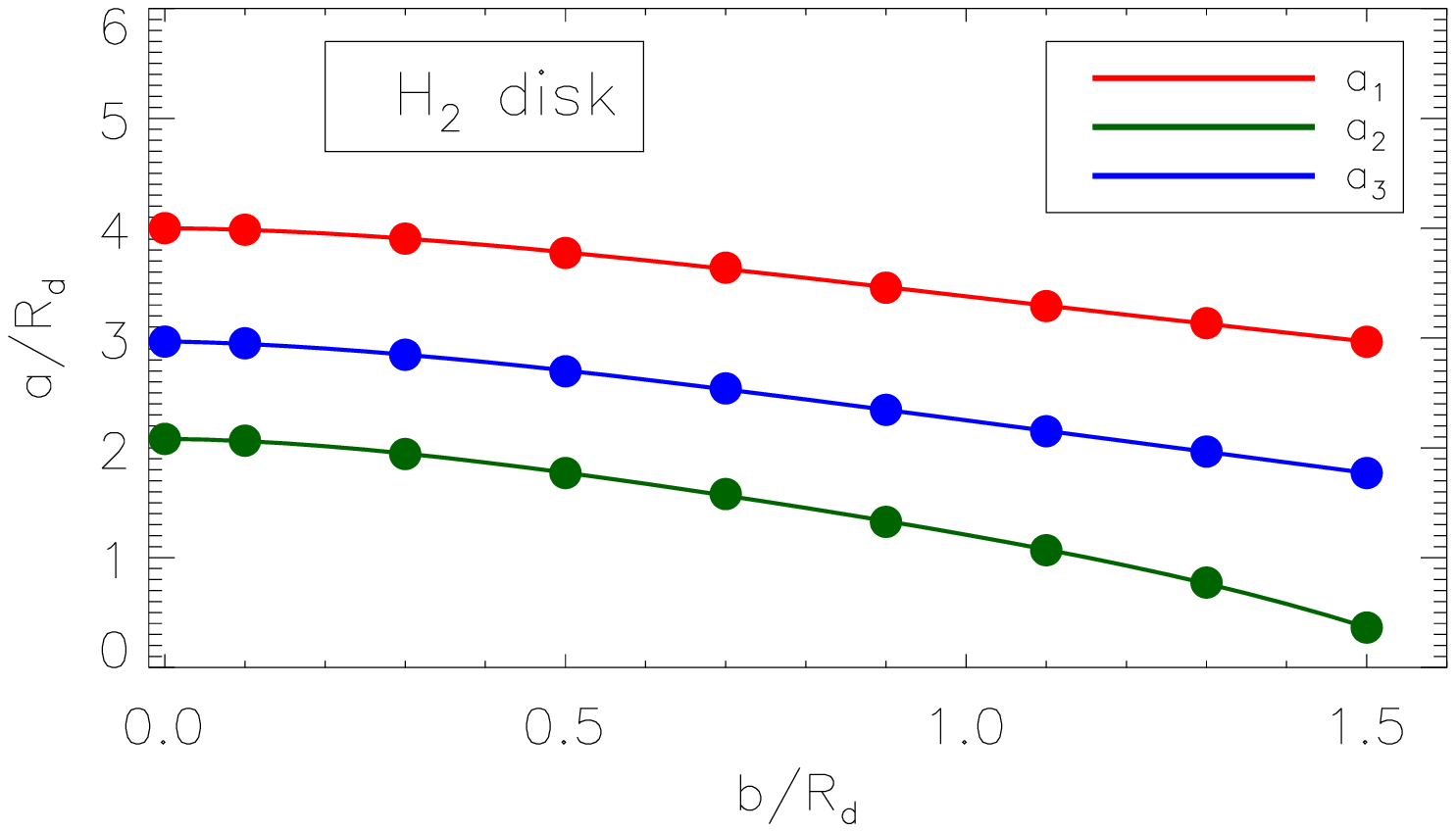}
\caption{The same as Fig.~\ref{fig:disksMN_thin}, but for the modelling of the H$_2$ disk.}
\label{fig:disksMN_H2}
\end{figure*}

Analogously, the relations between ($M_{i}/M_{\mathrm{d}}$; $a_{i}/R_{\mathrm{d}}$) and $b/R_{\mathrm{d}}$ for the thick stellar disk, the H\,{\scriptsize I} disk, and the H$_2$ disk are shown in Figs.~\ref{fig:disksMN_thick},~\ref{fig:disksMN_HI}, and~\ref{fig:disksMN_H2}, respectively. The solid lines in these figures also represent 4th-order polynomial fits to the distribution of points shown as filled circles, and are all also written in the form of Eq.~\ref{eq:fit_poly4}. The coefficients $c_j$ are given in Tables~\ref{tab:coeff_poly_thick},~\ref{tab:coeff_poly_HI} and~\ref{tab:coeff_poly_H2}, for the thick disk, the H\,{\scriptsize I} disk and the H$_2$ disk, respectively.

\begin{table*}
\caption{Coefficients of the fourth-order polynomial fits (Eq.~\ref{eq:fit_poly4_thickness}) to the variations of $b/R_{\mathrm{d}}$ as a function of $h_{z}/R_{\mathrm{d}}$ for the modelling of each disk subcomponent, as shown in Fig.~\ref{fig:disksMN-hzxb}.}             
\label{tab:coeff_poly_thickness}      
\centering          
\begin{tabular}{c c c c c c}
\hline\hline       

{\bf Component} & $k_0$ & $k_1$ & $k_2$ & $k_3$ & $k_4$ \\ 
\hline                   
thin disk & 0.0132 & 0.9995 & 0.4828 & 0.0752 & -0.1284 \\
thick disk & 0.0078 & 1.0648 & 0.6067 & -0.3934 & 0.0660 \\
H\,{\scriptsize I} disk & -0.0064 & 1.1169 & -0.0776 & -0.1604 & 0.0631 \\
H$_2$ disk & -0.0003 & 1.2694 & 0.2919 & 0.1023 & -0.0935 \\
\hline                  
\end{tabular}
\end{table*}

\begin{table*}
\caption{Coefficients of the fourth-order polynomial fits (Eq.~\ref{eq:fit_poly4}) to the variation of each of the six parameters ($M_{i}/M_{\mathrm{d}}$ and $a_{i}/R_{\mathrm{d}}$) as a function of $b/R_{\mathrm{d}}$ for the modelling of the {\bf thin stellar disk} based on the 3 MN-disks model shown in Fig.~\ref{fig:disksMN_thin}.}             
\label{tab:coeff_poly_thin}      
\centering          
\begin{tabular}{c c c c c c}
\hline\hline       

{\bf Parameter} & $c_0$ & $c_1$ & $c_2$ & $c_3$ & $c_4$ \\ 
\hline                   
$M_{1}/M_{\mathrm{d}}$ & 0.8211 & 0.3237 & -1.0203 & 1.1239 & -0.2535 \\
$M_{2}/M_{\mathrm{d}}$ & 0.8887 & -0.2588 & 0.8416 & -1.2136 & 0.4836 \\
$M_{3}/M_{\mathrm{d}}$ & -0.7007 & 0.1463 & -0.0123 & 0.0789 & -0.1921 \\
$a_{1}/R_{\mathrm{d}}$ & 1.8576 & -0.8626 & 0.8327 & -1.3109 & 0.6661 \\
$a_{2}/R_{\mathrm{d}}$ & 3.9650 & 2.1244 & -4.3408 & 3.4060 & -1.4135 \\
$a_{3}/R_{\mathrm{d}}$ & 1.5509 & -1.3112 & 2.0039 & -3.2060 & 1.3294 \\
\hline                  
\end{tabular}
\end{table*}

\begin{table*}
\caption{Coefficients of the fourth-order polynomial fits (Eq.~\ref{eq:fit_poly4}) to the variation of each of the six parameters ($M_{i}/M_{\mathrm{d}}$ and $a_{i}/R_{\mathrm{d}}$) as a function of $b/R_{\mathrm{d}}$ for the modelling of the {\bf thick stellar disk} based on the 3 MN-disks model shown in Fig.~\ref{fig:disksMN_thick}.}             
\label{tab:coeff_poly_thick}      
\centering          
\begin{tabular}{c c c c c c}
\hline\hline       

{\bf Parameter} & $c_0$ & $c_1$ & $c_2$ & $c_3$ & $c_4$ \\ 
\hline                   
$M_{1}/M_{\mathrm{d}}$ & 0.1014 & -0.0327 & 0.0942 & -0.0304 & 0.0047 \\
$M_{2}/M_{\mathrm{d}}$ & 6.6861 & -0.3237 & 0.5063 & -0.3897 & 0.0989 \\
$M_{3}/M_{\mathrm{d}}$ & -5.7721 & 0.4544 & -1.2467 & 1.0481 & -0.2958 \\
$a_{1}/R_{\mathrm{d}}$ & 0.4853 & -0.5214 & -0.5232 & 0.4506 & -0.1299 \\
$a_{2}/R_{\mathrm{d}}$ & 2.2102 & -0.1937 & -0.1880 & 0.0157 & 0.0265 \\
$a_{3}/R_{\mathrm{d}}$ & 2.5712 & -0.2637 & 0.1422 & -0.3246 & 0.1365 \\
\hline                  
\end{tabular}
\end{table*}

\begin{table*}
\caption{Coefficients of the fourth-order polynomial fits (Eq.~\ref{eq:fit_poly4}) to the variation of each of the six parameters ($M_{i}/M_{\mathrm{d}}$ and $a_{i}/R_{\mathrm{d}}$) as a function of $b/R_{\mathrm{d}}$ for the modelling of the {\bf H\,{\scriptsize I} disk} based on the 3 MN-disks model shown in Fig.~\ref{fig:disksMN_HI}.}             
\label{tab:coeff_poly_HI}      
\centering          
\begin{tabular}{c c c c c c}
\hline\hline       

{\bf Parameter} & $c_0$ & $c_1$ & $c_2$ & $c_3$ & $c_4$ \\ 
\hline                   
$M_{1}/M_{\mathrm{d}}$ & 1.7159 & 0.3717 & -0.5278 & 0.8551 & -0.2456 \\
$M_{2}/M_{\mathrm{d}}$ & 1.8429 & -0.3901 & 1.9922 & -1.6618 & 0.5041 \\
$M_{3}/M_{\mathrm{d}}$ & -2.5705 & -0.1235 & -0.8102 & 0.2631 & -0.0957 \\
$a_{1}/R_{\mathrm{d}}$ & 1.8188 & -0.4340 & 0.0823 & -0.3249 & 0.1405 \\
$a_{2}/R_{\mathrm{d}}$ & 1.8468 & -0.5471 & 0.2117 & -0.3611 & 0.1336 \\
$a_{3}/R_{\mathrm{d}}$ & 1.5671 & -0.4595 & -0.0567 & -0.1456 & 0.0719 \\
\hline                  
\end{tabular}
\end{table*}

\begin{table*}
\caption{Coefficients of the fourth-order polynomial fits (Eq.~\ref{eq:fit_poly4}) to the variation of each of the six parameters ($M_{i}/M_{\mathrm{d}}$ and $a_{i}/R_{\mathrm{d}}$) as a function of $b/R_{\mathrm{d}}$ for the modelling of the {\bf H}$_{\mathbf 2}$ {\bf disk} based on the 3 MN-disks model shown in Fig.~\ref{fig:disksMN_H2}.}             
\label{tab:coeff_poly_H2}      
\centering          
\begin{tabular}{c c c c c c}
\hline\hline       

{\bf Parameter} & $c_0$ & $c_1$ & $c_2$ & $c_3$ & $c_4$ \\ 
\hline                   
$M_{1}/M_{\mathrm{d}}$ & 4.0738 & 0.1725 & -0.1944 & 0.3135 & -0.0961 \\
$M_{2}/M_{\mathrm{d}}$ & 0.7103 & 0.1239 & -0.1807 & 0.1250 & -0.0410 \\
$M_{3}/M_{\mathrm{d}}$ & -3.6689 & -0.2371 & 0.2933 & -0.3647 & 0.1202 \\
$a_{1}/R_{\mathrm{d}}$ & 4.0004 & -0.0632 & -1.0066 & 0.5599 & -0.1112 \\
$a_{2}/R_{\mathrm{d}}$ & 2.0821 & -0.0541 & -1.6461 & 1.2543 & -0.4278 \\
$a_{3}/R_{\mathrm{d}}$ & 2.9702 & -0.1452 & -1.0342 & 0.5826 & -0.1227 \\
\hline                  
\end{tabular}
\end{table*}


\section{Radial and vertical components of the gradients of the gravitational potential models for each Galactic component}
\label{app2}

Here we give the explicit forms for the radial and vertical components of the gradients of the gravitational potential expressions modelled for each Galactic mass component, and which can be used in the construction of the gravitational force-field model of the Galaxy.

\subsection*{\bf The Miyamoto-Nagai disks}

In the following, we write the components of the gradients for
the three models of Miyamoto-Nagai gravitational potential, which are expressed by Eqs.~\ref{eq:Phi_MN1},~\ref{eq:Phi_MN2} and~\ref{eq:Phi_MN3} in the main text. In all expressions, we use the identity $\zeta=\sqrt{z^{2}+b^{2}}$.

\subsubsection*{Model 1 (Eq.~\ref{eq:Phi_MN1})}

\begin{equation}
\label{eq:dPhi_MN1_dR}
\frac{\partial\Phi_{\mathrm{MN}_{1}}}{\partial R}=\frac{GMR}{\left[R^{2}+\left(a+\zeta\right)^{2}\right]^{3/2}}
\end{equation}

\begin{equation}
\label{eq:dPhi_MN1_dz}
\frac{\partial\Phi_{\mathrm{MN}_{1}}}{\partial z}=\frac{GM\,z\left(a+\zeta\right)}{\zeta\left[R^{2}+\left(a+\zeta\right)^{2}\right]^{3/2}}
\end{equation}

\subsubsection*{Model 2 (Eq.~\ref{eq:Phi_MN2})}

\begin{equation}
\label{eq:dPhi_MN2_dR}
\frac{\partial\Phi_{\mathrm{MN}_{2}}}{\partial R}=\frac{GMR}{\left[R^{2}+\left(a+\zeta \right)^{2}\right]^{5/2}}\left[R^{2}+\left(a+\zeta\right)^{2}+3a\left(a+\zeta\right)\right]
\end{equation}

\begin{multline}
\label{eq:dPhi_MN2_dz}
\frac{\partial\Phi_{\mathrm{MN}_{2}}}{\partial z}=\frac{GM\,z}{\zeta\left[R^{2}+\left(a+\zeta \right)^{2}\right]^{5/2}}\left[\zeta R^{2}+\left(a+\zeta\right)^{3}+\right.\\
+\left. 2a\left(a+\zeta\right)^{2}\right]
\end{multline}

\subsubsection*{Model 3 (Eq.~\ref{eq:Phi_MN3})}

\begin{multline}
\label{eq:dPhi_MN3_dR}
\frac{\partial\Phi_{\mathrm{MN}_{3}}}{\partial R}=\frac{GMR}{\left[R^{2}+\left(a+\zeta \right)^{2}\right]^{7/2}}\left\{\left[R^{2}+\left(a+\zeta\right)^{2}\right]^{2}+3a\zeta R^{2}+ \right.\\
+ \left. 3a\left(a+\zeta\right)^{3}+4a^{2}\left(a+\zeta\right)^{2}+2a^{2}R^{2} \right\}
\end{multline}

\begin{multline}
\label{eq:dPhi_MN3_dz}
\frac{\partial\Phi_{\mathrm{MN}_{3}}}{\partial z}=\frac{GM\,z}{\zeta\left[R^{2}+\left(a+\zeta \right)^{2}\right]^{7/2}}\left\{\left\{\left[R^{2}+\left(a+\zeta\right)^{2}\right]^{2}-2a^{2}R^{2}+ \right.\right.\\
+ \left.\left. a\zeta R^{2}\right\}(a+\zeta)+2a^{2}\left(a+\zeta\right)^{3}+2a\left(a+\zeta\right)^{4}-aR^{4} \right\}
\end{multline}


\

\subsection*{{\bf The bulge} (Eq.~\ref{eq:Phi_bulge})}

\begin{equation}
\label{eq:dPhib_dR}
\frac{\partial\Phi_{\mathrm{b}}}{\partial R}=\frac{GM_{\mathrm{b}}}{\left(\sqrt{R^{2}+z^{2}}+a_{\mathrm{b}}\right)^{2}}\frac{R}{\sqrt{R^{2}+z^{2}}}
\end{equation}

\begin{equation}
\label{eq:dPhib_dz}
\frac{\partial\Phi_{\mathrm{b}}}{\partial z}=\frac{GM_{\mathrm{b}}}{\left(\sqrt{R^{2}+z^{2}}+a_{\mathrm{b}}\right)^{2}}\frac{z}{\sqrt{R^{2}+z^{2}}}
\end{equation}


\

\subsection*{{\bf The dark halo} (Eq.~\ref{eq:Phi_halo})}

\begin{equation}
\label{eq:dPhih_dR}
\frac{\partial\Phi_{\mathrm{h}}}{\partial R}=\frac{v_{\mathrm{h}}^{2}\,R}{R^{2}+z^{2}+r_{\mathrm{h}}^{2}}
\end{equation}

\begin{equation}
\label{eq:dPhih_dz}
\frac{\partial\Phi_{\mathrm{h}}}{\partial z}=\frac{v_{\mathrm{h}}^{2}\,z}{R^{2}+z^{2}+r_{\mathrm{h}}^{2}}
\end{equation}


\

\subsection*{{\bf The ring potential} (Eq.~\ref{eq:pot_ring})}

With the functions $\varphi_{R}(R)$ and $\varphi_{z}(z)$ given in Eq.~\ref{eq:pot_ring}, we have:

\begin{equation}
\label{eq:dPhiring_dR}
\frac{\partial\Phi_{ring}}{\partial R}=\frac{A_{ring}\,\beta_{ring}\,\varphi_{z}(z)}{R}\,\mathrm{sech}\,\phi\left[ \tanh^{2}\phi - \mathrm{sech}^{2}\phi \right]
\end{equation}

\begin{equation}
\label{eq:dPhiring_dz}
\frac{\partial\Phi_{ring}}{\partial z}=-\frac{\varphi_{R}(R)}{h_{z_{ring}}}\mathrm{sech}\left(\frac{z}{h_{z_{ring}}}\right)\tanh\left(\frac{z}{h_{z_{ring}}}\right)
\end{equation}
with the argument $\phi=\ln \left(\dfrac{R}{R_{ring}}\right)^{\beta_{ring}}$ in Eq.~\ref{eq:dPhiring_dR}


\end{document}